\newif\ifdraft\draftfalse
\newif\iflong\longtrue

\iflong
\documentclass[10pt,numbers,nocopyrightspace]{sigplanconf}
\else
\documentclass[10pt,numbers]{sigplanconf}
\usepackage[final]{oopsla2016}
\fi

\usepackage{listings}       \usepackage{graphicx}
\usepackage{amsmath}
\usepackage{amsthm}
\usepackage{array}
\usepackage{tikz}
\usetikzlibrary{positioning}

\usepackage{pervasives}
\usepackage{enumitem}
\usepackage{flushend}

\begin{document}
\toappear{}

\iflong
\title{Type Inference for Static Compilation of JavaScript (Extended Version)}
\fi

\newcolumntype{K}[1]{>{\centering\arraybackslash}p{#1}}
\authorinfo{Satish Chandra${}^*$\and Colin S. Gordon${}^\dagger$\and
                 Jean-Baptiste Jeannin${}^*$\and Cole Schlesinger${}^*$\\
                 Manu Sridharan${}^*$\and Frank Tip${}^\ddag$\and Youngil Choi${}^\S$}
  {
  \begin{tabular}{K{8cm} K{8cm}}
    ${}^*$Samsung Research America, USA &
    ${}^\dagger$Drexel University, USA
    \vspace{2pt}\\
    {\sf \{schandra,jb.jeannin,cole.s,m.sridharan\}@samsung.com} &
    {\sf csgordon@cs.drexel.edu}
    \vspace{5pt}\\
    ${}^\ddag$Northeastern University, USA &
    ${}^\S$Samsung Electronics, South Korea
    \vspace{2pt}\\
    {\sf f.tip@northeastern.edu} &
    {\sf duddlf.choi@samsung.com}
  \end{tabular}
  }
  {}

\maketitle

\begin{abstract}

  We present a type system and inference algorithm for a rich subset
  of JavaScript equipped with objects, structural subtyping, prototype inheritance, and first-class methods.  The type system supports
  abstract and recursive objects, and is expressive enough to
  accommodate several standard benchmarks with only minor workarounds.
  The invariants enforced by the types enable an ahead-of-time
  compiler to carry out optimizations typically beyond the reach of
  static compilers for dynamic languages.  Unlike previous inference
  techniques for prototype inheritance, our algorithm uses a
  combination of lower and upper bound propagation to infer types and
  discover type errors in \emph{all} code, including uninvoked
  functions.  The inference is expressed in a simple constraint
  language, designed to leverage off-the-shelf fixed point solvers.
  We prove soundness for both the type system and inference algorithm.
  An experimental evaluation showed that the inference is powerful,
  handling the aforementioned benchmarks with no manual type annotation, and
  that the inferred types enable effective static compilation.

\end{abstract}



\makeatletter{}
\section{Introduction}\label{sec:intro}

JavaScript is one of the most popular programming languages currently in
use~\cite{redmonk}. It has become the de facto standard in web programming, and
its growing use in large-scale, real-world applications---ranging from servers
to embedded devices---has sparked significant interest in JavaScript-focused program analyses and
type systems in both the academic research community and in
industry.

In this paper, we report on a type inference algorithm for JavaScript developed
as part of a larger, ongoing effort to enable type-based \textit{ahead-of-time}
compilation of JavaScript programs.  Ahead-of-time compilation has the potential to enable lighter-weight execution,
compared to runtimes that rely on just-in-time optimizations~\cite{jsCore,v8},
without compromising performance.  This is particularly relevant for
resource-constrained devices such as mobile phones where both performance and
memory footprint are important.  Types are key to doing
effective optimizations in an ahead-of-time compiler.

JavaScript is famously dynamic; for example, it contains \texttt{eval} for
runtime code generation and supports introspective behavior, features that are
at odds with static compilation. Ahead-of-time compilation of unrestricted
JavaScript is \textit{not} our goal. Rather, our goal is to compile a subset that is rich enough for idiomatic use by JavaScript developers.
Although JavaScript code that uses
highly dynamic features does exist~\cite{Richards:2010},
data shows that
the majority of application code does not require that flexibility.
With growing interest in using JavaScript across a range of devices,
including resource-constrained devices, it is important to examine the
tradeoff between language flexibility and the cost of implementation.

The JavaScript compilation scenario imposes several desiderata for a
type system.  First, the types must be \emph{sound}, so they can be
relied upon for compiler transformations.  Second, the types must
impose enough restrictions to allow the compiler to generate code with
good, predictable performance for core language constructs
(\autoref{sec:background} discusses some of these optimizations).  At
the same time, the system must be expressive enough to type check
idiomatic coding patterns and make porting of
mostly-type-safe JavaScript code easy.  Finally, in keeping with the
nature of the language, as well as to ease porting of existing code,
we desire powerful type inference.
To meet developer
expectations, the inference must infer types and
discover type errors in \emph{all} code, including uninvoked functions
from libraries or code under development.

No existing work on JavaScript type systems and inference meets our needs entirely.  Among
the recently developed type systems, TypeScript~\cite{typescript}
and Flow~\cite{flow} both have rich type systems that focus on programmer
productivity at the expense of soundness. TypeScript relies heavily on programmer annotations to be
effective, and both treat inherited properties too imprecisely for efficient code generation.
Defensive JavaScript~\cite{djs} has a sound type system and type inference,
and Safe TypeScript~\cite{safets} extends TypeScript with a sound type system,
but neither supports prototype inheritance.  TAJS~\cite{JensenMT09} is
sound and handles prototype inheritance precisely, but it does not
compute types for uninvoked functions; the classic work on type
inference for \textsc{Self}~\cite{SelfTypeInference} has the same drawback.
Choi \etal~\cite{DBLP:conf/sas/ChoiCNS15} present a JavaScript
type system targeting ahead-of-time compilation that forms the basis
of our type system, but their work does not have inference and instead relies on programmer annotations.
See \autoref{sec:related} for further discussion of related work.

This paper presents a type system and inference algorithm for a rich
subset of JavaScript
that achieves our goals.
Our type system builds upon that of Choi et
al.~\cite{DBLP:conf/sas/ChoiCNS15}, adding support for abstract objects, first-class methods, and recursive objects, each of which we found crucial for handling real-world JavaScript idioms; we prove these extensions sound.
The type system supports a number
of additional features such as polymorphic arrays, operator overloading, and
intersection types in manually-written interface descriptors for library code, which is important for building GUI applications.

Our type inference technique builds on existing literature (e.g.,~\cite{pottier-ic-01, SelfTypeInference,
iogti-RCH12}) to handle a complex combination of language features, including
structural subtyping, prototype inheritance, first-class methods,
and recursive types; we are unaware of any single previous type inference technique that soundly handles these features in combination.

We formulate type inference as a constraint satisfaction problem over
a language  composed primarily of subtype constraints over standard row
variables.  Our formulation shows that various aspects of the
type system, including source-level subtyping, prototype inheritance,
and attaching methods to objects, can all be reduced to these
simple subtype constraints.  Our constraint solving algorithm first
computes lower and upper bounds of type variables through a
propagation phase (amenable to the use of efficient, off-the-shelf
fixed-point solvers), followed by a straightforward error-checking and
ascription phase.  Our use of both lower \emph{and} upper bounds
enables type inference and error checking for uninvoked functions,
unlike previous inference techniques supporting prototype
inheritance~\cite{JensenMT09,SelfTypeInference}.  As shown in
\autoref{sec:complications}, sound inference for our type system is
non-trivial, particularly for uninvoked functions; we prove
that our inference algorithm is sound.

Leveraging inferred types, we have built a backend that compiles
type-checked JavaScript programs to optimized native binaries for both PCs and
mobile devices. We have compiled slightly-modified versions of six of the
Octane benchmarks~\cite{Octane}, which ranged from 230 to 1300 LOC,
using our compiler. The modifications needed for the programs to type
check were minor (see \autoref{sec:cases} for details).

Preliminary data suggests that for resource-constrained devices, trading off some language
flexibility for static compilation is a compelling proposition.  With ahead-of-time compilation
(AOTC), the six Octane programs incurred a \textit{significantly} smaller memory footprint compared
to running the same JavaScript sources with a just-in-time optimizing engine. The execution
performance is not as fast as JIT engines when the programs run for a large number of iterations,
but is acceptable otherwise, and vastly better than a non-optimizing interpreter (details in \autoref{sec:cases}).

We have also created six GUI-based
applications for the Tizen~\cite{tizen} platform,
reworking from existing web applications; these programs ranged between
250 to 1000 lines of code. In all cases, all types in the user-written JavaScript code were inferred, and \emph{no} explicit annotations were required.
We do require annotated signatures
of library functions, and we have created these for many of the JavaScript standard
libraries as well as for the Tizen platform API.  Experiences with and
limitations of our system are discussed in \autoref{sec:cases}.

\mypara{Contributions:}
\begin{itemize}
\item We present a type system, significantly extending previous
  work~\cite{DBLP:conf/sas/ChoiCNS15}, for typing common JavaScript
  inheritance patterns, and we prove the type system sound.  Our system strikes a useful balance between allowing common coding patterns and enabling ahead-of-time compilation.

\item We present an inference algorithm for our type system and prove
  it sound.  To our best knowledge, this algorithm is the first to
  handle a combination of structural subtyping, prototype inheritance,
  abstract types, and recursive types, while also inferring types for
  uninvoked functions.  This inference algorithm may be of
  independent interest, for example, as a basis for software
  productivity tools.

\item We discuss our experiences with applying an ahead-of-time
  compiler based on our type inference to several existing benchmarks.
  We found that  our inference could infer
  all the necessary types for these benchmarks automatically with only
  slight modifications.  We also
  found that handling a complex combination of type system features
  was crucial for these programs.
  Experimental data points to the promise of ahead-of-time compilation for running JavaScript on resource-constrained devices.
       
\end{itemize}

\makeatletter{}
\section{Overview}
\label{sec:overview}

Here we give an overview of our
type system and inference.  We illustrate some requirements and
features of typing and type inference by way of a simple example.  We
also highlight some challenges in inference, and show in more detail
why previous techniques are insufficient for our needs.

\subsection{Type System Requirements}
\label{sec:background}

\begin{figure}
\begin{lstlisting}
var v1 = { d : 1, // o1 /*@\label{li:o1-alloc}@*/
      m : function (x) { this.a = x + this.d }}/*@\label{li:m-attach}@*/
var v2 = { a : 2 } proto v1; // o2 /*@\label{li:o2-alloc}@*/
v2.m(3); /*@ \label{li:m-call-int} @*/
v2.m("foo"); // type error in our system /*@ \label{li:m-call-str} @*/
var v3 = { b : 4 } proto v2; // o3 /*@ \label{li:o3-alloc} @*/
v3.m(4);      // type error in our system /*@ \label{li:call-on-abstr} @*/
\end{lstlisting}
\caption{An example program to illustrate our type system.
}
\label{fig:ts-ex-code}
\end{figure}

Our type system prevents certain dynamic JavaScript behaviors that can
compromise performance in an AOTC
scenario.  In many cases, such behaviors also reflect latent program
bugs.  Consider the example of \autoref{fig:ts-ex-code}.
We refer to the object literals as \code{o1}, \code{o2}, and
\code{o3}.  To keep our examples readable, we use a
syntactic sugar for prototype inheritance: the expression
\lstinline|{a : 2} proto o1| makes \code{o1} the prototype parent of
the \lstinline|{a: 2}| object, corresponding to the following
JavaScript:
\begin{lstlisting}[numbers=none]
function C() { this.a = 2 } // constructor
C.prototype = o1; new C()
\end{lstlisting}

In JavaScript, the \lstinline{v2.m("foo")} invocation
(\autoref{li:m-call-str}) runs without error, setting \code{v2.a} to
\code{"foo1"}.  In SJS, we do not allow this operation, as \code{v2.a} was
initialized to an integer value; such restrictions are standard with static typing.

\begin{figure}
\includegraphics[width=\linewidth,trim=60 350 0 0,clip=true]{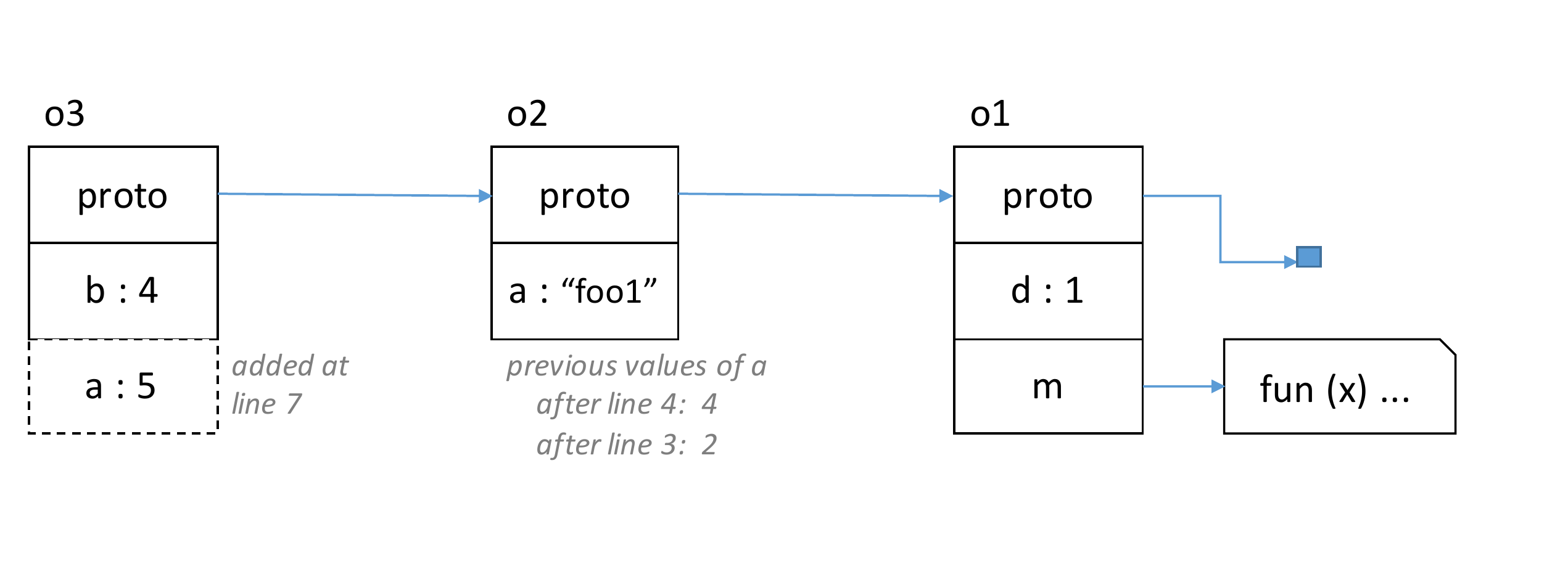}
\vspace{-15pt}
\caption{Runtime heap for \autoref{fig:ts-ex-code} at \autoref{li:o3-alloc}.
}
\label{fig:heap}
\end{figure}

JavaScript field accesses also present a challenge for AOTC.  \autoref{fig:heap} shows a runtime heap layout of the three objects allocated
in \autoref{fig:ts-ex-code} (after \autoref{li:o3-alloc}). In JavaScript, a field read \code{x.f} first
checks \code{x} for field \code{f}, and continues up \code{x}'s
prototype chain until \code{f} is found.  If \code{f} is not found, the read evaluates to
\code{undefined}.  Field writes \code{x.f  = y} are peculiar.  If \code{f} exists
in \code{x}, it is updated in place.  If not, \code{f} is created in \code{x},
\emph{even if} \code{f} is available up the prototype chain.  This peculiarity
is often a source of bugs.  For our example, the write to
\lstinline{this.a} within the invocation
\lstinline{v3.m(4)} on \autoref{li:call-on-abstr} creates a new
slot in \code{o3} (dashed box in \autoref{fig:heap}), rather than updating \code{o2.a}.

Besides being a source of bugs, this field write behavior prevents a compiler from
optimizing field lookups.  If the set of fields in every object is fixed at the
time of allocation---a \textit{fixed object layout}~\cite{DBLP:conf/sas/ChoiCNS15}---then the compiler can use a constant indirection table for field offsets.\footnote{The compiler may even be able to allocate a field in the same position
in all containing objects, eliminating the indirection table.}
Fixed layout also establishes
the availability of fields for reading / writing, obviating the need for runtime
checks.\footnote{When dynamic addition and deletion of fields is necessary, a
map rather than an object is more suited; see \autoref{sec:practical-appendix}.}

In summary, our type system must enforce the following properties:
\begin{itemize} \item \textbf{Type compatibility}, e.g., integer and
string values cannot be assigned to the same variable.
\item \textbf{Access safety} of object fields:
fields that are neither available locally nor in the prototype chain cannot be read; and
fields that are not locally available cannot be written.
\end{itemize}
These properties promote good programming practices and make code more amenable to compilation.  Note that detection of errors that require flow-sensitive reasoning, like \code{null} dereferences, is out of scope for our type system; extant systems like TAJS~\cite{JensenMT09} can be applied to find such issues.

\subsection{The Type System}
\label{sec:typesystem}

\mypara{Access safety.} In our type system, the fields in an object type $O$ are maintained as
two rows (maps from field names to types), $\sr{O}$ for
\emph{readable} fields and $\sw{O}$ for \emph{writeable} fields.
Readable fields are those present either locally or in the prototype
chain, while writeable fields must be present locally (and hence must also be readable).  Since
\code{o1} in \autoref{fig:ts-ex-code} only has local fields $d$ and $m$, we have
$\sr{O_1} = \sw{O_1} = \rowtype{d,m}$.\footnote{For
  brevity, we elide the field types here, as the discussion focuses on
  which fields are present.}   For \code{o2}, the readable fields $\sr{O_2}$
include local fields $\rowtype{a}$ and fields $\rowtype{d,m}$ inherited from \code{o1}, so we have $\sr{O_2} = \rowtype{d,m,a}$ and $\sw{O_2} = \rowtype{a}$.  Similarly,
$\sr{O_3} = \rowtype{d,m,a,b}$ and
$\sw{O_3} = \rowtype{b}$.
The type system rejects writes to
read-only fields; e.g., \lstinline{v2.d = 2} would be rejected.

Detecting that the call \lstinline{v3.m(4)} on
\autoref{li:call-on-abstr} violates access safety is less
straightforward.  To handle this case, the type system tracks two
additional rows for certain object types: the fields that attached
methods may read ($\smr{O}$), and those that methods may write
($\smw{O}$).  The typing rules ensure that such \emph{method-accessed
  fields} for an object type include the fields of the receiver types
for all attached methods.  Let $T_m$ be the receiver type for
method \code{m} (\autoref{li:m-attach}).  Based on the uses of
\code{this} within \code{m}, we have $\sr{T_m} = \rowtype{d,a}$ and
$\sw{T_m} = \rowtype{a}$ (again, writeable fields must be readable).
Since \code{m} is the only method attached to \code{o1}, we have $\smr{O_1} = \sr{T_m} = \rowtype{d,a}$ and $\smw{O_1} = \sw{T_m} = \rowtype{a}$.  Since \code{o2} and \code{o3} inherit \code{m} and have no other methods, we also have $\smr{O_3} = \smr{O_2} = \rowtype{d,a}$ and $\smw{O_3} = \smw{O_2} = \rowtype{a}$.

With these types, we have \( \mathtt{a} \in \smw{O_3} \) and
\( \mathtt{a} \not\in \sw{O_3} \): i.e., a method of $O_3$ can write field
field \code{a}, which is not locally present.  Hence, the method call
\lstinline{v3.m(4)} is unsafe.  The type system considers $O_3$ to be
\emph{abstract}, and method invocations on abstract types are
rejected.  (Types for which method invocations are safe are
\emph{concrete}.) Similarly, $O_1$ is also abstract.
Note that rejecting abstract types completely is too
restrictive: JavaScript code often has prototype objects that are abstract, with methods
referring to fields declared only in inheritors.

The idea of tracking method-accessed fields follows the type system of Choi et
al.~\cite{DBLP:conf/sas/ChoiCNS15,SJSTechReport}, but they did not distinguish
between \kw{mr} and \kw{mw}, essentially placing all accessed fields in \kw{mw}.
Their treatment would reject the safe call at \autoref{li:m-call-int}, whereas with \kw{mr}, we are able to type it.\footnote{Throughout the paper, we call out extensions we made to enhance the power of Choi et al.'s type system.}

\mypara{Subtyping.} A type system for JavaScript must also support structural
subtyping between object types to handle common idioms.
But, a conflict arises between structural subtyping and tracking of
method-accessed fields.  Consider the following code:
\begin{lstlisting}
p = cond()
    ? { m : fun() { this.f = 1 }, f: 2 } // o1
    : { m : fun() { this.g = 2 }, g: 3 } // o2
p.m();
\end{lstlisting}
Both \code{o1} and \code{o2} have concrete types, as they contain all
fields accessed by their methods.  Since \code{m} is the only common
field between \code{o1} and \code{o2}, by structural subtyping,
\code{m} is the only field in the type of \code{p}.  But what
should the method-writeable fields of \code{p} be?  A sound approach
of taking the union of such fields from \code{o1} and
\code{o2} yields $\rowtype{f,g}$.  But, this makes the type of
\code{p} abstract (neither $f$ nor $g$ is present in \code{p}),
prohibiting the safe call of \code{p.m()}.

To address this issue, we adopt ideas from previous
work~\cite{DBLP:conf/sas/ChoiCNS15,DBLP:journals/iandc/PalsbergZ04}
and distinguish \emph{prototypal types}, suitable for prototype
inheritance, from \emph{non-prototypal types}, suitable for structural
subtyping.  Non-prototypal types elide method-accessed fields,
thereby avoiding bad interactions with structural
subtyping.  For the example above, we can assign \code{p} a non-prototypal
concrete type, thereby allowing the \code{p.m()} call.  However, an
expression \lstinline|{...} proto p| would be disallowed: without
method-accessed field information for \code{p}, inheritance
cannot be soundly handled.  For further details, see
\autoref{sec:subtyping}.

\subsection{Inference Challenges}\label{sec:complications}

As noted in \autoref{sec:intro}, we found that no extant type
inference technique was suitable for our needs.  The closest
techniques are those that reason about prototype inheritance
precisely, like type inference for
\textsc{Self}~\cite{SelfTypeInference} and the TAJS system for
JavaScript~\cite{JensenMT09}.  Both of these systems work by tracking
which values may flow to an operation (a ``top-down'' approach), and
then ensuring the operation is legal for those values.  They also gain
significant scalability by only analyzing reachable code, as
determined by the analysis itself.  But, this approach cannot infer
types or find type errors in \emph{unreachable} code, e.g., a function
under development that is not yet invoked.  Consider this example:
\begin{lstlisting}
function f(x) {
  var y = -x;
  return x[1];
}
f(2);
\end{lstlisting}
Without the final call \code{f(2)}, the previous techniques would
not find the (obvious) type error within \code{f}.  This limitation
is unacceptable, as developers expect a compiler to report errors in
\emph{all} code.

An alternative inference approach is to compute types based on how
variables/expressions are used (a ``bottom-up'' approach), and then
check any incoming values against these types.  Such an approach is
standard in unification-style inference algorithms, combined with
introduction of parametric polymorphism to generalize types as
appropriate~\cite{damas1982principal}.  Unfortunately, since our type system has subtyping, we
cannot apply such unification-based techniques, nor can we easily
infer parametric polymorphism.

Instead, our inference takes a hybrid approach, tracking value flow in
\emph{lower bounds} of type variables and uses in \emph{upper bounds}.
Both bounds are \emph{sets} of types, and the final ascribed type must be a 
subtype of all upper bound types and a supertype of all lower bound types. 
Upper bounds enable type inference and error discovery for uninvoked functions, 
e.g., discovery of the error within \code{f} above.

If upper bounds 
alone under-constrain a type, lower bounds
provide a further constraint to inform ascription.  For example, given the identity
function \lstinline|id(x) { return x; }|, since no operations are
performed on \code{x}, upper bounds give no information on its type.
However, if there is an invocation \code{id("hi")}, inference can use the lower
bound information from \code{"hi"} to ascribe the type
$\funt{\stringt}{\stringt}$.  Note that as in other
systems~\cite{typescript,flow}, we could combine our inference
with checking of user-written polymorphic types, e.g., if the user
provided a type $\funt{T}{T}$ ($T$ is a type variable) for
\code{id}.

Once all upper and lower bounds are computed, an assignment needs
to be made to each type variable. A natural choice is the greatest lower bound
of the upper bounds, with a further check that the result 
is a supertype of all lower bound types. However,
if upper and lower bounds are based solely on value flow and uses,
type variables can be \emph{partially} constrained, with $\emptyset$
as the upper bound (if there are no uses) or the lower bound (if no values flow in).
In the first case, since our type system does not include a top type,
\footnote{We exclude $\top$ from the type system to detect more errors; see discussion in \autoref{sec:type-ascription}.} 
it is
not clear what assignment to make.
This is usually not a concern in unification-based
analyses, which flow information across assignments symmetrically, but
it \emph{is} an issue in subtyping-based analyses such as ours.

Particular care thus needs to be taken to soundly assign type variables whose 
upper bound is empty. A sound choice would be to simply fail in inference, 
but this would be too restrictive.
We could compute an assignment based on the lower bound types, e.g., their least upper bound.  
But this scheme is unsound, 
as shown by the following example:
\begin{lstlisting}[numbers=none]
function f(x) {
  var y = x; y = 2; return x.a+1;
}
\end{lstlisting}
Assume \code{f} is uninvoked.  Using a graphical notation (edges reflect subtyping), the relevant constraints for this code are:
\begin{center}
\begin{tikzpicture}[node distance=1.2cm]
\node(xr){$\sr{X}$};
\node(aintrow)[right of=xr,xshift=0.25cm]{$\rowtype{a: \intt}$};
\node(yr)[right of=aintrow]{$\sr{Y}$};
\node(int)[right of=yr]{$\intt$};
\draw [->] (xr)--(aintrow);
\draw [->] (xr) to [out=30,in=150] (yr);
\draw [->] (int)--(yr);
\end{tikzpicture}
\end{center}
\code{x} has no incoming value flow, but it is used as an object with an integer $a$ field (shown as the $\sr{X} \longrightarrow \rowtype{a: \intt}$ edge).  For \code{y}, we see no uses, but the integer \code{2} flows into it (shown as the $\sr{Y} \longleftarrow \intt$ edge).  A technique based solely on value flow and uses would compute the upper bound of $\sr{X}$ as $\set{\rowtype{a: \intt}}$, the lower bound of $\sr{Y}$ as $\set{\intt}$, and the lower bound of $\sr{X}$ and upper bound of $\sr{Y}$ as $\emptyset$.  But, ascribing types based on these bounds would be \emph{unsound}: they do not capture the fact that if \code{x} is ascribed an object type, then \code{y} must also be an object, due to the assignment \code{y = x}.

Instead, our inference \emph{strengthens} lower
bounds based on upper bounds, and vice-versa.  For the above case, bound strengthening yields the following constraints (edges due to strengthening are dashed):
\begin{center}
\begin{tikzpicture}[node distance=1.2cm]
\node(botrow){$\botrow$};
\node(xr)[right of=botrow]{$\sr{X}$};
\node(aintrow)[right of=xr,xshift=0.25cm]{$\rowtype{a: \intt}$};
\node(yr)[right of=aintrow]{$\sr{Y}$};
\node(int)[right of=yr]{$\intt$};
\node(emprow)[right of=int]{$\emptyrow$};
\draw [dashed,->] (botrow)--(xr);
\draw [->] (xr)--(aintrow);
\draw [->] (xr) to [out=30,in=150] (yr);
\draw [->] (int)--(yr);
\draw [dashed,->] (yr) to [out=30,in=150](emprow);
\end{tikzpicture}
\end{center}
Given the type $\rowtype{a: \intt}$ in the upper bound of $\sr{X}$, we
strengthen $\sr{X}$'s lower bound to $\botrow$ (a subtype of all
rows), as we know that any type-correct value flowing into \code{x}
must be an object.  As $\sr{Y}$ is now reachable from $\botrow$,
$\botrow$ is added to $\sr{Y}$'s lower bound.  With this bound, the algorithm
strengthens $\sr{Y}$'s upper bound to $\emptyrow$, a supertype of all
rows.  Given these strengthened bounds, inference tries to ascribe an
object type to \code{y}, and detects a type error with $\intt$ in $\sr{Y}$'s
lower bound, as desired.  
Apart from aiding in
correctness, bound strengthening simplifies ascription, as any type
variable can be ascribed the greatest-lower bound of its upper bound
(details in \autoref{sec:type-ascription}).

\makeatletter{}
\section{Terms, Types, and Constraint Generation}
\label{sec:types}

This section details the terms and types for a core calculus based on
that of Choi \etal~\cite{SJSTechReport}, modelling a JavaScript
fragment equipped with integers, objects, prototype inheritance, and
methods.  The type system includes structural subtyping, abstract
types, and recursive types.  As this paper focuses on inference,
rather than presenting the typing relation here, we show the
constraint generation rules for inference instead, which also capture
the requirements for terms to be well-typed.
\iflong
\autoref{app:type_system_metatheory} presents the full typing relation.
\fi

\begin{figure}
\[ \begin{array}{l@{~~}r@{~~}c@{~~}l}

\textbf{fields} & \f & \in & \Fields \\

\textbf{expressions} & e & ::= &
          n \mid \letin{x}{e_1}{e_2} \mid x \mid \ass{x}{e_1} \\
\multicolumn{4}{l}{
\quad \mid \obj{\cdot} \mid \proto{\obj{\f_1:e_1, \ldots, \f_n:e_n}}{e_p} \mid \mynull \mid \this}\\
\multicolumn{4}{l}{
\quad  \mid e.\f  \mid \ass{e_1.\f}{e_2 } \mid \fun{x}{e_1} \mid \app{e_1.\f}{e_2}} \\

\end{array} \]
\caption{Syntax of terms.
}
\label{fig:term_syntax}
\end{figure}

\subsection{Terms}

\autoref{fig:term_syntax} presents the syntax of the calculus.  The metavariable
$\f$ ranges over a finite set of fields $\Fields$, which describe the fields of
objects.  Expressions $e$ include base terms $n$ (which we take to be
integers), and variable declaration ($\letin{x}{e_1}{e_2}$), use ($x$), and
assignment ($\ass{x}{e}$).  An object is either the empty object $\obj{\cdot}$ or a record of fields $\proto{\obj{\f_1:e_1, \ldots, \f_n:e_n}}{e_p}$, where $e_p$ is the object's prototype.
We also have the $\mynull$ and the receiver, $\this$.

Field projection $e.\f$ and assignment $\ass{e_1.\f}{e_2}$ take the expected
form.  The calculus includes first-class methods (declared with the $\mathsf{function}$ syntax, as
in JavaScript), which must be invoked with a receiver argument.  Our implementation also handles
first-class functions, but they present no additional complications for inference beyond methods,
so we omit them here for simplicity.
\iflong
Appendix \ref{app:type_system_metatheory} gives details.
\fi

\subsection{Types}\label{sec:the-types}

\begin{figure}
\[ \begin{array}{l@{~~}r@{~~}c@{~~}l}
\textbf{types} & \llap{$\tau,\sigma\in \Tau$} & ::= & \intt \mid \nu \mid \alpha \mid
\\
&& \mid & \metht{\nu}{\tau_1}{\tau_2} \mid \metht{\cdot}{\tau_1}{\tau_2} \\

\textbf{rows} & \meta{r},\meta{w},\meta{mr}, \meta{mw} & ::= & \rowtype{\f_1 : \tau_1, \ldots, \f_n : \tau_n} \\

\textbf{base types} & \rho & ::= & \objt{\meta{r}}{\meta{w}}\\

\textbf{object types} & \nu & ::= & \rho^q \mid \rect{\alpha}{\nu}\\

\textbf{qualifiers} & \qual & ::= & \prototypal{\meta{mr}}{\meta{mw}} \mid \notprotoconc \mid \notprotoabs \\

\end{array} \]

\caption{Syntax of types.
}
\label{fig:type_syntax}
\end{figure}

\autoref{fig:type_syntax} presents the syntax of types. Types $\tau$
include a base type (integers), objects ($\nu$), and two method types: unattached methods
($\metht{\tau_r}{\tau_1}{\tau_2}$), which retain the receiver type
$\tau_r$, and attached methods ($\metht{\cdot}{\tau_1}{\tau_2}$),
wherein the receiver type is elided and assumed to be the type of the
object to which the method is attached.  (If $e_1.a := e_2$
assigns a new method to $e_1.a$, $e_2$ is typed as an unattached method. Choi et al~\cite{DBLP:conf/sas/ChoiCNS15} restricted $e_2$ to method literals, whereas our treatment is more general.)

Object types comprise a base type, $\rho$, and a qualifier,
$\qual$.  The base type is a pair of rows (finite maps
from names to types), one for the readable fields \meta{r} and one for
the writeable fields \meta{w}.\footnote{Note that row types cannot be
  ascribed to terms directly; they only appear as part of object
  types.}  Well-formedness for object types (detailed in
\autoref{sec:subtyping}) requires that writeable fields are
also readable.
We choose to repeat the fields of $w$ into $r$ in this way because it enables a
simpler mathematical treatment based on row subtyping.
Object types also contain recursive object types
$\rect{\alpha}{\nu}$, where $\alpha$ is bound in $\nu$ and may appear in
field types.

\sloppypar Object qualifiers $\qual$ describe the field accesses performed by the
methods in the type, required for reasoning about access safety (see \autoref{sec:typesystem}).  A
\emph{prototypal} qualifier $\prototypal{\meta{mr}}{\meta{mw}}$ maintains the information explicitly with two rows, one for fields readable by methods of
the type (\meta{mr}), and another for method-writeable fields
(\meta{mw}).  At a method call, the type system ensures that all method-readable fields are readable on the base object, and similarly for method-writeable fields.
The $\notprotoconc$ and $\notprotoabs$ qualifiers are used to enable structural subtyping on object types, and are discussed further in \autoref{sec:subtyping}.

\subsection{Subtyping and Type Equivalence}\label{sec:subtyping}

\begin{figure}
\begin{mathpar}
\inferrule*[left=S-Row]{
  \forall \f \in \dom{\meta{r'}}. \f \in \dom{\meta{r}} \wedge \meta{r}[\f] \equiv \meta{r'}[\f]
} {
  {\rowsub{\meta{r}}\meta{r'}} 
}
\and \inferrule*[left=S-NonProto]{
  \rowsub{\meta{r_1}}{\meta{r_2}} \quad \rowsub{\meta{w_1}}{\meta{w_2}} \quad
  k = \notprotoconc \vee k = \notprotoabs
} {
  {\subt{\objt{\meta{r_1}}{\meta{w_1}}^k}{\objt{\meta{r_2}}{\meta{w_2}}^k}} 
}
\and \inferrule*[left=S-Proto]{
  \meta{r_1}\equiv\meta{r_2} \quad \meta{w_1}\equiv\meta{w_2} \quad
  \meta{mr_1}\equiv\meta{mr_2} \quad \meta{mw_1}\equiv\meta{mw_2}
} {
  {\subt
  {\objt{\meta{r_1}}{\meta{w_1}}^\prototypal{\meta{mr_1}}{\meta{mw_1}}}
  {\objt{\meta{r_2}}{\meta{w_2}}^\prototypal{\meta{mr_2}}{\meta{mw_2}}}   }
}
\and \inferrule*[left=S-ProtoConc]{
  \rowsub{\meta{r}}{\meta{mr}} \quad \rowsub{\meta{w}}{\meta{mw}}
} {
  {\subt{\objt{\meta{r}}{\meta{w}}^\prototypal{\meta{mr}}{\meta{mw}}}{\objt{\meta{r}}{\meta{w}}^\notprotoconc}} 
}
\and \inferrule*[left=S-ProtoAbs]{
} {
  {\subt{\objt{\meta{r}}{\meta{w}}^\prototypal{\meta{mr}}{\meta{mw}}}{\objt{\meta{r}}{\meta{w}}^\notprotoabs}}
}
\and \inferrule*[left=S-ConcAbs]{
} {
  {\subt{\objt{\meta{r}}{\meta{w}}^\notprotoconc}{\objt{\meta{r}}{\meta{w}}^\notprotoabs}} 
}
\and \inferrule*[left=S-Method]{
} {
  {\subt{\metht{\tau}{\tau_1}{\tau_2}}{\metht{\cdot}{\tau_1}{\tau_2}}}
}
\and \inferrule*[left=S-Trans]{
  \subt{\tau_1}{\tau_2} \quad
  \subt{\tau_2}{\tau_3}
} {
  \subt{\tau_1}{\tau_3} 
}
\and \inferrule*[left=S-Refl]{ } { \subt{\tau}{\tau} }
\end{mathpar}
\vspace{-6pt}
\hrulefill
\vspace{4pt}
\begin{mathpar}
\yourinfer{WF-NonObject}{\textrm{$\tau$ is not an object type or a type variable}}{\ctxtwf{\Delta}{\tau}}
\and
\yourinfer{WF-NC}{\rowsub{\meta{r}}{\meta{w}}\quad
\forall\f\in\dom{\meta{r}}.\ \ctxtwf{\Delta}{{\meta{r}}[\f]}}
{\ctxtwf{\Delta}{\objt{\meta{r}}{\meta{w}}^\notprotoconc}}
\and
\yourinfer{WF-NA}{\rowsub{\meta{r}}{\meta{w}}\quad
\forall\f\in\dom{\meta{r}}.\ \ctxtwf{\Delta}{{\meta{r}}[\f]}}
{\ctxtwf{\Delta}{\objt{\meta{r}}{\meta{w}}^\notprotoabs}}
\and
\yourinfer{WF-P}{\rowsub{\meta{r}}{\meta{w}}\quad
\forall\f\in\dom{\meta{r}}.\ \ctxtwf{\Delta}{{\meta{r}}[\f]}\\
\rowsub{\meta{mr}}{\meta{mw}}\quad
\forall\f\in\dom{\meta{mr}}.\ \ctxtwf{\Delta}{{\meta{mr}}[\f]}\\
\forall \f \in \dom{\meta{mr}} \cap \dom{\meta{r}}.\ \meta{mr}[a] \equiv \meta{r}[a]}
{\ctxtwf{\Delta}{\objt{\meta{r}}{\meta{w}}^\prototypal{\meta{mr}}{\meta{mw}}}}
\and
\yourinfer{WF-Rec}{\ctxtwf{\Delta,\alpha}{\nu}}
{\ctxtwf{\Delta}{\rect{\alpha}{\nu}}}
\and
\yourinfer{WF-Var}{ }
{\ctxtwf{\Delta, \alpha}{\alpha}}
\end{mathpar}
\caption{Subtyping and object-type well-formedness.}
\label{fig:subtyping}
\end{figure}

Any realistic type system for JavaScript must support structural subtyping for object types.
\autoref{fig:subtyping} presents the subtyping rules for our type
system.
In the premises, we sometimes write $\tau_1 \equiv \tau_2$ as a shorthand for
$\subt{\tau_1}{\tau_2}\land\subt{\tau_2}{\tau_1}$,
and similarly $r_1 \equiv r_2$ as a shorthand for
$\subt{r_1}{r_2}\land\subt{r_2}{r_1}$.
The \myrule{S-Row} rule enables width subtyping on rows and row reordering,
and the \myrule{S-NonProto} rule lifts those properties to nonprototypal objects (ignore the qualifier $k$ for the moment).
Note from \myrule{S-Row} that overlapping field types must be
equivalent, disallowing depth subtyping---such subtyping is known to
be unsound with mutable fields~\cite{fisher1995delegation,cardelli1996theory}.
Depth subtyping would be sound for read-only fields, but we disallow it
to simplify inference.

As discussed in \autoref{sec:typesystem}, there is no good way to preserve information about
method-readable and method-writeable fields across use of structural subtyping.
Hence, other than row reordering enabled by \myrule{S-Row} and \myrule{S-Proto}, there is \emph{no} subtyping between distinct prototypal types, which are the ones that carry method-readable and method-writeable information.  To employ structural subtyping, a prototypal type must first be converted to a \emph{non-prototypal} $\notprotoconc$ or $\notprotoabs$ type (distinction to be discussed shortly), using the \myrule{S-ProtoConc} or \myrule{S-ProtoAbs} rules.  After this conversion, structural subtyping is possible using \myrule{S-NonProto}.  Since non-prototypal types have no specific information about which fields are accessed by methods, they cannot be used for prototype inheritance or method updates; see \autoref{sec:constraint-generation}.

The type system also makes a distinction between \emph{concrete}
object types, on which method invocations are allowed, and
\emph{abstract} types, for which invocations are prohibited.
For prototypal types, concreteness can be checked directly, by ensuring that all method-readable fields are readable on the object and similarly for method-writeable fields, i.e., $\subt{\meta{r}}{\meta{mr}}$ and $\subt{\meta{w}}{\meta{mw}}$ (the assumptions of the \myrule{S-ProtoConc} rule).
For non-prototypal types, we employ separate qualifiers $\notprotoconc$ and $\notprotoabs$ to distinguish concrete from abstract.
Rule \myrule{S-ProtoConc} only allows concrete prototypal types to be
converted to an $\notprotoconc$ type, whereas rule \myrule{S-ProtoAbs}
allows any prototypal type to be converted to an $\notprotoabs$
type.   The type system only allows a method call
if the receiver type can be converted to an $\notprotoconc$ type (see
\autoref{sec:constraint-generation}).  The \myrule{S-ConcAbs}
rule allows any $\notprotoconc$ type to be converted to the
corresponding $\notprotoabs$ type, as this only removes the ability to invoke methods.

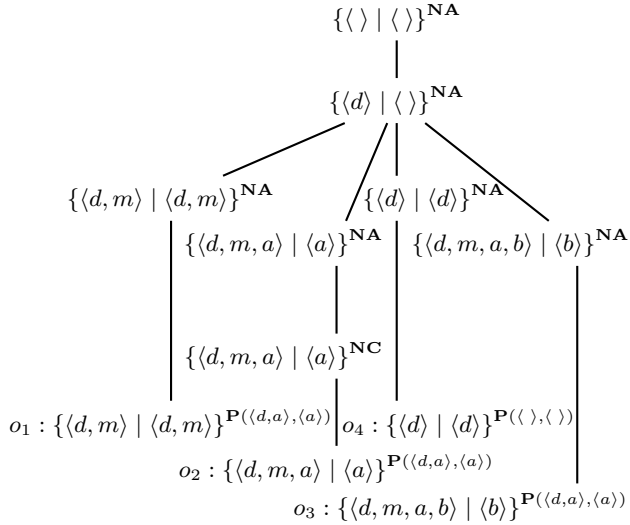
\begin{figure}
\begin{tikzpicture}[node distance=1.5cm]
\node(dmproto) {$o_1:\objt{\rowtype{d,m}}{\rowtype{d,m}}^\prototypal{\rowtype{d,a}}{\rowtype{a}}$};
\node(dmaproto) [right of=dmproto,xshift=.7cm,yshift=-0.6cm] {$o_2:\objt{\rowtype{d,m,a}}{\rowtype{a}}^\prototypal{\rowtype{d,a}}{\rowtype{a}}$};
\node(dproto) [right of=dmproto,xshift=1.5cm] {$o_4:\objt{\rowtype{d}}{\rowtype{d}}^\prototypal{\emptyrow}{\emptyrow}\hspace{-1.65cm}$};
\node(dmabproto) [right of=dmaproto,xshift=1.7cm,yshift=-0.5cm] {$o_3:\objt{\rowtype{d,m,a,b}}{\rowtype{b}}^\prototypal{\rowtype{d,a}}{\rowtype{a}}\hspace{3.1cm}$};
\node(dmnc) [above of=dmproto] {}; \node(dmanc) [above of=dmaproto] {$\objt{\rowtype{d,m,a}}{\rowtype{a}}^\notprotoconc\hspace{1.4cm}$};
\node(dmna) [above of=dmnc] {$\objt{\rowtype{d,m}}{\rowtype{d,m}}^{\notprotoabs}$};
\node(dmana) [above of=dmanc] {$\objt{\rowtype{d,m,a}}{\rowtype{a}}^\notprotoabs\hspace{1.4cm}$};
\node(dmabna) [right of=dmana,xshift=1.7cm] {$\objt{\rowtype{d,m,a,b}}{\rowtype{b}}^\notprotoabs\hspace{1.5cm}$};
\node(ddna) [right of=dmna,xshift=1.5cm] {$\objt{\rowtype{d}}{\rowtype{d}}^\notprotoabs\hspace{-1cm}$};
\node(dna) [above of =ddna,yshift=-.2cm] {$\objt{\rowtype{d}}{\emptyrow}^\notprotoabs$};
\node(empna) [above of=dna,yshift=-.4cm] {$\objt{\emptyrow}{\emptyrow}^\notprotoabs$};
\draw [thick] (dmproto)--(dmna)--(dna)--(empna);
\draw [thick] (dmaproto)--(dmanc)--(dmana)--(dna);
\draw [thick] (dmabproto)--(dmabna)--(dna);
\draw [thick] (dproto)--(ddna)--(dna);
\end{tikzpicture}
\caption{Lattice of object types}
\label{fig:lattice}
\end{figure}

Revisiting the example in \autoref{fig:ts-ex-code}, here are the types for objects $O_1$, $O_2$ and $O_3$.
\begin{small}
\begin{align*}
O_1 &: \objt{\rowtype{d: \intt, m:
      \metht{\cdot}{\intt}{\voidt}}}{\rowtype{d,m}}^{\prototypal{\rowtype{d,a}}{\rowtype{a}}} \\
O_2 &: \objt{\rowtype{d: \intt, m:
      \metht{\cdot}{\intt}{\voidt}, a: \intt}}{\rowtype{a}}^{\prototypal{\rowtype{d, a}}{\rowtype{a}}} \\
O_3 &: \objt{\rowtype{d: \intt, m:
      \metht{\cdot}{\intt}{\voidt}, a: \intt, b: \intt}}{\rowtype{b}}^{\prototypal{\rowtype{d, a}}{\rowtype{a}}}
\end{align*}
\end{small}
(We omit writing the types of fields in rows duplicatively.)
In view of the subtyping relation presented above,
the conversion of the prototypal type of $O_2$ to a $\notprotoconc$ type is allowed (by \myrule{S-ProtoConc}), so a method call at line \autoref{li:m-call-int} is allowed. By contrast, the conversion of the prototypal type of $O_3$ to a $\notprotoconc$ type is not allowed (because the condition $\rowsub{\meta{w}}{\meta{mw}}$ is not satisfied in \myrule{S-ProtoConc}), and so the method call at \autoref{li:call-on-abstr} is disallowed.
\autoref{fig:lattice} gives a graphical view of the associated type lattice,
showing the order between prototypal,
$\notprotoconc$ and $\notprotoabs$ types.
\makeatletter{}

The $\mathbf{NA}$ qualifier aids in expressivity.  Consider extending \autoref{fig:ts-ex-code} as follows:
\begin{lstlisting}[firstnumber=8]
  var v4 = cond() ? /*@\label{li:v4-def}@*/
              v3 :
              { d : 2 } // o4
\end{lstlisting}
The type of $O_4$ is
\(
 \objt{\rowtype{d:\intt}}{\rowtype{d}}^{\prototypal{\emptyrow}{\emptyrow}}
\).  To ascribe a type to \lstinline{v4}, we need to find a common supertype of the types of $O_3$ and $O_4$.  We cannot simply upcast $O_3$ to the type of $O_4$ because there is no subtyping on prototypal types; \myrule{S-NonProto} does not apply.  We also cannot apply \myrule{S-ProtoConc} to $O_3$, as $O_3$ is not concrete.  However, we \textit{can} use \myrule{S-ProtoAbs} and \myrule{S-NonProto}, in that order, to upcast the type of $O_3$ to \( \objt{\rowtype{d:\intt}}{\emptyrow}^{\notprotoabs} \) (see \autoref{fig:lattice}).  This type is also a supertype of the type of $O_4$, and therefore, a suitable type to be ascribed to \lstinline{v4}.\footnote{While the type system of Choi et al.~\cite{DBLP:conf/sas/ChoiCNS15} restricts subtyping on prototypal types, their system does not have the notion of $\notprotoabs$, and hence cannot type the example above.}
$\notprotoabs$ serves as a top element in the object type lattice, which also simplifies type ascription as we will illustrate in \autoref{sec:type-ascription}.

Rule \myrule{S-Method} introduces a limited form of method subtyping to allow
an unattached method to be attached to an object, thereby losing its receiver type.  This stripping of receiver types is important for object subtyping.  In the subtyping example in \autoref{sec:typesystem}, without attached methods, \code{o1} and \code{o2} would not have a common supertype with \code{m} present, as the receiver types for their \code{m} methods would differ; this would make \code{p.m()} a type error.
We exclude any other form of method subtyping, as we have not encountered a need for it in practice. More general function/method subtyping poses additional challenges for inference, due to contravariance, but extant techniques could be adopted to handle these issues~\cite{pottier-icfp-98,pottier-ic-01}; we plan to do so when a practical need arises.

Subtyping is reflexive and transitive. Recursive types are
equi-recursive and admit $\alpha$-equivalence, that is:
\begin{align*}
{\rect{\alpha}{\nu}} & \ \mathsf{<:}\
\override{\nu}{\alpha}{\rect{\alpha}{\nu}}\\
\override{\nu}{\alpha}{\rect{\alpha}{\nu}} & \ \mathsf{<:}\
{\rect{\alpha}{\nu}}\\
{\rect{\alpha}{\nu}} & \ \mathsf{<:}\
{\rect{\beta}{(\nu[\alpha\mapsto\beta])}}\\
\end{align*}
which directly implies:
\begin{align*}
{\rect{\alpha}{\nu}} & \equiv
\override{\nu}{\alpha}{\rect{\alpha}{\nu}}\\
{\rect{\alpha}{\nu}} & \equiv
{\rect{\beta}{(\nu[\alpha\mapsto\beta])}}\\
\end{align*}
Note that it is
possible to expand a recursive type and then apply
rule \myrule{S-NonProto} or \myrule{S-Proto}
to achieve a form of width subtyping.

\autoref{fig:subtyping} also shows the well-formedness
for types, $\ctxtwf{\Delta}{\tau}$, in the context $\Delta$ representing a set of bound variables.
 All non-object non-variable types are well-formed (rule \myrule{WF-NonObject}).
For object types, well-formedness
requires that any writeable field is also readable, and that all field types are
also well-formed (rules \myrule{WF-NC} and \myrule{WF-NA}).
In prototypal types, well-formedness further requires that
method-writeable fields are also method-readable, and that
 for any field $\f$ that is both readable and method-readable, the
\meta{mr} and \meta{r} rows agree on $\f$'s type (rule \myrule{WF-P}).
Finally, rules \myrule{WF-Rec} and \myrule{WF-Var} respectively introduce and eliminate
type variables to enable well-formedness of recursive types.

\subsection{Constraint Language}\label{sec:constraint-language}

Here, we present the constraint language used to express our type
inference problem.  Constraints primarily operate over
\emph{families of row variables}, rather than directly constraining more complex
source-level object types.
\autoref{sec:constraint-generation} reduces inference for
the source type system to this constraint language, and
\autoref{sec:constraint-solving} gives an algorithm for solving such
constraints.  
\begin{figure}
\[ \begin{array}{l@{~~}r@{~~}c@{~~}l}

\textbf{type variables} & \X, \Y & & \textit{range over source types} \\
\textbf{variable sorts} & s & ::= & \kw{r} \mid \kw{w} \mid \kw{mr} \mid \kw{mw} \mid \kw{all} \\
\textbf{row variables} & \X^s, \Y^s & & \textit{range over row / non-object types} \\
\textbf{literals} & L & ::= & \intt \mid \botrow \mid \rowtype{\ldots,a:X,\ldots} \\
&&&\mid \metht{\Xr}{\X_1}{\X_2} \\
\textbf{constraints} & C & ::= &
\rowsub{L}{\svar{X}} \mid \rowsub{\svar{X}}{L} \mid \rowsub{\svar{X}}{\svar{Y}}\\
&&& \mid C\land C \\
&&& \mid \rowsub{\svar{X}}{\rowexcl{\svar{Y}}{a_1,\ldots,a_n}} \\
&&& \mid \protocon{\X} \mid \concrete{\X} \\
&&& \mid \strip{\X} \\
&&& \mid \handleMethod{\X_b}{\X_f}{\X_v} \\
\textbf{acceptance criteria\hspace{-1cm}} & A & ::= &
\notmethod{X} \mid \notprotocon{X}
\end{array} \]
\hrulefill
\[ \begin{array}{rl}
 \textbf{well-formedness} & \sall{\X}\ \rowsubop\ \sr{\X}\ \rowsubop\ \sw{\X} \\
 & \wedge\ \sall{\X}\ \rowsubop\ \smr{\X}\ \rowsubop\ \smw{\X}
 \end{array} \]
\caption{Constraint language.  We give the language syntax above the line, and well-formedness
constraints below.}
\label{fig:inference_syntax}
\end{figure}

\autoref{fig:inference_syntax} defines the constraint language syntax.
The language distinguishes \emph{type variables}, which represent
source-level types, and \emph{row variables}, which represent the
various components of a source-level object type.  Each type variable
$\X$ has five corresponding row variables:
\sr{\X}, \sw{\X}, \smr{\X}, \smw{\X}, and \sall{\X}. The first four
 correspond directly to the \meta{r}, \meta{w},
\meta{mr}, and \meta{mw} rows from an object type.  To enforce the
condition $\wf{\objt{\meta{r}}{\meta{w}}}^q$ (\autoref{fig:subtyping}) on
all types, we impose the well-formedness conditions in
\autoref{fig:inference_syntax} on all $X$.  The last variable, $\sall{X}$, is used to
ensure that types of fields in both \meta{r} and \meta{mr} are equivalent;
if ascription fails for $\sall{X}$, there must be some inconsistency
between $\sr{X}$ and $\smr{X}$.

Type literals include $\intt$, unattached methods, and rows.  The $\botrow$ type ensures a complete row subtyping lattice and is used in type propagation (see \autoref{sec:type-propagation}).
To handle non-object types, row variables are ``overloaded'' and can
be assigned non-object types as well.  Our constraints ensure that if any row
variable for $\X$ is assigned a non-row type $\tau$,
then all row variables for $\X$ will be assigned $\tau$, and hence
$\X$ should map to $\tau$ in the final ascription.

The first three constraint types introduced in \autoref{fig:inference_syntax}
express subtyping over literals and row variables.
We write $\svar{X}\equiv L$ as a shorthand for
$\rowsub{L}{\svar{X}}\land\rowsub{\svar{X}}{L}$ and
$\svar{X}\equiv \tvar{Y}$ as a shorthand for
$\rowsub{\svar{X}}{\tvar{Y}}\land\rowsub{\tvar{Y}}{\svar{X}}$.
Constraints can be composed together using the $\land$ operator.
A constraint
$\rowsub{\svar{X}}{\rowexcl{\svar{Y}}{a_1,\ldots,a_n}}$ means that
$\svar{X}$ must be a subtype of the type obtained by removing the
fields $a_1,\ldots,a_n$ from $\svar{Y}$.  Such constraints are needed
for handling prototype inheritance, discussed further in
\autoref{sec:constraint-generation}.

The $\protocon{\X}$ and $\concrete{\X}$ constraints enable inference
of object type qualifiers. Constraint $\protocon{\X}$ means the ascribed type for $\X$ must be
prototypal, while $\concrete{\X}$ means the type for $\X$ must be a subtype of an $\notprotoconc$ type.
The $\strip{X}$ constraint ensures $X$ is assigned an \emph{attached} method type, with no receiver type.  Conversion of unattached method types to attached occurs during ascription (\autoref{sec:type-ascription}), so the constraint syntax only includes unattached method types.

The constraint $\handleMethod{\X_b}{\X_f}{\X_v}$ in \autoref{fig:inference_syntax} handles method attachment to objects.  For a field assignment $\ass{e_1.\f}{e_2}$, $X_b$, $X_f$, and $X_v$ respectively represent the type of $e_1$, the type of $\f$ in $e_1$'s (object) type, and the type of $e_2$.
Intuitively, this constraint ensures the following condition:
\begin{align*}
(\subt{\sr{X_v}}{\metht{\Xr}{\_}{\_}}) & \implies (\protocon{X_b}\ \wedge\\
\subt{\smr{X_b}}{\sr{\Xr}} & \wedge \subt{\smw{X_b}}{\sw{\Xr}} \wedge \strip{\X_f})
\end{align*}
That is, when $X_v$ is an unattached method type with receiver $\Xr$, then $X_b$ is prototypal, its method-readable and method-writeable fields must respectively include the readable and writeable fields of $\Xr$, and $X_f$ is an attached method type.
Note that $\handleMethod{\X_b}{\X_f}{\X_v}$ is \emph{not} a macro for the above condition, as we do not directly support an implication operator in the constraint language.  Instead, the condition is enforced directly during constraint propagation (\autoref{fig:propagation}, \autoref{it:attach}).

The acceptance criteria in \autoref{fig:inference_syntax} are additional conditions on solutions that need only be checked after the constraints have been solved.  The two possible criteria are checking that a variable is not assigned a method type, $\notmethod{X}$, and ensuring a variable is not assigned a prototypal type, $\notprotocon{X}$.

\subsection{Constraint Generation}\label{sec:constraint-generation}

Constraint generation takes the form of a judgement
\[ \gen{\Xr, \Gamma}{e}{\X}{\C}, \]
to be read as:
\emph{in a context with receiver type $\Xr$ and inference environment $\Gamma$,
expression $e$ has type $\X$ such that constraints in $\C$ are
satisfied.}
\autoref{fig:constraint_generation} presents rules for constraint generation; see \autoref{sec:type-propagation} for an example.

\renewcommand{\MathparLineskip}{\lineskiplimit=6pt\lineskip=6pt}
\begin{figure*}
\begin{mathpar}
\fbox{\gen{\Xr, \Gamma}{e}{\X}{\C}} \quad
\inferrule*[left=C-Int]{\fresh{\X} } { \gen{\Xr, \Gamma}{n}{\X}{} \sr{\X} \equiv \intt }
\and \inferrule*[left=C-Var]{
  \Gamma(x) = \X
} {
  \gen{\Xr, \Gamma}{x}{\X}{\emptyset}
}
\and
\inferrule*[left=C-This]{
}{
  \gen{\Xr, \Gamma}{\this}{\Xr}{}  \emptyset \\
}
\and \inferrule*[left=C-VarDecl]{
  \gen{\Xr, \override{\Gamma}{x}{\X_1}}{e_1}{\Y_1}{\C_1} \\
  \fresh{\X_1}
  \\
  \gen{\Xr, \override{\Gamma}{x}{\X_1}}{e_2}{\X}{\C_2}
}{
  \gen{\Xr, \Gamma}{\letin{x}{e_1}{e_2}}{\X}{} 
    \C_1 \wedge \C_2 
   \wedge \subt{\sr{\Y_1}}{\sr{\X_1}}
   \wedge \subt{\sw{\Y_1}}{\sw{\X_1}}
}
\and \inferrule*[left=C-VarUpd]{
  x:\X_1 \in \Gamma \\
  \gen{\Xr, \Gamma}{e_1}{\X}{\C_1}
}{
          \gen{\Xr, \Gamma}{\ass{x}{e_1}}{\X}{} \C_1 \wedge \rwsubt{\X}{\X_1} 
}
\and \inferrule*[left=C-Null]{
  \fresh{\X}
}{\gen{\Xr, \Gamma}{\mynull}{\X}{\subt{\sw{\X}}{\emptyrow}}
}
\and \inferrule*[left=C-MethDecl]{
  \fresh{\Yr, \Y_1, \X} \quad
  \hasthis{e} \quad
  \gen{\Yr, \override{\Gamma}{x}{\Y_1}}{e}{\Y_2}{\C}
} {
  \gen{\Xr, \Gamma}{\fun{x}{e}}{\X}{}  \C \wedge \subt{\sw{\Yr}}{\emptyrow}
 \wedge  \concrete{\Yr} \wedge \notprotocon{\Yr} 
 \wedge  \sr{\X} \equiv (\metht{\Yr}{\Y_1}{\Y_2})
}
\and \myinferAleft{C-MethApp}{
  \fresh{\Xm, \Yr, \X_3, \X} \\
  \gen{\Xr, \Gamma}{e_1}{\X_1}{\C_1} \qquad
  \gen{\Xr, \Gamma}{e_2}{\X_2}{\C_2}
} {
  \gen{\Xr, \Gamma}{\app{e_1.\f}{e_2}}{\X}{} & &
     \C_1 \wedge \C_2 \wedge \rowsub{\sr{\X_1}}{\rowtype{\f : \Xm}}
   \wedge  \sr{\Xm} \equiv (\metht{\Yr}{\X_3}{\X}) \wedge \strip{\Xm} 
   \wedge  \concrete{\X_1} \arcr & \wedge & \concrete{\Yr} 
    \wedge  \subt{\sw{\Yr}}{\emptyrow} \wedge \notprotocon{\Yr}
   \wedge \rwsubt{\X_2}{\X_3}  
}
\and \inferrule*[left=C-ObjEmp]{
} {
  \gen{\Xr, \Gamma}{\obj{\cdot}}{\X}{}  \protocon{\X}  \wedge
                                            \sr{\X} \equiv \emptyrow
                                            \wedge \smr{\X} \equiv \emptyrow
}
\quad \inferrule*[left={C-Attr}]{
  \fresh{\X} \quad \gen{\Xr, \Gamma}{e}{\X_1}{\C}
} {
  \gen{\Xr, \Gamma}{e.\f}{\X}{}  \C \wedge \rowsub{\sr{\X_1}}{\rowtype{a: \X}} 
   \wedge  \notmethod{\X} 
}
\and \inferrule*[left=C-AttrUpd]{
  \fresh{\X_f} \quad
  \gen{\Xr, \Gamma}{e_1}{\X_b}{\C_1} \quad
  \gen{\Xr, \Gamma}{e}{\X_v}{\C_2}
} {
  \gen{\Xr, \Gamma}{\ass{e_1.\f}{e}}{\X_v}{}  \C_1 \wedge \C_2 \wedge \rowsub{\sw{\X_b}}{\rowtype{\f:\X_f}} 
   \wedge  \rwsubt{X_v}{X_f} 
  \wedge  \handleMethod{\X_b}{\X_f}{\X_v} 
}
\and
\myinferAleft{C-ObjLit}{
  \fresh{X} \qquad
  \forall i \in 1..n.~\fresh{\X_{i}} \qquad
  \forall i \in 1..n.~\gen{\Xr, \Gamma}{e_i}{\Y_i}{\C_i} \quad
  \gen{\Xr, \Gamma}{e_p}{\X_p}{\C_p}
} {
  \gen{\Xr, \Gamma}
      {\proto{\obj{\f_1:e_1, \ldots, \f_n:e_n}}{e_p}}
      {\X}
      {} &        & \C_p \wedge \bigwedge_i (\C_i \wedge \rwsubt{\Y_i}{\X_i} \wedge \handleMethod{\X}{\X_i}{\Y_i})  \arcr
      & \wedge & \sw{\X} \equiv \rowtype{a_1: \X_1,\ldots,a_n: \X_n}
                 \wedge \rowsub{\sr{\X}}{\sr{\X_p}} \wedge \rowsub{\sr{\X_p}}{\rowexcl{\sr{\X}}{a_1,\ldots,a_n}} \arcr
      & \wedge & \protocon{\X} \wedge \protocon{\X_p} \wedge \rowsub{\smr{\X}}{\smr{\X_p}}
      \wedge \rowsub{\smw{\X}}{\smw{\X_p}} 
}
\end{mathpar}\vspace{-20pt}
\caption{Constraint generation.
}
\label{fig:constraint_generation}
\end{figure*}
\renewcommand{\MathparLineskip}{\lineskiplimit=3pt\lineskip=3pt}

Rules \myrule{C-Int} and \myrule{C-Var} generate
straightforward constraints.  The constraints for \myrule{C-ObjEmp}
ensure the empty object is assigned type
$\objt{\cdot}{\cdot}^{\prototypal{\cdot}{\cdot}}$.  The rule
\myrule{C-This} is the only rule directly using the carrier's type $\Xr$.
The constraint $\subt{\sw\X}{\emptyrow}$
in rule \myrule{C-Null} ensures that $X$ is assigned an object type.
(Recall that $\subt{\sr\X}{\sw\X}$.)

The rule for variable declaration \myrule{C-VarDecl} passes on the
constraints generated by its subexpressions ($C_1, C_2$), with
additional constraints $\rwsubt{\Y_1}{\X_1}$, which are sufficient to ensure
that the type $Y_1$ of the expression $e_1$ is a subtype of the fresh inference
variable $\X_1$ ascribed to $x$ in the environment (no constraint on $\smr{Y_1}$,
$\smr{X_1}$, $\smw{Y_1}$ or $\smw{X_1}$ is needed).  Constraining both the \meta{r} and
\meta{w} rows is consistent with the \myrule{S-NonProto} subtyping rule (\autoref{fig:subtyping}).
We put $x$ in the initialization scope of $e_1$ in order to allow for the definition of
recursive functions.

The \myrule{C-MethDecl} rule constrains the type of the
body $e$ using fresh variables $\Y_1$ and $\Yr$ for the parameter and
receiver types.  $\Yr$ is constrained to be non-prototypal and
concrete, as in any legal method invocation, the receiver type must be
a subtype of an $\notprotoconc$ type. (Recall that prototypal types,
if they are concrete, can be safely cast to $\notprotoconc$.)
The rule for method application \myrule{C-MethApp} ensures that the
type $\X_1$ of $e_1$ is concrete, and that its field $\f$ has a method
type $\Xm$ with appropriate argument type $\X_3$ and return type $\X$.
The $\strip{\Xm}$ constraint ensures $\Xm$ is an \emph{attached} method type.
Note that a relation between $\X_1$ and $\Yr$ is ensured by an $\mathsf{attach}$ constraint
when method $\f$ is attached to object $e_1$, following \myrule{C-AttrUpd} or \myrule{C-ObjLit}.

The last three rules deal more directly with objects.  Constraint generation
for attribute use \myrule{C-Attr} applies to non-methods (for methods,
\myrule{C-MethApp} is used instead); the rule generates constraints requiring
that $e$ has an object type $\X_1$ with a readable field $\f$, such that
$\f$ does not have a method type (preventing detaching of methods).
The attribute update rule \myrule{C-AttrUpd} constrains $\f$ to be a
\emph{writable} field of $e_1$ ($\rowsub{\sw{\X_b}}{\rowtype{\f:\X_f}}$), and ensures that $\X_f$ is a supertype of $e_2$'s type $X_v$.  Finally, it
uses the $\kw{attach}$ constraint to handle a possible method
update.

Finally, the rule \myrule{C-ObjLit} imposes constraints governing object
literals with prototype inheritance.  Its constraints dwarf those of other
rules, as object literals encompass potential method
attachment for each field (captured by $\handleMethod{\X_l}{\X_i}{\Y_i}$) in
addition to prototype inheritance.
For the literal type $\X$, the constraints ensure that the writeable
fields $\sw{\X}$ are precisely those declared in the literal.  The
readable fields must include those inherited from the prototype
($\rowsub{\sr{\X}}{\sr{\X_p}}$); note that $\subt{\sr{X}}{\sw{X}}$ is imposed by well-formedness.
Furthermore, the constraint
$\rowsub{\sr{\X_p}}{\rowexcl{\sr{\X}}{a_1,\ldots,a_n}}$ ensures that
additional readable fields do not appear ``out of thin air,'' by
requiring that any fields in $\sr{\X}$ apart from the
locally-present $\f_1,\ldots,\f_n$ be present in the prototype.
Finally, we ensure both $\X$ and $\X_p$ are prototypal, and that any
method-accessed fields from $\X_p$ are also present in $\X$.

\newcommand{\wfedge}{\draw [->]}
\newcommand{\litedge}{\draw [blue,->]}
\newcommand{\bidirlitedge}{\draw [blue,<->]}
\newcommand{\litpath}{\path [blue,->]}
\newcommand{\assignedge}{\draw [blue,->]}
\newcommand{\bidirproto}{\draw [red, densely dotted,<->]}
\newcommand{\attachedge}{\draw [brown,dashed,->]}
\newcommand{\attredge}{\draw [blue,->]}

\begin{figure}
\begin{tikzpicture}[node distance=0.5cm and 1cm]
\node(o2all){$\sall{O_2}$};
\node(o2r) [above=of o2all]{$\sr{O_2}$};
\node(o2w) [above=of o2r] {$\sw{O_2}$};
\node(o2mr) [below=of o2all]{$\smr{O_2}$};
\node(o2mw) [below=of o2mr] {$\smw{O_2}$};

\node(v1all) [right=of o2all] {$\sall{V_1}$};
\node(v1r) [above=of v1all]{$\sr{V_1}$};
\node(v1w) [above=of v1r] {$\sw{V_1}$};
\node(v1mr) [below=of v1all]{$\smr{V_1}$};
\node(v1mw) [below=of v1mr] {$\smw{V_1}$};

\node(o1all) [right=of v1all] {$\sall{O_1}$};
\node(o1r) [above=of o1all]{$\sr{O_1}$};
\node(o1w) [above=of o1r] {$\sw{O_1}$};
\node(o1mr) [below=of o1all]{$\smr{O_1}$};
\node(o1mw) [below=of o1mr] {$\smw{O_1}$};

\node(obj1) [above=of o1w,xshift=1cm] {$\rowtype{d: \intt, m: M}$};
\node(obj2) [above=of o2w,xshift=1cm] {$\rowtype{a: \intt}$};

\node(thisr) [right=of o1mr] {$\sr{\Yr}$};
\node(thisw) [right=of o1mw] {$\sw{\Yr}$};
\node(thisobjr) [right=of thisr] {$\rowtype{d: D}$};
\node(thisobjw) [right=of thisw] {$\rowtype{a: A}$};

\wfedge (o2all)--(o2mr);
\wfedge (o2all)--(o2r);
\wfedge (o2r)--(o2w);
\wfedge (o2mr)--(o2mw);

\wfedge (v1all)--(v1mr);
\wfedge (v1all)--(v1r);
\wfedge (v1r)--(v1w);
\wfedge (v1mr)--(v1mw);

\wfedge (o1all)--(o1mr);
\wfedge (o1all)--(o1r);
\wfedge (o1r)--(o1w);
\wfedge (o1mr)--(o1mw);

\bidirlitedge (o2w)--(obj2);
\bidirlitedge (o1w)--(obj1);
\litedge (o2r)--(v1r);
\litedge (o2mr)--(v1mr);
\litedge (o2mw)--(v1mw);
\litpath (v1r) edge [bend right=30] node [above] {$\backslash\set{a}$} (o2r);

\assignedge (o1w)--(v1w);
\assignedge (o1r)--(v1r);

\bidirproto (v1all)--(o1all);
\bidirproto (v1mr)--(o1mr);
\bidirproto (v1mw)--(o1mw);
\bidirproto (v1r) to [out=30,in=150] (o1r);
\bidirproto (v1w) to [out=30,in=150] (o1w);

\attachedge (o1mr)--(thisr);
\attachedge (o1mw)--(thisw);

\attredge (thisr)--(thisobjr);
\attredge (thisw)--(thisobjw);

\end{tikzpicture}
\caption{Selected constraints for the example of \autoref{fig:ts-ex-code}.}
\label{fig:new-solving-example}
\end{figure}
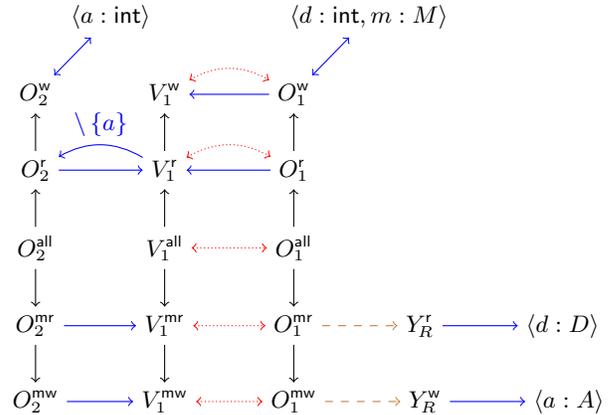

\mypara{Example}
\autoref{fig:new-solving-example} shows a graph representation of some constraints for \code{o1}, \code{o2}, and \code{v1} from lines~\ref{li:o1-alloc}--\ref{li:o2-alloc} of \autoref{fig:ts-ex-code}.  Nodes represent row variables and
type literals, with variable names matching the corresponding program
entities ($\Yr$ corresponds to \lstinline{this} on line~\ref{li:m-attach}).
Each edge $\svar{X} \rightarrow \tvar{Y}$ represents a constraint
$\subt{\svar{X}}{\tvar{Y}}$.
Black solid edges represent well-formedness constraints (\autoref{fig:inference_syntax}), while blue solid edges represent constraints generated from the code (\autoref{fig:constraint_generation}).  Dashed or dotted edges are added
during constraint solving, and will be discussed in \autoref{sec:type-propagation}.

We first discuss constraints for the body of the method declared on line~\ref{li:m-attach}.  For the field read \code{this.d}, the \myrule{C-Attr} rule generates $\sr{\Yr} \rightarrow
\rowtype{d: D}$.  Similarly, the \myrule{C-AttrUpd} rule generates $\sw{\Yr}
\rightarrow \rowtype{a: A}$ for the write to \code{this.a}.

For the containing object literal \code{o1}, the \myrule{C-ObjLit} rule
creates row variables for type $O_1$ and edges $\sw{O_1}
\leftrightarrow \rowtype{d: \intt, m: M}$ (due to type equality).  It also generates a
constraint $\handleMethod{O_1}{M}{F}$ (not shown in \autoref{fig:new-solving-example}) to handle method
attachment to field $m$ ($F$ is the type of the line~\ref{li:m-attach} function); we shall return to this constraint in \autoref{sec:type-propagation}.  The assignment to \lstinline{v1} yields the $V_1$ row variables and the $\sr{O_1} \rightarrow
\sr{V_1}$ and $\sw{O_1} \rightarrow \sw{V_1}$ edges via
\myrule{C-VarDecl}.\footnote{The code uses JavaScript \lstinline{var}
  syntax rather than \textsf{let} from the calculus.}

For \code{o2} on line~\ref{li:o2-alloc}, \myrule{C-ObjLit} yields the $\sw{O_2}
\leftrightarrow \rowtype{a: \intt}$ edges for the declared \code{a} field.  
The use of \lstinline{v1} as a prototype yields the constraints
$\sr{O_2} \rightarrow \sr{V_1}$, $\smr{O_2} \rightarrow \smr{V_1}$,
and $\smw{O_2} \rightarrow \smw{V_1}$,
capturing inheritance.  We also have $\sr{V_1} \xrightarrow{\backslash\set{a}} \sr{O_2}$ to prevent "out of thin air" readable fields on $O_2$.  Finally, we generate $\protocon{V_1}$ (not shown) to ensure $V_1$ gets a prototypal type.

\makeatletter{}
\section{Constraint Solving}\label{sec:constraint-solving}

Constraint solving proceeds in two phases.  First, \emph{type propagation} computes lower and upper bounds for every row variable, extending techniques from previous work~\cite{iogti-RCH12,pottier-icfp-98}.  Then, \emph{type ascription} checks for type errors, and, if none are found, computes a satisfying assignment for the type variables.  
\subsection{Type Propagation}\label{sec:type-propagation}

\begin{figure}[t]
\raggedright
\begin{enumerate}[label=(\roman*)]
\item
$\mathcal{C}\subseteq\mathcal{C'}$; \label{it:incl}
\end{enumerate}
\emph{Well-formedness}
\begin{enumerate}[resume,label=(\roman*)]
\item
$\subt{\sall{X}}{\subt{\sr{X}}{\sw{X}}}\cond$ and
$\subt{\sall{X}}{\subt{\smr{X}}{\smw{X}}}\cond$; \label{it:wf}
\end{enumerate}
\emph{Subtyping}
\begin{enumerate}[resume,label=(\roman*)]
\item
if $\subt{\svar{X}}{L}\cond$, then $L\in\ub{\svar{X}}$;\label{it:ub}
\item
if $\subt{L}{\svar{X}}\cond$, then $L\in\lb{\svar{X}}$;\label{it:lb}
\item
if $\subt{\svar{X}}{\tvar{Y}}\cond$, then
   $\lb{\svar{X}}\subseteq\lb{\tvar{Y}}$ and
   $\ub{\tvar{Y}}\subseteq\ub{\svar{X}}$;\label{it:subt}
\item
if $\subt{\svar{X}}{\rowexcl{Y^t}{a_1,\ldots,a_n}}\ \in\mathcal{C}'$,
then for any $\rowtype{F}\in\ub{Y^t}$,
add $\rowtype{\rowexcl{F}{a_1,\ldots,a_n}}$ to $\ub{\svar{X}}$;\label{it:subtminus}
\end{enumerate}
\emph{Bound strengthening}
\begin{enumerate}[resume,label=(\roman*)]
\item
 if $L\in\lb{\svar{X}}$, then $\kw{top}(L)\in\ub{\svar{X}}$;\label{it:top}
\item
 if $L\in\ub{\svar{X}}$, then $\kw{bot}(L)\in\lb{\svar{X}}$;\label{it:bot}
\end{enumerate}
\emph{Prototypalness and concreteness}
\begin{enumerate}[resume,label=(\roman*)]
\item
 if $\protocon{Y}\cond$, $\subt{\sr{X}}{\sr{Y}}\cond$ and
 $\subt{\sw{X}}{\sw{Y}}\cond$, then $\protocon{X}\cond$,
$\sr{X}\equiv\sr{Y}\cond$, $\sw{X}\equiv\sw{Y}\cond$,
$\smr{X}\equiv\smr{Y}\cond$ and $\smw{X}\equiv\smw{Y}\cond$;
\label{it:proto}
\item
 if $\concrete Y\cond$, $\subt{\sr{X}}{\sr{Y}}\cond$, and $\subt{\sw{X}}{\sw{Y}}\cond$, then
$\concrete X\cond$;
\label{it:conc}
\item
 if $\protocon X\cond$ and $\concrete X\cond$ then $\subt{\sr{X}}{\smr{X}}\cond$ and
$\subt{\sw{X}}{\smw{X}}\cond$;
\label{it:protoconc}
\end{enumerate}
\emph{Attaching methods}
\begin{enumerate}[resume,label=(\roman*)]
\item
if $\handleMethod{\X_b}{\X_f}{\X_v}\cond$ and $\metht{\Xr}{\Y_1}{\Y_2}\in\ub{\sr X_v}$, then
$\protocon{\X_b}\cond$, $\subt{\smr{X_b}}{\sr{\Xr}}\cond$, $\subt{\smw{X_b}}{\sw{\Xr}}\cond$, and
 $\strip{\X_f}\cond$.
\label{it:attach}
\item
 if $\strip X\cond$, $\subt{\sr{X}}{\sr{Y}}\cond$, and $\subt{\sw{X}}{\sw{Y}}\cond$, then
$\strip Y\cond$;
\label{it:strip}
\end{enumerate}
\emph{Inferring equalities (not essential for soundness)}
\begin{enumerate}[resume,label=(\roman*)]
\item
 if $\rowtype{f_1:F_1,\ldots,f_n:F_n,\ldots}\in\lb{\svar{X}}$
and $\rowtype{f_1:G_1,\ldots,f_n:G_n}\in\!\ub{\svar{X}}$, then
$\forall s.\set{\svar{F_1}\equiv\svar{G_1},\ldots,\svar{F_n}\equiv\svar{G_n}} \subseteq \mathcal{C'}$;\label{it:uplo}
\item
 if $\rowtype{f_1:F_1,\ldots,f_n:F_n,\ldots}\in\ub{\svar{X}}$
and $\rowtype{f_1:G_1,\ldots,f_n:G_n,\ldots}\in\ub{\svar{X}}$, then
$\forall s.\set{\svar{F_1}\equiv\svar{G_1},\ldots,\svar{F_n}\equiv\svar{G_n}} \subseteq \mathcal{C'}$;\label{it:upup}
\item
 if $\metht{\Xr}{\X_1}{\X_2}\in\ub{\svar X}$ and $\metht{\Yr}{\Y_1}{\Y_2}\in\ub{\svar X}$,
 then $\forall s.\set{\svar{\X_1}\equiv\svar{\Y_1}, \svar{\X_2}\equiv\svar{\Y_2}} \subseteq \mathcal{C'}$.\label{it:upmethod}
\end{enumerate}
\caption{Propagation rules.
}
\label{fig:propagation}
\end{figure}

Type propagation computes a lower bound $\lb{\svar{\X}}$ and upper bound  $\ub{\svar{\X}}$ for each row variable $\svar{\X}$ appearing in the constraints, with each bound represented as a \emph{set} of types.
Intuitively, $\svar{\X}$ must be ascribed a type between its lower and upper bound in the subtype lattice.
\autoref{fig:propagation} shows the rules for type propagation.  Given initial constraints $\mathcal{C}$, propagation computes the smallest set of constraints $\mathcal{C'}$, and the smallest sets of types $\ub{\svar{X}}$ and $\lb{\svar{X}}$ for each variable $\svar{X}$, verifying the rules of \autoref{fig:propagation}.
In practice, propagation starts with $\mathcal{C'} = \mathcal{C}$ and $\ub{\svar{X}} = \lb{\svar{X}} = \emptyset$ for all $\svar{X}$.  It then iteratively grows $\mathcal{C'}$ and the bounds to satisfy the rules of \autoref{fig:propagation} until all rules are satisfied, yielding a least fixed point.

\autoref{it:wf} adds the standard well-formedness rules for object types.
Rules~\ref{it:ub}--\ref{it:subtminus} show how to update bounds for the core subtype constraints.  \autoref{it:subt} states that if we have $\subt{\svar{X}}{\tvar{Y}}$, then any upper bound of $\tvar{Y}$ is an upper bound of $\svar{X}$, and vice-versa for any lower bound of $\svar{X}$.  \autoref{it:subtminus} propagates upper bounds in a similar way for constraint $\subt{\svar{X}}{\rowexcl{\svar{Y}}{a_1,\ldots,a_n}}$, but it removes fields $\set{a_1,\ldots,a_n}$ from each upper bound before propagation.  Lower bounds are \emph{not} propagated in \autoref{it:subtminus}, as the right-hand side of the constraint is not a type variable.

Rules~\ref{it:top} and~\ref{it:bot} perform \emph{bound strengthening}, a crucial step for ensuring soundness (see \autoref{sec:complications}).  The rules leverage predicates $\kw{top}(L)$ and $\kw{bot}(L)$, defined as follows:
\begin{align*}
\kw{top}(L) &=
  \begin{cases}
    \emptyrow, & \text{if}\ L\ \text{is a row type} \\
    L & \text{otherwise}\\
  \end{cases}
\end{align*}
\begin{align*}
\kw{bot}(L) &=
  \begin{cases}
    \botrow, & \text{if}\ L\ \text{is a row type} \\
    L & \text{otherwise}\\
  \end{cases}
\end{align*}
The rules ensure that any lower bound $\lb{\svar{X}}$ includes the best type information that can be inferred from $\ub{\svar{X}}$, and vice-versa.

Rules~\ref{it:proto}--\ref{it:protoconc} handle the constraints for prototypalness and concreteness.
Recall from \autoref{sec:subtyping} that a prototypal type is only related to itself by subtyping
(modulo row reordering).
So, if we have $\protocon{Y}$ and $\subt{X}{Y}$, it must be true that $X \equiv Y$ and also
$\protocon{X}$ (to handle transitive subtyping).  \autoref{it:proto} captures this logic at the
level of row variables. The subtyping rules (\autoref{fig:subtyping}) show that for any concrete (\textbf{NC}) type $Y$, if $\subt{X}{Y}$, then $X$ must also be concrete, either as an \textbf{NC} type (\myrule{S-NonProto}) or a concrete prototypal type (\myrule{S-ProtoConc}); \autoref{it:conc} captures this logic.  Finally, if we have both $\protocon{X}$ and $\concrete{X}$, \autoref{it:protoconc} imposes the assumptions from the \myrule{S-ProtoConc} rule of \autoref{fig:subtyping}, ensuring any method-accessed field is present in the type.

Rules~\ref{it:attach} and~\ref{it:strip} handle method attachment.  \autoref{it:attach} enforces the meaning of $\mathsf{attach}$ as discussed in \autoref{sec:constraint-language}.  To understand \autoref{it:strip}, say that $X$ and $Y$ are both \emph{unattached} method types such that $\subt{X}{Y}$.  If we add $\strip{Y}$ to make $Y$ an attached method, $\subt{X}{Y}$ still holds, by the \myrule{S-Method} subtyping rule (\autoref{fig:subtyping}).  However, if $\strip{X}$ is introduced, then $\strip{Y}$ must also be added, or else $\subt{X}{Y}$ will be violated.

Rules~\ref{it:uplo}--\ref{it:upmethod} introduce new type equalities that enable the inference to succeed in more cases (the rules are not needed for soundness).  \autoref{it:uplo} equates types of shared fields for any rows $r_1 \in \lb{\svar{X}}$ and $r_2 \in \ub{\svar{X}}$; the types must be equal since $\subt{r_1}{r_2}$ and the type system has no depth subtyping.  \autoref{it:upup} imposes similar equalities for two rows in the same upper bound, and \autoref{it:upmethod} does the same for methods.

\mypara{Example.}
We describe type propagation for the example of \autoref{fig:new-solving-example}.  For the graph, type propagation ensures that if there is a path from row variable $X^s$ to
type $L$ in the graph, then
$L \in \ub{X^s}$.  E.g., given the path $\sr{O_2} \rightarrow \sw{O_2} \rightarrow \rowtype{a: \intt}$, propagation ensures that $\set{\rowtype{a: \intt}} \subseteq \ub{\sr{O_2}}$.
The new subtype / equality constraints added to $\mathcal{C'}$ in the rules in \autoref{fig:propagation} correspond to adding new edges to the graph.
For the example,
the \myrule{C-MethDecl} rule generates a constraint $\sr{F} \equiv
\metht{\Yr}{Y_1}{Y_2}$ (not shown in \autoref{fig:new-solving-example}) for the method literal on \autoref{li:m-attach} of \autoref{fig:ts-ex-code}.  Once propagation adds $\metht{\Yr}{Y_1}{Y_2}$ to $\ub{\sr{F}}$,
handling of the $\handleMethod{O_1}{M}{F}$ constraint (\autoref{it:attach}) constrains the method-accessible fields of $O_1$ to accommodate
receiver $\Yr$.  Specifically, the solver adds the brown dashed edges
$\smr{O_1} \rightarrow \sr{\Yr}$ and $\smw{O_1} \rightarrow \sw{\Yr}$.

The $\protocon{V_1}$ constraint, combined with
$\subt{\sr{O_1}}{\sr{V_1}}$, leads the solver to equate all
corresponding row variables for $O_1$ and $V_1$ (\autoref{it:proto}).
This leads to the addition of the red dotted edges in
\autoref{fig:new-solving-example}.  These new red edges make all the literals reachable from $\sall{O_2}$; e.g., we have path $\sall{O_2} \rightarrow \sr{O_2} \rightarrow \sr{V_1} \rightarrow \sr{O_1} \rightarrow \sw{O_1} \rightarrow \rowtype{d: \intt, m: M}$. So, propagation yields:
\begin{equation*}
\set{\rowtype{a: \intt}, \rowtype{d: \intt, m: M}, \rowtype{d: D}, \rowtype{a: A}} \subseteq \ub{\sall{O_2}}
\end{equation*}
Via \autoref{it:upup}, the types of $a$ and $d$ are equated across the rows, yielding $A \equiv D \equiv \intt$.  Hence, the inference discovers \code{this.a} and \code{this.d} on line~\ref{li:m-attach} both have type $\intt$, \emph{without} observing the invocations of \code{m}.

\mypara{Implementation.}
Our implementation computes type propagation using the iterative fixed-point solver available in WALA~\cite{wala}.  WALA's solver accommodates generation of new constraints during the solving process, a requirement for our scenario.
WALA's solver includes a variety of optimizations, including sophisticated worklist ordering heuristics and machinery to only revisit constraints when needed.  By leveraging this solver, these optimizations came for free and saved significant implementation work.  As the sets of types and fields in a program are finite, the fixed-point computation terminates.

\subsection{Type Ascription}\label{sec:type-ascription}

\algnewcommand\algorithmicforeach{\textbf{for each}}
\algdef{S}[FOR]{ForEach}[1]{\algorithmicforeach\ #1\ \algorithmicdo}

\begin{algorithm}
 \begin{algorithmic}[1]
 \Procedure{AscribeType}{$\X$}
 \If{$\strip \X\in\mathcal C'$}{ strip receivers in $\ub{\svar X}$, $\lb{\svar X}$}\EndIf\label{as:strip}
 \ForEach {$\svar{X}$}
   \If{$\ub{\svar{X}} = \emptyset$} $\Phi(\svar{X}) \gets \defaultt$ \label{as:default}
   \Else
   \State $\Phi(\svar{X}) \gets {\sf glb}(\ub{\svar{X}})$ \Comment{Fails if no glb} \label{as:glb}
   \ForEach {$L \in \lb{\svar{X}}$}
     \If{$\nsubt{L}{\Phi(\svar{X})}$} fail \label{as:lb}
     \EndIf
   \EndFor
   \EndIf
 \EndFor
 \If{$\Phi(\sr{X}) = \intt \vee \Phi(\sr{X}) = \defaultt$} \label{as:checkbase}
   \State $\Phi(X) \gets \Phi(\sr{X})$ \label{as:base}
 \ElsIf{$\Phi(\sr{X})$ is method type} \label{as:method}
   \If{$\notmethod{\X} \in \mathcal{C'}$} fail \label{as:nsubt}
   \EndIf
   \State $\Phi(X) \gets \Phi(\sr{X})$ \label{as:unattached}
 \Else \Comment{$\Phi(\sr{X})$ must be a row}
 \State $\rho \gets \objt{\Phi(\sr{X})}{\Phi(\sw{X})}$ \label{as:object}
 \If{$\protocon{X} \in \mathcal{C'}$} \label{as:proto}
   \If{$\notprotocon{X} \in \mathcal{C'}$} fail \label{as:notproto}\EndIf
   \State $\Phi(X) \gets \rho^{\prototypal{\Phi(\smr{X})}{\Phi(\smw{X})}}$ \label{as:ascribe-proto}
 \ElsIf{$\concrete{X} \in \mathcal{C'}$}
   $\Phi(X) \gets \rho^\notprotoconc$ \label{as:nc}
 \Else\ $\Phi(X) \gets \rho^\notprotoabs$  \label{as:na}
 \EndIf
 \EndIf
 \EndProcedure
 \end{algorithmic}
 \caption{Type ascription.}
 \label{alg:type-ascription}
\end{algorithm}

\autoref{alg:type-ascription} shows how to ascribe a type to variable $X$,
given bounds for all row variables $\svar{X}$ and the implied constraints
$\mathcal{C'}$.  Here, we assume each type variable can be ascribed
independently, for simplicity; \iflong \autoref{app:type_system_metatheory}
\fi gives a slightly-modified ascription
algorithm that handles variable dependencies and recursive types
.

If required by a $\strip{X}$ constraint, line~\ref{as:strip} handles stripping the receiver type
in all method literals of $\ub{\svar X}$ and $\lb{\svar X}$ .
For each $\svar{X}$, we check if its upper bound is empty, and if so assign it the $\defaultt$ type.  For soundness, the same default type must be used everywhere in the final ascription; our implementation uses $\intt$.
Conceptually, an empty set upper bound corresponds to a $\top$ (top) type.
However we do not allow $\top$ in our system, as it would hide problems like objects and ints flowing into the same (unused) location, e.g., \verb+x = { }; x = 3+.

If the upper bound is non-empty, we compute its \emph{greatest lower bound}
(glb) (line~\ref{as:glb}).  The glb of a set of row
types is a row containing the union of their fields, where each common
field must have the same type in all rows.  For example:
\begin{align*}
\kw{glb}(\set{\rowtype{a: \intt}, \rowtype{b: \mathsf{string}}}) &= \rowtype{a: \intt, b: \mathsf{string}} \\
\kw{glb}(\set{\rowtype{a: \intt}, \rowtype{a: \mathsf{string}}}) &\ \textrm{is undefined}
\end{align*}
If no glb exists for two upper bound types, ascription fails with a type error.\footnote{We compute glb over a semi-lattice excluding $\botrow$, to get the desired failure with conflicting field types.}   Given a glb, the algorithm then checks that every type in the lower bound is a subtype of the glb (line~\ref{as:lb}).  If this does not hold, then some use in the program may be invalid for some incoming value, and ascription fails (examples forthcoming).

Once all glb checks are complete, lines~\ref{as:checkbase}--\ref{as:na} compute a type for $X$ based on its row variables.  If $\Phi({\sr{X}})$ is an integer, method, or $\defaultt$ type, then $X$ is assigned $\Phi({\sr{X}})$.  Otherwise, an object type for $X$ is computed based on its row variables.  The appropriate qualifier is determined based on the presence of $\protocon{X}$ or $\concrete{X}$ constraints in $\mathcal{C'}$, as seen in lines~\ref{as:proto}--\ref{as:na}.  The algorithm also checks the acceptance criteria (\autoref{sec:constraint-language}), ensuring ascription failure if they apply (they are introduced by the \myrule{C-MethDecl} and \myrule{C-Attr} rules in \autoref{fig:constraint_generation}).

Notice that $\notprotoabs$ is crucial to enable ascription based exclusively on glb of upper bounds.  Absent $\notprotoabs$, if an object of abstract type $\tau$ flows from \code{x} to \code{y}, the types of \code{x} and \code{y} must be \emph{equal}, as $\tau$ would have no supertypes in the lattice.  Hence, qualifiers would have to be considered when deciding which fields should appear in object types, losing the clean separation in \autoref{alg:type-ascription}.  Note also, abstractness is not syntactic (in \autoref{fig:ts-ex-code}, \code{v3} is only abstract because of inheritance), so even computing abstractness could require another fixed point loop.

\mypara{Example.} Returning to $O_2$ in the example of \autoref{fig:new-solving-example}, $\ub{\sr{O_2}} = \set{\rowtype{a: \intt}, \rowtype{d: \intt, m: M}}$ after type propagation.  Given type $\metht{\cdot}{\intt}{\voidt}$ for $M$, ${\sf glb}(\ub{\sr{O_2}}) = \rowtype{a: \intt, d: \intt, m: \metht{\cdot}{\intt}{\voidt}}$.  $\Phi(\sw{O_2})$, $\Phi(\smr{O_2})$, and $\Phi(\smw{O_2})$ are computed similarly.  Since we have $\protocon{O_2}$ (by \myrule{C-ObjLit}, \autoref{fig:constraint_generation}), at line~\ref{as:ascribe-proto} ascription assigns $O_2$ the following type, shown previously in \autoref{sec:subtyping}:
\begin{equation*}
\objt{\rowtype{d: \intt, m: \metht{\cdot}{\intt}{\voidt}, a: \intt}}{\rowtype{a}}^{\prototypal{\rowtype{d, a}}{\rowtype{a}}}
\end{equation*}

Using glb of upper bounds for ascription ensures a type captures what is needed from the term, rather than what is available.  In \autoref{fig:ts-ex-code}, note that \code{v3} is only used to invoke method \code{m}.  Hence, only \code{m} will appear in the upper bound of $\sr{V_3}$, and the type of \code{v3} will only include \code{m}, despite the other fields available in object \code{o3}.

\mypara{Type error examples.} We now give two examples to illustrate detection of type errors.  The expression \lstinline!({a: 3} proto {}).b! erroneously reads a non-existent field \code{b}.  For this code, the constraints are:

\begin{tikzpicture}[node distance=1.5cm]
\node(emprow){$\emptyrow$};
\node(empobjr)[right of=emprow]{$\sr{E}$};
\node(obj1r)[right of=empobjr]{$\sr{O}$};
\node(obj1w)[right of=obj1r,yshift=0.3cm]{$\sw{O}$};
\node(brow)[right of=obj1r,yshift=-0.3cm,xshift=0.25cm]{$\rowtype{b: B}$};
\node(arow)[right of=obj1w,xshift=0.25cm]{$\rowtype{a: \intt}$};
\draw [<->] (emprow)--(empobjr);
\draw [->] (obj1r)--(empobjr);
\draw [<->] (obj1w)--(arow);
\draw [->] (obj1r)--(obj1w);
\draw [->] (obj1r)--(brow);
\path [->] (empobjr) edge [bend left=30] node [above] {$\backslash\set{a}$} (obj1r);
\end{tikzpicture}\\
$E$ is the type of the empty object, and $O$ the type of the parenthesized object literal.  The $\rowtype{} \leftrightarrow \sr{E}$ edges are generated by the \myrule{C-ObjEmp} rule.  As $O$ inherits from the empty object, we have $\sr{O} \rightarrow \sr{E}$, modeling inheritance of readable fields, and also 
$\sr{E} \xrightarrow{\backslash\set{a}} \sr{O}$, ensuring any readable field of $O$ except $a$ is inherited from $E$.  Since $E$ is the empty object, these constraints ensure $a$ is the only readable field of $O$.

Propagation and ascription detect the error as follows.  $\rowtype{a: \intt}$ is \emph{not} added to $\ub{\sr{E}}$, though it is reachable, due to the $\backslash\set{a}$ filter on the edge from $\sr{E}$ to $\sr{O}$.  Instead, we have $\set{\rowtype{b: B}} \subseteq \ub{\sr{E}}$: intuitively, since $b$ is not present locally in $O$, it can only come from $E$.  Further, we have $\set{\emptyrow} \subseteq \lb{\sr{E}}$.  Since $\nsubt{\emptyrow}{\rowtype{b: B}}$, line~\ref{as:lb} of \autoref{alg:type-ascription} reports a failure.

As a second example, consider:
\begin{lstlisting}[numbers=none]
({m: fun () { this.f = 3; }}).m()
\end{lstlisting}
The invocation is in error, since the object literal \code{o} is abstract (it has no \code{f} field).  Our constraints are:

\begin{tikzpicture}[node distance=1.5cm]
\node(row1){$\rowtype{m: M}$};
\node(obj1w)[right of=row1]{$\sw{O}$};
\node(obj1mw)[right of=obj1w]{$\smw{O}$};
\node(tw)[right of=obj1mw]{$\sw{\Yr}$};
\node(row2)[right of=tw]{$\rowtype{f: \intt}$};
\draw [<->] (row1)--(obj1w);
\draw [red,densely dotted,->] (obj1w)--(obj1mw);
\draw [brown,dashed,->] (obj1mw)--(tw);
\draw [->] (tw)--(row2);
\end{tikzpicture}

\noindent As in \autoref{fig:new-solving-example}, the brown dashed edge stems from method attachment.  From the invocation and \myrule{C-MethApp}, we have $\concrete{O}$.  We also have $\protocon{O}$ (from \myrule{C-ObjLit}), leading (via \autoref{it:protoconc}) to the dotted edge from $\sw{O}$ to $\smw{O}$.  Now, we have a path from $\rowtype{m: M}$ to $\sw{O}$, and from $\sw{O}$ to $\rowtype{f: \intt}$.  Since $\nsubt{\rowtype{m: M}}{\rowtype{f: \intt}}$, line~\ref{as:lb} will again report an error.

\subsection{Soundness of Type Inference}

We prove soundness of type inference, including soundness of constraint
generation, constraint propagation, and type ascription. We also prove our type
system sound.  Our typing judgment and proofs can be found in \iflong
\autoref{app:type_system_metatheory}\fi.

Our proof of soundness of type inference relies on three lemmas on constraint propagation and ascription, subtyping constraints, and well-formedness of ascripted types.

\begin{definition}[Constraint satisfaction]
We say that a typing substitution $\Phi$, which maps fields in $\Fields$ to
types in $\Tau$, satisfies the constraint $\C$ if, after substituting for
inference variables in $\C$ according to $\Phi$, the resulting constraint
holds.
\end{definition}

\begin{lemma}[Soundness of constraint propagation and ascription]
\label{lem:ascription}
For any set of constraints $\mathcal C$ generated by the rules of \autoref{fig:constraint_generation}, 
on variables $X_1,\ldots,X_n$ and their associated row
variables, if constraint propagation and ascription succeeds with assignment $\Phi$, then
$\forall i,\forall s,\satisfies{\Phi(X_i)}{\mathcal C}$ and $\satisfies{\Phi(X_i^s)}{\mathcal C}$.
\end{lemma}

\begin{lemma}[Soundness of subtyping constraints]
\label{lem:subt}
For a set of constraints $\mathcal C$ containing the constraints $\subt{\sr X}{\sr Y}$ and
$\subt{\sw X}{\sw Y}$, if constraint generation and ascription succeeds with assignment
$\Phi$, then $\subt{\Phi(X)}{\Phi(Y)}$.
\end{lemma}

\begin{lemma}[Well-formedness of ascripted types]
\label{lem:wf}
For a set of constraints $\mathcal C$ containing constraints on variable $X$,
if constraint generation and ascription succeeds with assignment $\Phi$, then
$\wf\Phi(X)$
\end{lemma}

\begin{theorem}[Soundness of type inference]
\label{thm:soundness}
For all terms $e$, receiver types $\Xr$, and contexts $\Gamma$, if $\gen{\Xr,
\Gamma}{e}{\X}{\C}$ and $\satisfies{\Phi}{\C}$, then
$\types{\Phi(\Xr), \Phi(\Gamma)}{e}{\Phi(\X)}$.
\end{theorem}

\makeatletter{}

\makeatletter{}
\section{Evaluation}
\label{sec:cases}

We experimented with a number of standard benchmarks (\autoref{tab:bench}),
among them a selection
from the Octane suite~\cite{Octane} (the same ones used in recent papers on
TypeScript~\cite{safets} and ActionScript~\cite{iogti-RCH12}), several from the SunSpider suite~\cite{SunSpider}, and \bench{cdjs} from Jetstream~\cite{jetstream}.\footnote{For SunSpider, we chose all benchmarks that did not make use of \lstinline{Date} and \lstinline{RegExp} library routines, which we do not support.  For Octane, we chose all benchmarks with less than 1000 LOC.}
In all cases, our compiler relied on the
inferred types to drive optimizations. A separate developer team also created six apps for
the Tizen mobile OS (further details in \autoref{sec:workarounds}).  In all these programs, inference took between 1 and 10 seconds.
We have used type inference on additional programs as well,
which are not reported here;
our regression suite runs over a hundred programs.

\begin{table}[t]
  \scalebox{0.9}{
\begin{tabular}{|l|c||l|c|}
\hline
benchmark & size  & benchmark & size \\ \hline
\bench{access-binary-trees} & 41  & \bench{splay} & 230  \\
\bench{access-fannkuch} & 54  & \bench{crypto} & 1296  \\
\bench{access-nbody} & 145 & \bench{richards} & 290  \\
\bench{access-nsieve} & 33  & \bench{navier} & 355  \\
\bench{bitops-3bit-bits-in-byte} & 19  & \bench{deltablue} & 466  \\
\bench{bitops-bits-in-byte} & 20 & \bench{raytrace} & 672  \\
\bench{bitops-bitwise-and} & 7  & \bench{cdjs} & 684 \\
\cline{3-4}
\bench{bitops-nsieve-bits} & 29  & \bench{calc} & 979   \\
\bench{controlflow-recursive} & 22  & \bench{annex} & 688  \\
\bench{math-cordic} & 59  & \bench{tetris} & 826   \\
\bench{math-partial-sums} & 31  &  \bench{2048} & 507  \\
\bench{math-spectral-norm} & 45  & \bench{file} & 278  \\
\bench{3d-morph} & 26 & \bench{sensor} & 266  \\
\bench{3d-cube} & 301 & &   \\ \hline
\end{tabular}
}
\caption{Size is non-comment non-blank lines of code.
Programs from the Sunspider suite appear on the left, those from Octane and Jetstream on the top right,
and the Tizen apps on the bottom right.
}
\label{tab:bench}
\end{table}

All the features our type inference supports---structural subtyping,
prototype inheritance, abstract types, recursive object types, etc.---were necessary in even this small sampling of programs.
As one example, the \bench{raytrace} program from Octane stores items of two different
types in a single array; when read from the array, only an implicit ``supertype'' is
assumed.  Our inference successfully infers the common supertype. We also found the ability to infer types and find type errors in uninvoked functions
to be useful in writing new code as well as typing legacy code.

\subsection{Practical Considerations}

Our implementation goes beyond the core calculus to support a number of features needed to
handle real-world JavaScript programs.  For user code, the primary additional features are support
for constructors and prototype initialization (see discussion in \autoref{sec:workarounds}) and
support for polymorphic arrays and heterogeneous maps.  The implementation also supports manually-written type declarations for external libraries: such declarations are used to give types for JavaScript's built-in operators and standard libraries, and also for native platform bindings.  These type declaration files can include more advanced types that are not inferred for user-written functions, specifically types with parametric polymorphism and intersection types.  We now give further details regarding these extensions.
\makeatletter{}
\label{sec:practical-appendix}

\mypara{Maps and arrays}
JavaScript supports dictionaries, which are key-value pairs where keys are
strings (which can be constructed on the fly)\footnote{By
contrast, object fields are \textit{fixed} strings.} and values are of
heterogeneous types.  Our implementation supports maps, albeit with a homogeneous
polymorphic signature $\text{\lstinline{string}}\rightarrow \tau$, where $\tau$ is any type.
Our implementation permits array syntax (\lstinline{a[f]})
for accessing maps, but not for record-style objects.  Arrays are supported similarly, with the index type \lstinline{int} instead of \lstinline{string}.
Note that maps (and arrays) containing different types can exist in the same
program; we instantiate the $\tau$ at each instance appropriately.

\mypara{Constructors}
Even though we present object creation as allocation of object literals,
JavaScript programmers often use constructors.  A
constructor implicitly declares an object's fields via assignments to
fields of \lstinline{this}.  We handle constructors by distinguishing them syntactically (as functions with a capitalized name) and using syntactic analysis to discover which fields of \code{this} they write.

\mypara{Operator overloading}
JavaScript operators such as \lstinline{+} are heavily overloaded.
Our implementation includes a separate environment file with all permissible types for
such operators; the type checker selects the appropriate one, and the backend emits the required conversion.
Many of the standard functions are also overloaded in terms of
the number or types of arguments and are handled similarly.

\mypara{Generic and native functions}
Some runtime functions, such as an allocator for a
new array, are generic by nature.  Type inference instantiates
the generic parameter appropriately and ensures that arrays are
used consistently (per instance).
As this project arose from pursuing native performance for JavaScript
applications on mobile devices, we also support type-safe interfacing with
native platform functions via type annotations supplied in a separate environment file.

\subsection{Explanation of Workarounds}\label{sec:workarounds}
Our system occasionally requires workarounds for type inference to succeed.
The key workarounds needed for the Octane programs and \bench{cdjs} are
summarized in \autoref{tab:workaround};
our modified versions are available in
the supplementary materials for this paper. The SunSpider programs did
not require any major workaround.\footnote{A trivial workaround had to do with the current implementation requirement that only constructor names to begin with an uppercase letter.}
After these workarounds, types were inferred fully automatically.

\textbf{C} (Constructors).
JavaScript programs often declare a behavioral interface by
defining methods on a prototype, as follows:
\begin{lstlisting}
function C() { ... } // constructor
C.prototype.m1 = function () {...}
C.prototype.m2 = function () {...}
...
\end{lstlisting}
We support this pattern, provided that such field writes (including the write to the prototype field itself) appear immediately and contiguously after the constructor definition.
  Without this
restriction, we cannot ensure in a flow-insensitive type system  that the constructor is not invoked before all the prototype properties have been initialized.
The code refactoring required to accommodate this restriction is
straightforward (see~\autoref{fig:richards} for an example).
We did not see any cases in which the prototype was updated more than once.

\begin{figure}
  \small
\begin{lstlisting}[basicstyle=\ttfamily\footnotesize,commentstyle=\color{CommentColor}\ttfamily\footnotesize]
// Original
function TaskControlBlock(...) {
  this.link = link;
  this.id = id;
  this.priority = priority;
  this.queue = queue;
  this.task = task;
  ...
}
var STATE_RUNNING = 0;
...
TaskControlBlock.prototype.setRunning =
  function () {
    this.state = STATE_RUNNING;
  };
...
\end{lstlisting}

\begin{lstlisting}[basicstyle=\ttfamily\footnotesize,commentstyle=\color{CommentColor}\ttfamily\footnotesize]
// Refactored
var STATE_RUNNING = 0;
...
function TaskControlBlock(...) {
  this.link = link;
  this.id = id;
  this.priority = priority;
  this.queue = queue;
  this.task = task;
  ...
}
TaskControlBlock.prototype.setRunning =
  function () {
    this.state = STATE_RUNNING;
  };
...
\end{lstlisting}
\caption{Code fragment from \texttt{richards}. In the refactored code (below), we simply moved the constant declarations out of the way (\textbf{C}).}
\label{fig:richards}
\end{figure}

\begin{table}[t]
  \begin{small}
\begin{tabular}{|l|c|c|}
\hline
benchmark & workarounds & classes / types in TypeScript \\ \hline
\bench{splay} &  & 2 / 15\\
\bench{crypto} & \textbf{C},\textbf{U} & 8(1) / 142\\
\bench{richards} & \textbf{C} &7(1) / 30\\
\bench{navier}  &  &1(1) / 41\\
\bench{deltablue}  & \textbf{I}, \textbf{P} &12 / 61\\
\bench{raytrace} & \textbf{I} &14(1) / 48\\
\bench{cdjs} & \textbf{U}, \textbf{P} & --- \\\hline
\end{tabular}
\end{small}
\caption{Workarounds needed in selected Octane benchmarks and cdjs.
Each workaround impacted multiple lines of code.  For relevant benchmarks, the last column quotes from Rastogi et al.~\cite{safets} the number of classes (abstract ones in parentheses) and type annotations added to type check these programs in TypeScript.}
\label{tab:workaround}
\end{table}

\textbf{U} (Unions). Lack of flow sensitivity also precludes type checking (and inference) for
unions distinguished via a type test.  This feature is useful in JavaScript programs, and we encountered it in one of the Octane programs.
In the original \bench{crypto}, the \texttt{BigInteger} constructor may accept a number, or a
string and numeric base (arity overloading as well); we split the string case into a separate function, and updated call sites as appropriate.  For \bench{cdjs}, there were two places where the fields present in an object type could differ depending on the value of another field.  We changed the code to always have all fields present, to respect fixed object layout.

\textbf{P} (Polymorphism).  Although inferring polymorphic types is well understood in the context of
languages like ML, its limits are less well understood in a language
with mutable records and subtyping.  We do not attempt to
infer parametric polymorphism, although this feature
is known to be useful in JavaScript programs and did come up
in \bench{deltablue} and \bench{cdjs}.   We plan to support generic types via manual annotations, as we already do for environment functions.  For now, we worked around the issue with code duplication.  See~\autoref{fig:deltablue-p} for an example.

\begin{figure}
\begin{lstlisting}[basicstyle=\ttfamily\footnotesize,commentstyle=\color{CommentColor}\ttfamily\footnotesize]
Planner.prototype.removePropagateFrom =
  function (out) {
    out.determinedBy = null;
    out.walkStrength = Strength.prototype.WEAKEST;
    out.stay = true;
    var unsatisfied = new OrderedCollection();
    // Original
    // var todo = new OrderedCollection();
    var todo = new OrderedCollectionVariable(); /*@ \label{li:ocv} @*/
    todo.add(out);
  };
\end{lstlisting}

\caption{Excerpt from modified \bench{deltablue}.  \lstinline{OrderedCollection}s
were being populated with different types, which cannot be typed without parametric polymorphism.  As a workaround, a duplicate type \lstinline{OrderedCollectionVariable} was created, and appropriate sites (like
\autoref{li:ocv} above) were changed to use the new type.}
\label{fig:deltablue-p}
\end{figure}

\textbf{I} (Class-based Inheritance). Finally, JavaScript programs often use an \textit{ad hoc}
encoding of class-based inheritance: programmers develop their own shortcuts (or use libraries)
that use ``monkey patching''\footnote{``Monkey patching'' here refers to adding previously
non-existent methods to an object (violating fixed layout) or modifying the pre-existing methods of
global objects such as \texttt{Object.prototype} (making code difficult to read accurately, and
thwarting optimization of common operations).  Our system permits dynamic update of \emph{existing}
methods of developer-created objects, preserving fixed layout.
} and
introspection.  We cannot type these constructs,
but our type system can support class-based inheritance via prototypal
inheritance, with some additional verbosity (see \autoref{fig:deltablue-i}).
The latest JavaScript specification includes class-based inheritance, which obviates the need for encoding classes by other means.  We
intend to support the new class construct in the future.

\begin{figure}
\small
\begin{lstlisting}[basicstyle=\ttfamily\footnotesize,commentstyle=\color{CommentColor}\ttfamily\footnotesize]
// Original
Object.defineProperty(Object.prototype,
  "inheritsFrom", ...)
function EqualityConstraint(var1, var2, strength) {
  EqualityConstraint.superConstructor
    .call(this, var1, var2, strength);
}
EqualityConstraint.inheritsFrom(BinaryConstraint);
\end{lstlisting}
\begin{lstlisting}[basicstyle=\ttfamily\footnotesize,commentstyle=\color{CommentColor}\ttfamily\footnotesize]
// Refactored
function EqualityConstraintInheritor() {
  this.execute = null;
}
EqualityConstraintInheritor.prototype =
  BinaryConstraint.prototype;
function EqualityConstraint(var1, var2, strength) {
  this.strength = strength;
  this.v1 = var1;
  this.v2 = var2;
  this.direction = Direction.NONE;
  this.addConstraint();
}
EqualityConstraint.prototype =
  new EqualityConstraintInheritor();
\end{lstlisting}
\caption{Excerpt showing a change in \texttt{deltablue} to work around \emph{ad hoc} class-based inheritance. The refactored code (bottom) avoids monkey-patching \code{Object} with a new introspective
method \code{inheritsFrom}.}
\label{fig:deltablue-i}
\end{figure}

\mypara{Usability by developers.} With our inference system, developers
remain mostly unaware of the types being inferred, as the inference is
automatic and no explicit type ascription is generated.  For inference failures, 
we invested significant effort to provide useful error messages~\cite{lin16explanation} that
were understandable without knowledge of the underlying type theory.  
While some more complex concepts like
intersection types are needed to express types for certain library routines,
these types can be written by specialists, so developers solely interacting
with the inference need not deal with such types directly.

More concretely, the Tizen apps listed in \autoref{tab:bench} were created by a
team of developers who were not experts in type theory.  The apps required
porting of code from existing web applications (e.g., for \bench{tetris} and
\bench{2048}) as well as writing new UI code leveraging native Tizen APIs.  To
learn our subset of JavaScript, the developers primarily used a manual we wrote
that described the restrictions of the subset without detailing the type
inference system;
\iflong \autoref{app:developer} gives more details on this manual.
\fi

\subsection{More Problematic Constructs}

Certain code patterns appearing in common JavaScript frameworks make heavy use of JavaScript's dynamic typing and introspective features; such code is difficult or impossible to port to our typed subset.
As an example, consider the \bench{json2.js}
program,\footnote{\url{https://github.com/douglascrockford/JSON-js}} a variant of which appears in Crockford~\cite{goodparts}.  A core computation in the program, shown in \autoref{fig:example-json}, consists of a loop to traverse a JSON data structure and make in-place substitutions.  In JavaScript, arrays are themselves objects, and like objects, their contents can be traversed with a for-in loop.  Hence, the single loop at \autoref{li:forin} applies equally well to arrays and objects.  Also note that in different invocations of \lstinline{walk}, the variable \lstinline{v} may be an array, object, or some value of primitive type.

\begin{figure}
\small
\begin{lstlisting}[basicstyle=\ttfamily\footnotesize,commentstyle=\color{CommentColor}\ttfamily\footnotesize]
  function walk(k, v) {
      var i, n;
      if (v && typeof v === object) {
          for (i in v) { /*@ \label{li:forin} @*/
              n = walk(i, v[i]);
              if (n !== undefined) {
                  v[i] = n;
              }
          }
      }
      return filter(k, v);
  }
\end{lstlisting}
\caption{JSON structure traversal.}
\label{fig:example-json}
\end{figure}

Our JavaScript subset does not allow such code. We were able to write an equivalent routine in our subset only after significant refactoring to deal with maps and arrays separately, as shown in \autoref{fig:example-json-typed}; moreover, we had to ``box'' values of different types into a common type to enable the recursive calls to type check.  Clearly, this version loses the economy of expression of dynamically-typed JavaScript.

\begin{figure}
\small
\begin{lstlisting}[basicstyle=\ttfamily\footnotesize,commentstyle=\color{CommentColor}\ttfamily\footnotesize]
function JSONVal() {
    this.tag = ...
    this.a = null; // array
    this.m = null; // map
    this.intval = 0; // int value
    this.strval = ""; // string value
}

function walk(k, v) { // v instance of JSONVal
  var i, j, n;
  switch (v.tag) {
    case Constants.INT:
    case Constants.STR:
      break;
    case Constants.MAP:
      for (var i in v.m) {
        n = walk(i, v.m[i]);
        if (n !== undefined) { v.m[i] = n; }
      }
      break;
    case Constants.ARRAY:
      for (j = 0; j < v.a.length; j++) {
        // j+"" converts j to a string
        n = walk(j+"",v.a[j]);
        if (n !== undefined) { v.a[j] = n; }
      }
      break;
  }
  return filter(k,v);
}
\end{lstlisting}
\caption{JSON structure traversal in our subset of JavaScript.}
\label{fig:example-json-typed}
\end{figure}

\begin{figure}
\small
\begin{lstlisting}[basicstyle=\ttfamily\footnotesize,commentstyle=\color{CommentColor}\ttfamily\footnotesize]
Object.prototype.extend = function (dst, src) {
   for (var prop in src) {
      dst[prop] = src[prop];
   }
}
\end{lstlisting}
\caption{\lstinline{extend} in JavaScript}
\label{fig:extends}
\end{figure}

JavaScript code in frameworks (even non-web frameworks like \bench{underscore.js}\footnote{\url{http://underscorejs.org/}}) is often written in a highly introspective style, using constructs not supported in our subset.  One common usage is extending an object's properties in-place using the pattern shown in \autoref{fig:extends}.
The code treats all objects---including those meant to be used as structs---as maps.
Moreover, it also can add properties to \lstinline{dst} that may not have been present previously, violating fixed-object layout. We do not support such routines in our subset.

As mentioned before, the full JavaScript language includes constructs such as
\texttt{eval} that are fundamentally incompatible with ahead-of-time
compilation.  We also do not support adding or modifying behavior (aka ``monkey
patching'') of built-in library objects like \texttt{Object.prototype} (as is
done in \autoref{fig:extends}). 
The community considers such usage as bad practices~\cite{goodparts}.

Even if we take away these highly dynamic features, there is a price to be paid for obtaining type information for JavaScript statically: either a programmer stays within a subset that admits automatic inference, as explored in this paper and requiring the workarounds of the kinds described in \autoref{sec:workarounds}; or, the programmer  writes strong enough type annotations (the last column of \autoref{tab:workaround} shows the effort required in adding such annotations for the same Octane programs in~\cite{safets}).

Whether this price is worth paying ultimately depends on the value one attaches to the benefits offered by ahead-of-time compilation.

\subsection{The Promise of Ahead-of-Time Compilation}

As mentioned earlier, we have implemented a compiler that draws upon the information computed by type inference (\autoref{sec:background}) and generates optimized code.  The details of the compiler are outside the scope of the paper, but we present preliminary data to show that AOTC for JavaScript yields advantages for resource-constrained devices.

\begin{figure}
  \vspace{-10pt}
  \includegraphics[trim=165 185 180 170,clip,width=\columnwidth]{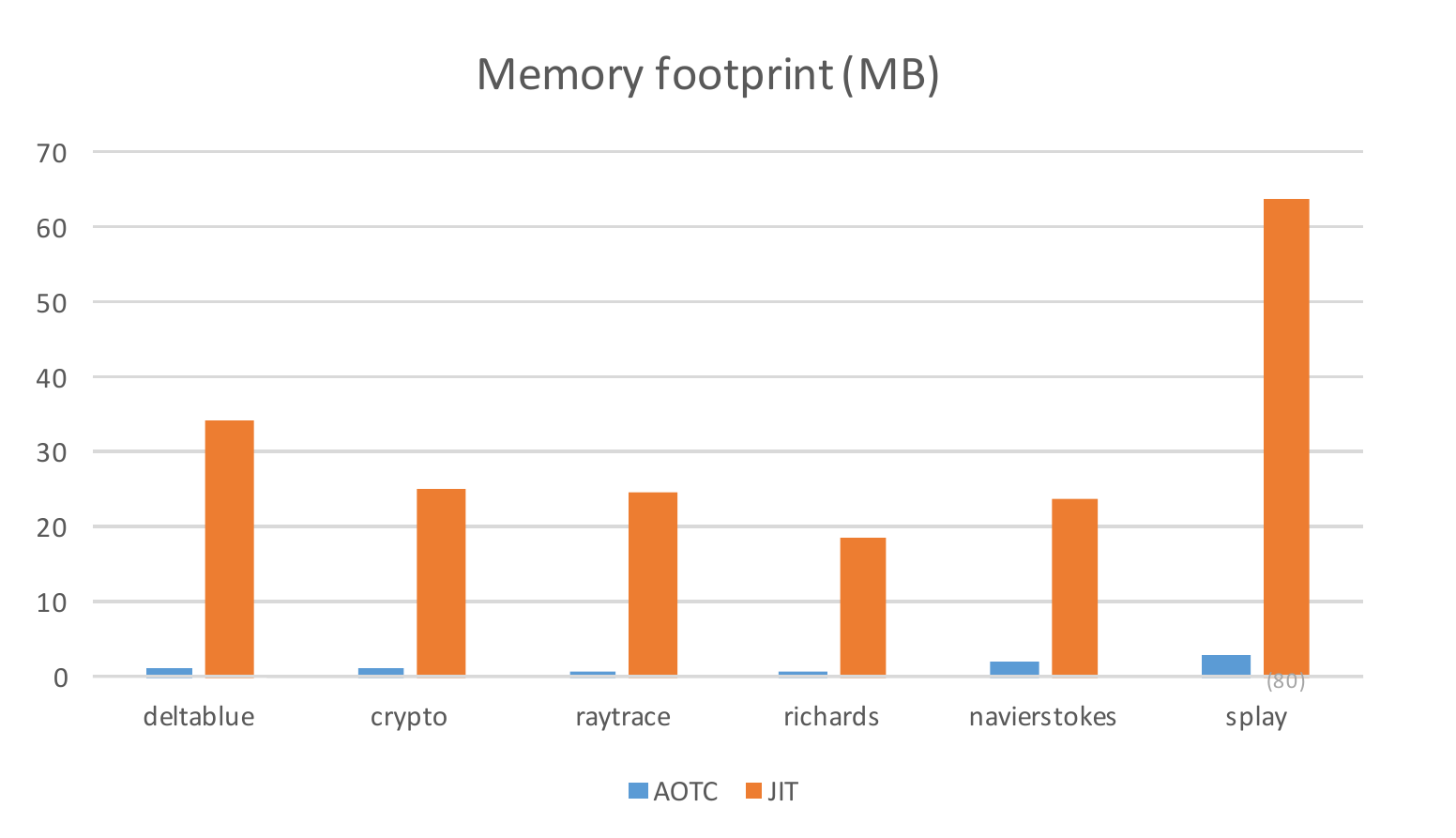}
  \caption{Memory comparison
    }
  \label{fig:memcmp}
  \end{figure}

We measured the space consumed by the compiled program against the space consumed by the program running on v8, a modern just-in-time compiler for JavaScript.  The comparative data is shown in \autoref{fig:memcmp}.  The Octane programs were run with their default parameters.\footnote{Except for \bench{splay}, which we ran for 80 as opposed to 8000 elements; memory consumption in \bench{splay} is dominated by program data.}  As the figure shows, ahead-of-time compilation yielded significant memory savings vs. just-in-time compilation.

\begin{figure}
  \vspace{-2pt}
  \includegraphics[trim=120 100 150 150,clip,width=0.8\columnwidth]{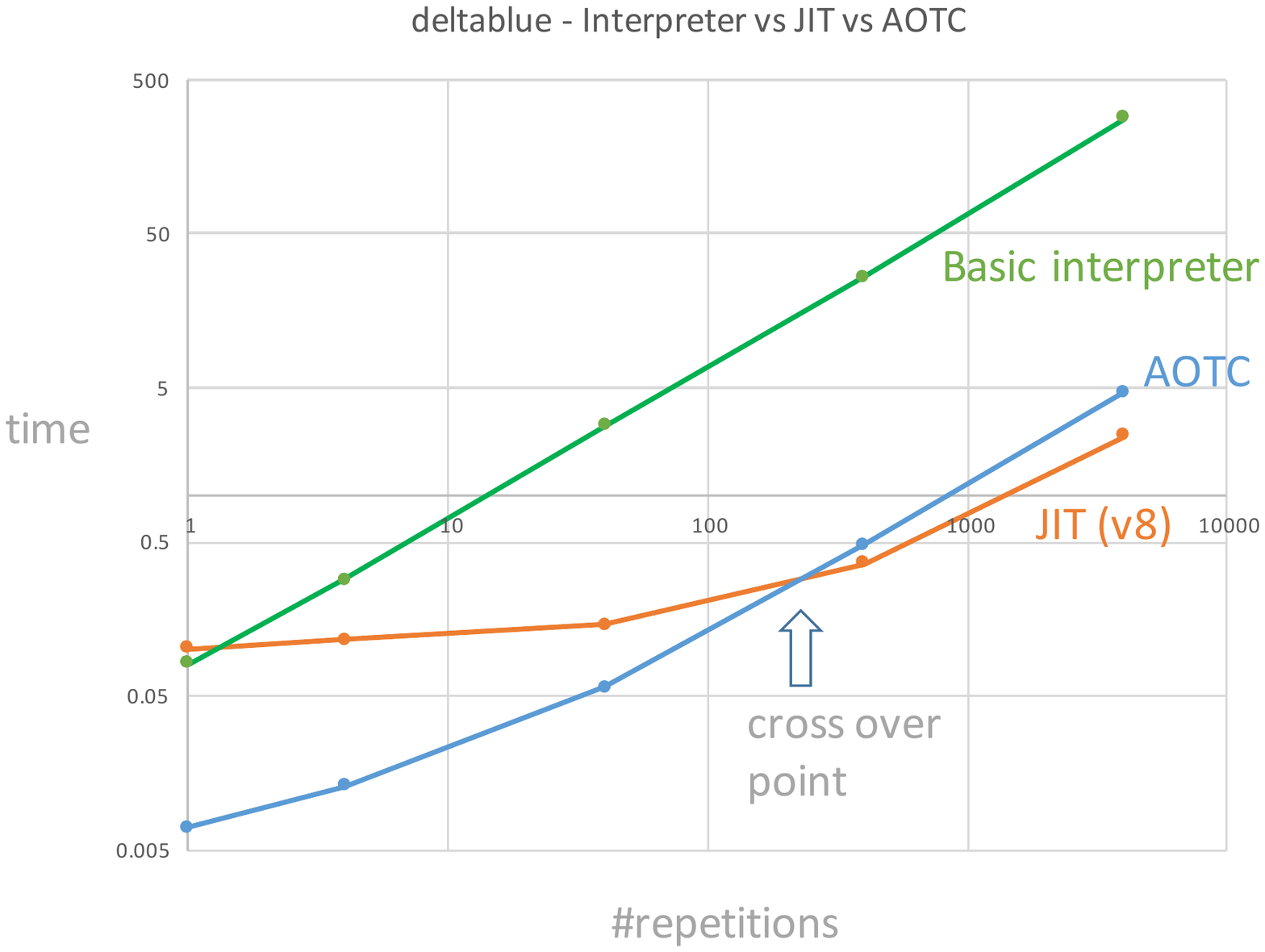}
  \caption{Running times for \bench{deltablue}. Note the log-log scale.
    }
  \label{fig:deltablue}
  \end{figure}

\begin{figure}
  \includegraphics[trim=20 0 180 0,clip,width=\textwidth]{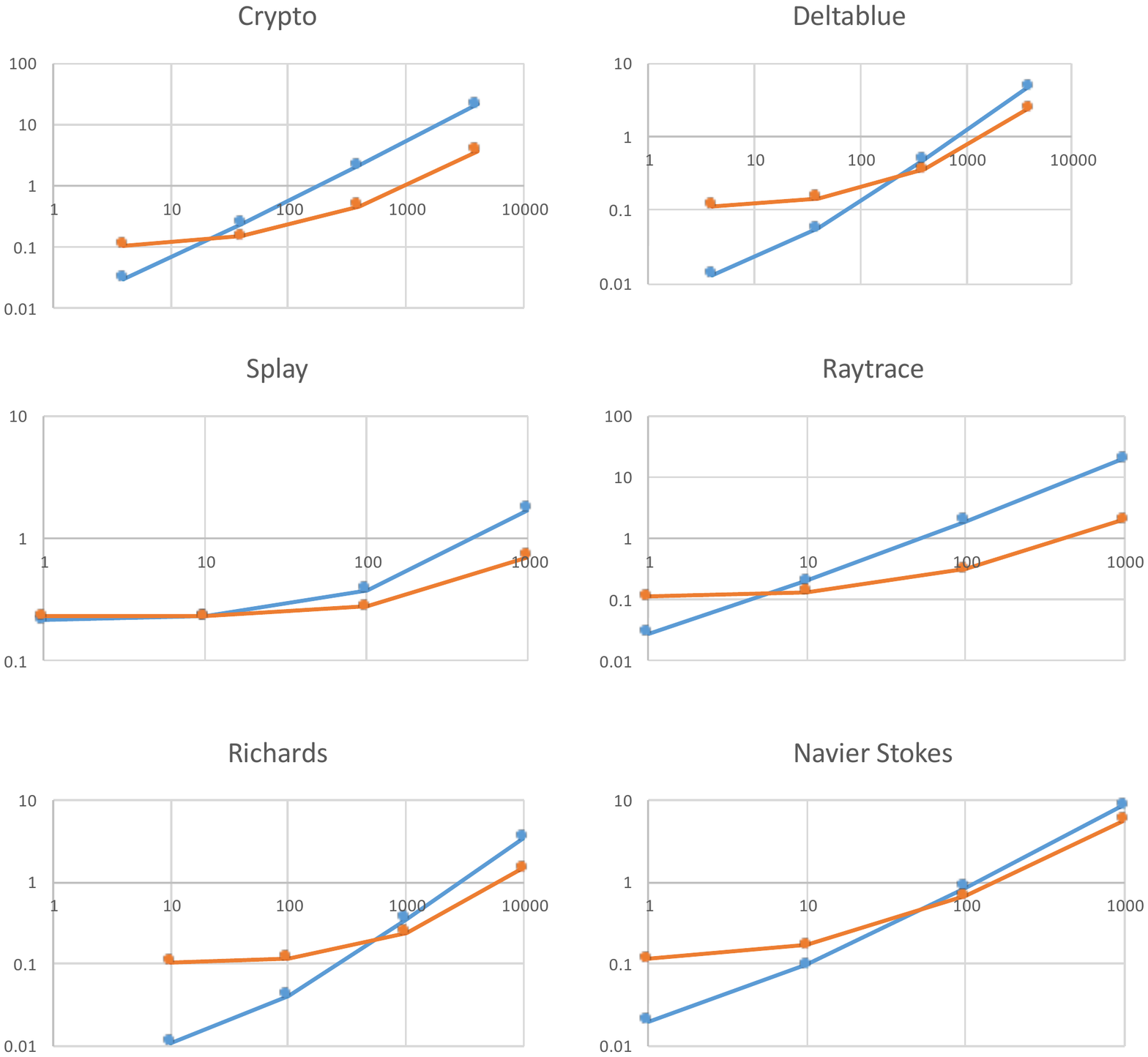}
  \caption{Crossover behavior of our AOTC system vs. the v8 runtime for the six Octane programs.}
\label{fig:six}
\end{figure}

We also timed these benchmarks for runtime performance on AOTC compiled binaries and the v8 engine.
\autoref{fig:deltablue} shows the results for one of the programs, \bench{deltablue}; the figure also includes running time on \textit{duktape}, a non-optimizing interpreter with a compact memory footprint.  We observe that (i) the non-optimizing interpreter is quite a bit slower than the other engines, and (ii) for smaller numbers of iterations, AOTC performs competitively with v8.  For larger iteration counts, v8 is significantly faster.  Similar behavior was seen for all six Octane programs (see \autoref{fig:six}).  The AOTC slowdown over v8 for the largest number of runs ranged from 1.5X (\bench{navier}) to 9.8X (\bench{raytrace}).  We expect significant further speedups from AOTC as we improve our optimizations and our garbage collector.
Full data for the six Octane programs, both for space and time, are presented
in
\iflong \autoref{sec:performance}.
\fi

\mypara{Interoperability.} In a number of scenarios, it would be useful for compiled code from our JavaScript subset to interoperate with unrestricted JavaScript code.  The most compelling case is to enable use of extant third-party libraries without having to port them, e.g., frameworks like jQuery\footnote{\url{http://jquery.com}} for the web\footnote{Note that running our compiled code in a web browser would require an implementation of the DOM APIs, which our current implementation does not support.} or the many libraries available for Node.js.\footnote{\url{http://nodejs.org}}  Additionally, if a program contains dynamic code like that of \autoref{fig:example-json} or \autoref{fig:extends}, and that code is not performance-critical, it could be placed in an unrestricted JavaScript module rather than porting it.

Interoperability with unrestricted JavaScript entails a number of interesting tradeoffs.  The
simplest scheme would be to invoke unrestricted JavaScript from our subset (and vice versa) via a
foreign function interface, with no shared heap.  But, this would impose a high cost on such calls,
due to marshalling of values, and could limit expressivity, e.g., passing functions would be
difficult.  Alternately, our JavaScript subset and unrestricted JavaScript could share the same
heap, with additional type checks to ensure that inferred types are not violated by the
unrestricted JavaScript.  The type checks could ``fail fast'' at any violation, like in other work
on gradual typing~\cite{safets,siek2015monotonic,vitousek2014design}.  But, this could lead to application behavior differing on our runtime versus a standard JavaScript runtime, as the standard runtime would not perform the additional checks.  Without ``fail fast,'' the compiled code may need to be deoptimized at runtime type violations, adding significant complexity and potentially slowing down code with no type errors.  At this point, we have a work-in-progress implementation of interoperability with a shared heap and ``fail fast'' semantics, but a robust implementation and proper evaluation of these tradeoffs remain as future work.

\makeatletter{}
\section{Related Work}
\label{sec:related}

Related work spans type systems and inference for JavaScript and dynamic languages in general, as
well as the type inference literature more broadly.

\mypara{Type systems and inference for JavaScript.}
Choi et al.~\cite{DBLP:conf/sas/ChoiCNS15,SJSTechReport} presented a typed
subset of JavaScript for ahead-of-time compilation.  Their work served as our
starting point, and we built on it in two ways.  First, our type system extends
theirs with features that we found essential for real code, most crucially
abstract types (see discussion throughout the paper).  We also present a
formalization and prove these extensions sound
\iflong (\autoref{app:type_system_metatheory}).
\fi
Second, whereas they relied on
programmer annotations to obtain types, we developed and implemented an
automatic type inference algorithm.

Jensen \etal~\cite{JensenMT09} present a type analysis for JavaScript based on
abstract interpretation. They handle prototypal inheritance soundly.  While their analysis could be adapted for compilation, it does not give a typing
discipline.  Moreover, their dataflow-based technique cannot handle partial programs, as discussed in \autoref{sec:complications}.

TypeScript~\cite{typescript} extends JavaScript with type annotations, aiming to expose bugs and improve developer productivity.  To minimize adoption costs, its type system is very expressive but deliberately unsound.  Further, it requires type annotations at function boundaries, while we do global inference.
Flow~\cite{flow} is another recent type system for JavaScript, with an emphasis on effective
flow-sensitive type inference.  Although a detailed technical description is unavailable at the
time of this writing, it appears that our inference technique has similarity to Flow's in its use
of upper- and -lower bound propagation~\cite{flow:personal}.  Flow's type language is similar to
that of TypeScript, and it also sacrifices strict soundness in the interest of usability.
It would be possible to create a sound gradually-typed version of Flow (i.e., one with
dynamic type tests that may fail), but this would not enforce fixed object layout.
For TypeScript, a sound gradually-typed variant already exists~\cite{safets}, which we discuss shortly.

Early work on type inference for JavaScript by Thiemann~\cite{Thiemann:2005} and Anderson \etal~\cite{AndersonGD05} ignored essential language features such as prototype
inheritance, focusing instead on dynamic operations such as property addition.
Guha \etal~\cite{guha2010essence} present a core calculus $\lambda_{\text{\it JS}}$ for JavaScript, upon which a number of type systems have been based.  TeJaS~\cite{tejas} is a
framework for building type checkers over $\lambda_{\text{\it JS}}$ using bidirectional type checking to
provide limited inference.
Politz \etal~\cite{Politz12} provide a type system enforcing access safety for a language with JavaScript-like dynamic
property access.

Bhargavan \etal~\cite{djs} develop a sound type system and inference for Defensive JavaScript (DJS), a JavaScript subset aimed at security embedding code in untrusted web pages.
Unlike our work, DJS forbids prototype inheritance, and their type inference technique is not described in detail.

\mypara{Gradual typing for JavaScript.}
Rastogi \etal~\cite{iogti-RCH12} give a constraint-based formulation of type inference for ActionScript,
a gradually-typed class-based dialect of JavaScript.  While they use many related techniques---their work and ours are inspired by Pottier~\cite{pottier-icfp-98}---their gradually-typed
setting leads to a very different constraint system.  Their (sound) inference
aims at proving runtime casts safe, so they need not validate upper bound constraints.
They do not handle prototype inheritance, relying on ActionScript classes.

Rastogi \etal~\cite{safets} present Safe TypeScript, a
\emph{sound, gradual} type system for TypeScript. After running TypeScript's (unsound) type inference, they run their (sound) type checker and
insert runtime checks to ensure type safety. Richards \etal~\cite{richards_et_al:LIPIcs:2015:5218} present StrongScript, another TypeScript extension with sound gradual typing.
They allow the programmer to enforce soundness of some (but not all) type annotations using
a specific type constructor, thus preserving some flexibility.
They also use sound types to improve compilation and performance. Being based on TypeScript, both systems require
type annotations, while we do not (except for signatures of external library functions).
Moreover, they do not support general prototype inheritance or mutable methods, but rather
rely on TypeScript's classes and interfaces.

\mypara{Type inference for other dynamic languages.}
Agesen \etal~\cite{SelfTypeInference} present inference for Self,
a key inspiration for JavaScript which includes prototype inheritance. Their constraint-based approach is inspired by
Palsberg and Schwartzbach~\cite{Palsberg:1991:OTI:117954.117965}.
However, their notion of type is a set of
values computed by data flow analysis, rather than syntactic typing discipline.

\mypara{Foundations of type inference and constraint solving.}
Type inference has a long history, progressing from early work~\cite{damas1982principal} through
record calculi and row variables~\cite{wand1987complete,wand1989type} through more modern
presentations.  Type systems for object calculi with object extension (e.g., prototype-based inheritance)
and incomplete (abstract) objects extends back to the late
1990s~\cite{bono1996lambda,bono1997subtyping,remy1998classes,fisher1995delegation}.  To our
knowledge, our system is the first to describe inference for a language with both abstract objects and prototype inheritance.

Trifonov and Smith~\cite{DBLP:conf/sas/TrifonovS96} describe constraint generation and
solving in a core type system where (possibly recursive)
types are generated by base types, $\bot$, $\top$ and $\rightarrow$ only.
They introduce techniques for removing redundant constraints and optimizing
constraint representation for faster type inference.
Building on their work, Pottier~\cite{pottier-icfp-98,pottier-ic-01} crisply describes
the essential ideas for subtyping constraint simplification and resolution in a similar
core type system. We do not know of any previous generalization of this work that handles prototype
inheritance.
In both of these systems, lower and upper bounds for each type variable are already defined while resolving and simplifying constraints.
Both lines of work support partial programs, producing schemas with arbitrary constraints rather than an established style of polymorphic type.

Pottier and R\'emy~\cite{pottier-remy-emlti} describe type inference for ML, including records, polymorphism, and references.
R\'emy and Vouillon~\cite{Remy-Vouillon/tapos} describe type inference for class-based
objects in Objective ML.
These approaches are based on row polymorphism rather than
subtyping, and they do not handle prototype inheritance or non-explicit subtyping.

Aiken~\cite{solving_set_constraints} gives an overview of program analysis in the
general framework of set constraints, with applications to dataflow analysis and simple
type inference.
Most of our constraints would fit in his framework with little adaptation, and
his resolution method also uses lower and upper bounds. His work is general and does not look into specific program construct details like objects,
or a specific language like JavaScript.

\section*{Acknowledgements} We thank the anonymous reviewers for their detailed feedback, which significantly improved the presentation of the paper.

 {}
\bibliographystyle{plainnat}

\iflong
\appendix

\clearpage
\makeatletter{}
\section{A developer-centric view}
\label{app:developer}

In this section, we give a developer-centric view of our JavaScript subset, without
resorting to details of type inference.  We produced similar
documentation for the developer team who created the native Tizen applications discussed in \autoref{sec:cases}.

From a developer's perspective, our subset restricts JavaScript in the following ways:
\begin{enumerate}
 \item Each variable, object property, or parameter can be assigned values only of a
 single type throughout the execution.  Here, \textit{type} denotes distinct
 categories of values, such as numbers, booleans, strings, and objects.

 \item  An object's type depends on the set of properties it contains.  It is
 acceptable to supply an object with more properties in a context in which
 an object type with fewer properties was expected, as long as the types of
 common properties are the same (akin to ``upcasting''); however, the other way
 around (``downcasting'') is not allowed.

 \item Although JavaScript makes little distinction between use of objects
 as records, maps, and arrays, our subset does.  Record usage is suitable
 when the set of properties is known ahead of time and fixed.  Map usage is suitable
 when the set of properties (or ``keys'') can be added and deleted; however,
 the values stored against those keys must all be of a common type.  Arrays are indexed
 with numeric indices, as opposed to maps which are indexed by string indices.
 We use the \lstinline{[]} accessor for maps and arrays, while the '\lstinline{.}' accessor
 is only for structs.  Moreover, iteration is only supported over maps and arrays; it is not possible to iterate over properties of a record using a for-in loop.

 \item When using objects as records, properties cannot be added or removed after creation.  This means that all the properties needed on that object throughout the execution
 of the program should be defined in the corresponding constructor or object literal.

 \item Properties inherited via prototype inheritance are \textit{read only}, that is, they cannot be modified using a reference to the inheritor.  (They can still be modified using a reference
 to the prototype.)  If a method in a prototype writes to a property, the property must be \textit{re-declared} in the inheritor.

 \item Prototype inheritance can be carried out only from a variable whose provenance
 does not include an intervening (implicit) upcast. That is to say, the object type of
 the variable should coincide with the actual runtime value that that variable holds.

 \item Although built-in types such as \lstinline{Array} and \lstinline{String} are available for use,
 their properties cannot be ``monkey patched'' using assignment to the corresponding
 prototypes.

 \item \texttt{eval} and \texttt{with} are not supported, and neither is the \texttt{arguments}
array.  Low-level ways of manipulating functions as objects are also not supported.\footnote{A complete list was provided in the user manual.}
\end{enumerate}

The above discipline has to be followed statically.  Code such
as \lstinline{var v = false ? "hello" : 1} is not acceptable, even though at runtime \lstinline{v} can only hold a numeric value.  The type checker will reject a program
if it cannot statically reason about the safety of the program, though sometimes it may appear to
be overly conservative.

\makeatletter{}
\section{Type system metatheory}
\label{app:type_system_metatheory}

This section presents the full type system in \autoref{fig:typing}, along with
a full proof of soundness of our type inference (proof of \autoref{thm:soundness}),
 an extension to handle recursive types, and a proof of soundness of the type system.

\autoref{fig:extra-generation} shows constraint generation rules for handling of first-class functions, which were omitted from
\autoref{fig:constraint_generation}.
Bound propagation and ascription can be extended to function types in a way similar
to method types.

\renewcommand{\MathparLineskip}{\lineskiplimit=6pt\lineskip=6pt}
\begin{figure*}
\begin{mathpar}
\inferrule*[left={T-Int}]{ }{ \types{R, \Gamma}{n}{\kw{int}} }
\and
\inferrule*[left={T-Var}]{
  \Gamma(x) = \tau
} {
  \types{R, \Gamma}{x}{\tau}
}
\and
\inferrule*[left={T-VarUpd}]{
  \Gamma(x) = \tau_1 \quad
  \types{R, \Gamma}{e}{\tau} \quad
  \subt{\tau}{\tau_1}
} {
  \types{R, \Gamma}{\ass{x}{e}}{\tau}
}
\and
\inferrule*[left={T-VarDecl}]{
  \types{R, \override{\Gamma}{x}{\tau_1}}{e_1}{\sigma_1} \quad
  \wf{\tau_1} \quad
  \subt{\sigma_1}{\tau_1} \\
  \types{R, \override{\Gamma}{x}{\tau_1}}{e_2}{\tau}
} {
  \types{R, \Gamma}{\letin{x}{e_1}{e_2}}{\tau}
}
\and
\myinferLeft{T-This}{
} {
  \types{\nu, \Gamma}{\this}{\nu}
}
\and
\myinferLeft{T-Fun}{
  \types{\top, \override{\Gamma}{x}{\tau_1}}{e}{\tau_2} \quad
  \wf{\tau_1}
} {
  \types{R, \Gamma}{\fun{x:\tau_1}{e}}{\funt{\tau_1}{\tau_2}}
}
\and
\myinferLeft{T-Meth}{
  \types{\nu, \override{\Gamma}{x}{\tau_1}}{e}{\tau_2} \quad
  \wf{\tau_1}
} {
  \types{R, \Gamma}{\fun{x:\tau_1}{e}}{\metht{\nu}{\tau_1}{\tau_2}}
}
\and
\myinferLeft{T-FCall}{
  \types{R, \Gamma}{e_1}{\funt{\tau_3}{\tau}} \\
  \types{R, \Gamma}{e_2}{\tau_2} \quad
  \subt{\tau_2}{\tau_3}
} {
  \types{R, \Gamma}{\app{e_1}{e_2}}{\tau}
}
\and
\myinferLeft{T-Null}{ \subt{\tau}{\objt{\emptyrow}{\emptyrow}^\notprotoabs}}{ \types{R, \Gamma}{\mynull}{\tau} }
\and
\myinferLeft{T-Attr}{
  \subt{\tau_1}{\objt{\rowtype{\f:\tau}}{\emptyrow}^\notprotoabs} \\
  \types{R, \Gamma}{e}{\tau_1} \quad
  \tau \neq \metht{\cdot}{\tau_2}{\tau_3}
} {
  \types{R, \Gamma}{e.\f}{\tau}
}
\and
\myinferLeft{T-AttrUpd}{
  \typesau{R, \Gamma}{\tau_b.\f:=e}{\tau} \quad
  \types{R, \Gamma}{e_1}{\tau_b}
} {
  \types{R, \Gamma}{\ass{e_1.\f}{e}}{\tau}
}
\and
\myinferLeft{T-MCall}{
  \types{R, \Gamma}{e_1}{\tau_1} \quad
  \types{R, \Gamma}{e_2}{\tau_2} \\
  \subt{\tau_1}{\objt{\rowtype{\f:\metht{\cdot}{\tau_3}{\tau}}}{\emptyrow}^\notprotoconc} \quad
  \subt{\tau_2}{\tau_3}
} {
  \types{R, \Gamma}{\app{e_1.\f}{e_2}}{\tau}
}
\and
\myinferLeft{T-Emp}{
} {
  \types{\R, \Gamma}{\obj{\cdot}}{\objt{\emptyrow}{\emptyrow}^\prototypal{\emptyrow}{\emptyrow}}
}
\and
\myinferLeft{T-ObjLit}{
  \wf{\rho^q} \\
  \rho = \objt{\meta{r}}{\meta{w}} \\
  \qual = \prototypal{\meta{mr}}{\meta{mw}} \\
  \types{\R, \Gamma}{e_p}{{\objt{\meta{r_p}}{\meta{w_p}}}^{\prototypal{\meta{mr_p}}{\meta{mw_p}}}}
    \\
  \meta{w} = \rowtype{\f_1:\tau_1,\ldots,\f_n:\tau_n} \quad \\
  \forall i \in 1..n.~\typesau{\R, \Gamma}{\ass{\rho^\qual.\f_i}{e_i}}{\sigma_i} \wedge
  \subt{\sigma_i}{\tau_i} \\
  {\meta{r_p}\cup\meta{w}}={\meta{r}} \\
  \subt{\meta{mr}}{\meta{mr_p}} \\
  \subt{\meta{mw}}{\meta{mw_p}}
} {
  \types{\R, \Gamma}{\proto{\obj{\f_1 : e_1, \ldots, \f_n : e_n}_{\rho^\qual}}{e_p}}{\rho^\qual} 
}
\and
\myinferLeft{T-AttrUpdV}{
  \subt{\tau_b}{\objt{\rowtype{\f: \tau_f}}{\rowtype{\f}}^\notprotoabs} \\
  \types{\R, \Gamma}{e}{\tau} \\
  \tau \neq \metht{\tau_x}{\tau_y}{\tau_z} \\
  \subt{\tau}{\tau_f}
} {
  \typesau{\R, \Gamma}{\tau_b.\f := e}{\tau}
}
\and
\myinferLeft{T-AttrUpdM}{
\subt{\tau_b}{\objt{\rowtype{\f : \metht{\cdot}{\tau_1}{\tau_2}}}{\rowtype{\f}}^\notprotoabs} \\
  \types{\R, \Gamma}{e}{\metht{\objt{\meta{r_r}}{\meta{w_r}}^{\notprotoconc}}{\tau_1}{\tau_2}} \\
  \tau_b = \rho^\qual \quad \qual = \prototypal{\meta{mr}}{\meta{mw}} \\
  \subt{\meta{mr}}{\meta{r_r}} \quad \subt{\meta{mw}}{\meta{w_r}}
} {
  \typesau{\R, \Gamma}{\tau_b.\f := e}{\metht{\objt{\meta{r_r}}{\meta{w_r}}^{\notprotoconc}}{\tau_1}{\tau_2}}
}
\end{mathpar}
\caption{Typing judgement. 
}
\label{fig:typing}
\end{figure*}

\begin{figure}
\begin{mathpar}
\myinferA{C-FunDecl}{
  \fresh{\X_1, \Y, \Xr, \X} \quad
  \kw{!}~\hasthis{e} \\
  \gen{\Xr, \override{\Gamma}{x}{\X_1}}{e}{\Y}{\C}
} {
  \gen{\R, \Gamma}{\fun{x}{e}}{\X}{} && \C \wedge \sr{\Xr} \equiv
                                          \toprow \arcr
  & \wedge & \sr{\X} \equiv (\funt{\X_1}{\Y})
}
\and
\myinferA{C-FunApp}{
  \fresh{\X_3, \X} \\
  \gen{\R, \Gamma}{e_1}{\X_1}{\C_1} \qquad
  \gen{\R, \Gamma}{e_2}{\X_2}{\C_2}
} {
  \gen{\R, \Gamma}{\app{e_1}{e_2}}{\X}{} & & \C_1 \wedge \C_2
    \wedge   \sr{\X_1} \equiv (\funt{\X_3}{\X}) \arcr
& & \wedge \rwsubt{\X_2}{\X_3}
}
\end{mathpar}
\caption{Constraint generation rules for functions.}
\label{fig:extra-generation}
\end{figure}

\subsection{Soundness of type inference}

\subsubsection{Proof of \autoref{lem:ascription}}
\begin{proof}
Let us consider a constraint $C\in\mathcal C$, by \autoref{it:incl}, $C\in\mathcal C'$.
The proof proceeds by case analysis over the form of the constraint $C$.
Note that Rules \ref{it:uplo}-\ref{it:upmethod}, not essential for soundness, are indeed not
used in the proof.

\begin{itemize}
\item If $C$ is of the form $\subt{\svar X}{L}$, then by \autoref{it:ub}, $L\in\ub{\svar X}$.
Therefore $\Phi(\svar X)=\subt{\kw{glb}(\ub{\svar X})}{L}$ (see \autoref{alg:type-ascription},
Line \ref{as:glb}).
\item If $C$ is of the form $\subt{L}{\svar X}$, then by \autoref{it:lb}, $L\in\lb{\svar X}$.
At type ascription (Line \ref{as:lb}) we checked that $\subt{L}{\Phi(X^s)}$.
\item If $C$ is of the form $\subt{X^s}{Y^t}$, then by \autoref{it:subt},
$\lb{X^s}\subseteq\lb{Y^t}$ and $\ub{Y^t}\subseteq\ub{\svar X}$. Note that by Rules
\ref{it:top} and \ref{it:bot}, either both $\ub{X^s}$ and $\lb{X^s}$ are empty, or both
are non-empty, and similarly for $\ub{Y^t}$ and $\lb{Y^t}$.
\begin{itemize}
\item If $\lb{X^s}=\ub{X^s}=\emptyset$ then $\ub{Y^t}=\emptyset$ thus
$\lb{Y^t}=\emptyset$ and
$\Phi(X^s)=\subt{\kw{default}}{\kw{default}}=\Phi(Y^t)$.
\item Symmetrically, if $\lb{Y^t}=\ub{Y^t}=\emptyset$ then $\lb{X^s}=\emptyset$ therefore
$\ub{X^s}=\emptyset$ and
$\Phi(X^s)=\subt{\kw{default}}{\kw{default}}=\Phi(Y^t)$.
\item Otherwise $\ub{X^s}$ and $\ub{Y^t}$ are both non-empty, thus
 $\Phi(X^s)=\subt{\kw{glb}(\ub{X^s})}{\kw{glb}(\ub{Y^t})}=\Phi(Y^t)$ since
$\ub{X^s}\subseteq\ub{Y^t}$.
\end{itemize}
\item If $C$ is of the form $\subt{X^s}{Y^t\backslash\{a_1,\ldots,a_n\}}$, then
$\toprow\in\ub{X^s}$ and $\toprow\in\ub{Y^t}$, i.e., $\Phi(X^s)$ and $\Phi(Y^t)$ are rows.
This is because such a constraint is always generated along with a constraint
$\subt{\sw Y}{\rowtype{a_1:X_1,\ldots,a_n:X_n}}$ (see Rule [C-ObjLit]),
and by Rules \ref{it:wf}, \ref{it:subt}, \ref{it:top}
and \ref{it:bot}.
But by \autoref{it:subtminus},
$\{\rowtype F\backslash\{a_1,\ldots,a_n\}\ |\ \rowtype F\in\ub{Y^t}\}\subseteq\ub{\svar X}$,
therefore
$\Phi(\svar X)=\subt{\kw{glb}(\ub{\svar X})}{\kw{glb}(\ub{Y^t})\backslash\{a_1,\ldots,a_n\}}
=\Phi(\svar Y)\backslash\{a_1,\ldots,a_n\}$.
\item If $C$ is of the form $\protocon X$, then $\toprow\in\ub{\sr X}$ since $\protocon X$
is always generated along with a row constraint on $\sr X$ (rules [C-ObjEmp] and [C-ObjLit];
also true if $\protocon X$ was generated using \autoref{it:proto}, by induction on the
number of uses of \autoref{it:proto}).
At type ascription (Lines~\ref{as:object} and \ref{as:ascribe-proto}), $\Phi(X)$ is a prototypal type.
\item If $C$ is of the form $\concrete X$, then $\toprow\in\ub{\sr X}$ since $\protocon X$
is always generated along with a row constraint on $\sr X$ (rules [C-MethApp];
also true if $\protocon X$ was generated using \autoref{it:conc}, by induction on the
number of uses of \autoref{it:conc}).
\begin{itemize}
\item if $\protocon X\cond$, then by \autoref{it:protoconc} and the case $\subt{\svar X}{Y^t}$,
$\subt{\Phi(\sr X)}{\Phi(\smr X)}$ and $\subt{\Phi(\sw X)}{\Phi(\smw X)}$, therefore
$\Phi(X)=\objt{\Phi(\sr{X})}{\Phi(\sw{X})}^\prototypal{\Phi(\smr{X})}{\Phi(\smw{X})}$
is concrete (Lines~\ref{as:object} and \ref{as:ascribe-proto}).
\item otherwise $\Phi(X)$ is of the form $\rho^\notprotoconc$, thus concrete
(Lines~\ref{as:object} and \ref{as:nc}).
\end{itemize}
\item if $C$ is of the form $\strip X$, then $\ub{\sr X}$ contains a method type since
$\strip X$ is always generated along with a method constraint on $\sr X$. By type ascription (Lines \ref{as:strip} and \ref{as:unattached}), $X$ is assigned an attached method type (of the form
$\metht{\cdot}{\tau_1}{\tau_2}$, i.e., without a receiver type),
 and therefore $\Phi$ satisfies constraint $C$.
\item If $C$ is of the form \handleMethod{\X_b}{\X_f}{\X_v}, then if $\Phi(\X_v)$ is not
a method type, the result is trivial. Otherwise, by definition of ascription
(Line~\ref{as:method}), there must exist
an $\metht{\Xr}{\Y_1}{\Y_2}\in\ub{\sr X_v}$. Using \autoref{it:attach},
$\protocon{\X_b}\cond$, $\subt{\smr{X_b}}{\sr{T}}\cond$, $\subt{\smw{X_b}}{\sw{T}}\cond$,
and $\strip{\X_f}\cond$. Therefore, using the previous cases
of this lemma, the assignment $\Phi$ satisfies constraint $C$.
\item if $C$ is any acceptance criterion ($\notmethod X$ or $\notprotocon X$),
 we do an explicit check during ascription (Lines \ref{as:nsubt} and \ref{as:notproto}),
thereby ensuring that $\Phi$ satisfies constraint $C$.
\end{itemize}
\end{proof}

\subsubsection{Proof of \autoref{lem:subt}}
\begin{proof}
We proceed by case analysis on $\Phi(\sr X)$, $\Phi(\sr Y)$, and the
possible constraints on $X$ and $Y$. By \autoref{lem:ascription},
$\subt{\Phi(\sr X)}{\Phi(\sr Y)}$ and $\subt{\Phi(\sw X)}{\Phi(\sw Y)}$.

If $\Phi(\sr Y)$ is a base or function type, then
$\Phi(\sr X)=\Phi(\sr Y)$ since there is no subtyping between those types
and by rules  \ref{it:subt}, \ref{it:top} and \ref{it:bot}.
By definition of ascription $\Phi(X)=\Phi(Y)$.

If $\Phi(\sr Y)$ is a method type $\metht{\tau_r}{\tau_1}{\tau_2}$,
then by definition of ascription, as well
as Rules \ref{it:subt}, \ref{it:top} and \ref{it:bot}, $\Phi(\sr X)=\Phi(\sr Y)$.
Three cases arise.
\begin{itemize}
\item If $\strip X\cond$ then by \autoref{it:strip}, $\strip Y\cond$. Therefore
by definition of ascription (Line \ref{as:strip} and \ref{as:unattached}),
$\Phi(\X)=\metht{\cdot}{\tau_1}{\tau_2}=\Phi(\Y)$.
\item Otherwise, if $\strip Y\cond$, then by definition of ascription
(Lines \ref{as:strip} and \ref{as:unattached}),
$\Phi(\X)=
\subt{\metht{\tau_r}{\tau_1}{\tau_2}}{\metht{\cdot}{\tau_1}{\tau_2}}=\Phi(\Y)$ by
subtyping rule [S-Method] (\autoref{fig:subtyping}).
\item Otherwise, by definition of ascription (Line \ref{as:unattached}),
$\Phi(\X)=\metht{\tau_r}{\tau_1}{\tau_2}=\Phi(\Y)$.
\end{itemize}
Otherwise $\Phi(\sr Y)$ is a row and $\toprow\in\ub{\sr Y}$.
\begin{itemize}
\item If $\protocon{Y}\cond$ then by \autoref{it:proto}$,
\protocon{X}\cond$ and $\forall s, X^s\equiv Y^s\cond$, hence by \autoref{lem:ascription},
 $\forall s, \Phi(X^s)=\Phi(Y^s)$. Therefore by
definition of ascription (Lines~\ref{as:object} and \ref{as:ascribe-proto}), $\Phi(X)=\Phi(Y)$.
\item Otherwise, if $\concrete{Y}\cond$, then by \autoref{it:conc}, we get 
$\concrete{X}\cond$. By definition of ascription,
$\Phi(Y)=\objt{\Phi(\sr Y)}{\Phi(\sw Y)}^\notprotoconc$, and
\begin{itemize}
\item
either
$\Phi(X)=\objt{\Phi(\sr X)}{\Phi(\sw X)}^\notprotoconc$;
\item
or $\Phi(X)=\objt{\Phi(\sr X)}{\Phi(\sw X)}^\prototypal{\Phi(\smr X)}{\Phi(\smw X)}$, with
$\subt{\Phi(\sr X)}{\Phi(\smr X)}$ and
$\subt{\Phi(\sw X)}{\Phi(\smw X)}$
by \autoref{it:protoconc} and \autoref{lem:ascription}).
\end{itemize}
In both cases $\subt{\Phi(X)}{\Phi(Y)}$.
\item Otherwise we have $\Phi(Y)=\objt{\Phi(\sr Y)}{\Phi(\sw Y)}^\notprotoabs$, and also
$\subt{\Phi(X)}{\objt{\Phi(\sr X)}{\Phi(\sw X)}^\notprotoabs}$, therefore $\subt{\Phi(X)}{\Phi(Y)}$.
\end{itemize}
\end{proof}

\subsubsection{Proof of \autoref{lem:wf}}
\begin{proof}
Following the definition of $\wf\tau$ in \autoref{fig:subtyping},
if $\Phi(X)$ is a non-object type then the result is trivial.

\sloppypar
Otherwise,
$\Phi(X)$ is of the form $\objt{\sf r}{\sf w}^q$. By \autoref{it:wf},
$\subt{\sall{X}}{\subt{\sr{X}}{\sw{X}}}\cond$ and
$\subt{\sall{X}}{\subt{\smr{X}}{\smw{X}}}\cond$, therefore by
\autoref{lem:ascription} and ascription, there exist rows
$\Phi(\sall X)$, $\Phi(\sr X)$, $\Phi(\sw X)$, $\Phi(\smr X)$ and $\Phi(\smw X)$
such that $\subt{\Phi(\sall{X})}{\subt{\Phi(\sr{X})}{\Phi(\sw{X})}}$ and
$\subt{\Phi(\sall{X})}{\subt{\Phi(\smr{X})}{\Phi(\smw{X})}}$. By type ascription
(Line~\ref{as:object}), ${\sf r} = \subt{\Phi(\sr{X})}{\Phi(\sw{X})} = {\sf w}$.

Moreover, if $q$ is of the form $\prototypal{\sf mr}{\sf mw}$, then by type
ascription (Line~\ref{as:ascribe-proto}),
${\sf mr} = \subt{\Phi(\smr{X})}{\Phi(\smw{X})} = {\sf mw}$. Finally,
$\subt{\Phi(\sall X)}{\Phi(\sr X)}={\sf r}$, therefore $\forall a\in\dom{{\sf r}},
{\sf r}[a]=\Phi(\sall X)[a]$. Similarly,
$\subt{\Phi(\sall X)}{\Phi(\smr X)}={\sf mr}$, therefore $\forall a\in\dom{{\sf mr}},
{\sf mr}[a]=\Phi(\sall X)[a]$.
Therefore, $\forall a\in \dom{{\sf mr}}\cap\dom{{\sf r}}, {\sf mr}[a]=\Phi(\sall X)[a]={\sf r}[a]$.
\end{proof}

\subsubsection{Proof of \autoref{thm:soundness}}
\begin{proof}

The proof proceeds by induction on the structure of the constraint-generating relation.
For ease of reference, we label the hypotheses as follows:
\begin{itemize}

\item{H1:} $\gen{\Xr, \Gamma}{e}{\X}{\C}$
\item{H2:} $\satisfies{\Phi}{\C}$

\end{itemize}
The cases \textbf{\myrule{C-Var}} and
\textbf{\myrule{C-This}} follow directly from H1 and H2, using typing rules
{\myrule{T-Var}} and {\myrule{T-This}}, respectively.

\mypara{\myrule{C-Int}:} Since $\sr X\equiv \kw{int}\cond$, by definition of ascription
(Line \ref{as:base}), $\Phi(X)=\kw{int}$ and we conclude using {\myrule{T-Int}}.

\mypara{\myrule{C-ObjEmp}:} Similarly, since $\protocon X\cond$, $\sr X\equiv\emptyrow\cond$ and
$\sw X\equiv\emptyrow\cond$, by \autoref{it:wf} and defintion of ascription (Line \ref{as:proto}),
$\Phi(X)=\objt{\emptyrow}{\emptyrow}^\prototypal{\emptyrow}{\emptyrow}$ and we conclude
using {\myrule{T-ObjEmp}}.

\mypara{\myrule{C-Null}:} Since $\subt{\sw X}{\emptyrow}\cond$, by definition of ascription
(Lines \ref{as:object}--\ref{as:na}), $\Phi(X)$ is an object type and we conclude
using {\myrule{T-Null}}, and the subtyping rules of \autoref{fig:subtyping}.

\mypara{\myrule{C-VarDecl}:}  From the induction hypothesis, we have that
\begin{itemize}

\item $\types{\Phi(\Xr),\Phi(\override{\Gamma}{x}{\Phi(\X_1)})}{e_1}{\Phi(\Y_1)}$ 
\item $\types{\Phi(\Xr),\Phi(\override{\Gamma}{x}{\Phi(\X_1)})}{e_2}{\Phi(\X)}$.

\end{itemize}
Moreover, since $\subt{\sr\Y_1}{\sr\X_1}\cond$ and $\subt{\sw\Y_1}{\sw\X_1}\cond$,
by \autoref{lem:subt} we get $\subt{\Phi(\Y_1)}{\Phi(\X_1)}$; and \autoref{lem:wf}
ensures that $\wf\Phi(X_1)$
Hence, \myrule{T-VarDecl} types $e$ at $\Phi(\X)$.

\mypara{\myrule{C-VarUpd}, \myrule{C-FunDecl}, \myrule{C-MethDecl},
\myrule{C-Attr}:}
As with \myrule{C-VarDecl}, these cases follow from a straightforward
application of the induction hypothesis, as well as
Lemmas \ref{lem:subt} and \ref{lem:wf}. 

\mypara{\myrule{C-FunApp} and \myrule{C-MethApp}:} Follow from the hypotheses (including the
induction hypothesis) and the definition of the subtyping relation,
as well as Lemmas \ref{lem:subt} and \ref{lem:wf}.

\mypara{\myrule{C-AttrUpd}:} There are two cases for attribute update: the
expression being attached is or is not ascribed a detached method type. This two cases
correspond respectively to typing rules \myrule{T-AttrUpdM} and \myrule{T-AttrUpdV}.

More
precisely, from the hypothesis of \myrule{C-AttrUpd} and induction hypothesis,
we have that $\types{\Phi(\Xr),\Phi(\Gamma)}{e_1}{\Phi(\X_b)}$ and
$\types{\Phi(\Xr),\Phi(\Gamma)}{e_2}{\Phi(\X_v)}$. 
From the constraints and \autoref{lem:subt}, we have that $\subt{\Phi(\X_v)}{\Phi(\X_f)}$;
and  by \autoref{lem:ascription} and definition of type ascription on constraint
$\subt{\sw\X_b}{\rowtype{a:\X_f}}$, we get
$$\subt{\Phi(X_b)}{\objt{\rowtype{a:\Phi(\X_f)}}{\rowtype{a}}}^\notprotoabs$$
Note that 
$\Phi(\X_v)$ may or may not be a detached method type.

\mypara{Subcase: $\Phi(\X_v)$ is not a detached method type.}
By rule \myrule{T-AttrUpdV}, we get
$$\typesau{\Phi(\Xr),\Phi(\Gamma)}{\Phi(\X_b).a:=e}{\Phi(\X_v)}$$
then we can apply \myrule{T-AttrUpd} and conclude.

\mypara{Subcase: $\Phi(\X_v)$ is a detached method type.}
The constraint $\handleMethod{\X_b}{\X_f}{\X_v}$ is not trivially true anymore,
and some detached method type is in $\ub{\X_v}$, therefore $\Phi(\X_v)$ is of
the form $\metht{\objt{\sr T}{\sw T}^\notprotoconc}{\tau_1}{\tau_2}$ 
for some type variable $T$.
(every receiver type can only be concrete
and nonprototypal, as it can only be introduced by rules \myrule{C-MethDecl} or
\myrule{C-MethApp}).

Moreover, since $\subt{\Phi(\X_v)}{\Phi(\X_f)}\cond$ and
$\strip{\X_f}\cond$, by ascription $\Phi(\X_f)=\metht{\cdot}{\tau_1}{\tau_2}$. 
Condition $\protocon{\X_b}$ ensures
that $\Phi(\X_b)$ is of the form $\rho^\prototypal{\Phi(\smr\X_b)}{\Phi(\smw\X_b)}$.
Finally, conditions $\subt{\smr\X_b}{\sr T}$ and $\subt{\smw\X_b}{\sw T}$
ensure that $\subt{\Phi(\smr\X_b)}{\Phi(\sr \T)}$ and $\subt{\Phi(\smw\X_b)}{\Phi(\sw \T)}$.
By rule \myrule{T-AttrUpdM}, we now get 
$$\typesau{\Phi(\Xr),\Phi(\Gamma)}{\Phi(\X_b).a:=e}{\Phi(\X_v)}$$
then we can apply \myrule{T-AttrUpd} and conclude.

\mypara{\myrule{C-ObjLit}:} We have the following, which satisfy the hypotheses
of \myrule{T-ObjLit}.  To ease the burden of notation, we elide the $\Phi$
substitution over the following terms.
\begin{itemize}

\item $\types{\Phi(\Xr),\Phi(\Gamma)}{e_p}
{\objt{\Phi(\sr \X_p)}{\sw \X_p}^\prototypal{\smr \X_p}{\smw \X_p}}$ follows
from induction hypothesis and condition $\protocon{\X_p}$; \\

\item $\Phi(\X)=\objt{\Phi(\sr \X)}{\sw \X}^\prototypal{\smr \X}{\smw \X}$ and
$\wf\Phi(\X)$ follow from condition $\protocon{\X}$ and \autoref{lem:wf}; \\

\item $\Phi(\sw\X)=\rowtype{a_1:\Phi(\X_1),\ldots,a_n:\Phi(\X_n)}$ follows from
condition $\sw\X\equiv\rowtype{a_1:\X_1,\ldots,a_n:\X_n}$; \\

\item $\subt{\Phi(\smr\X)}{\Phi(\smr\X_p)}$ and $\subt{\Phi(\smw\X)}{\Phi(\smw\X_p)}$ follow from 
conditions $\subt{\smr\X}{\smr\X_p}$ and $\subt{\smw\X}{\smw\X_p}$; \\

\item $\Phi(\sr\X_p)\cup\Phi(\sw\X)=\Phi(\sr\X)$ follows from conditions
$\sw{\X} \equiv \rowtype{a_1: \X_1,\ldots,a_n: \X_n}$, 
$ \rowsub{\sr{\X_p}}{\rowexcl{\sr{\X}}{a_1,\ldots,a_n}}$
and 
$\rowsub{\sr{\X}}{\sr{\X_p}}$, as well 
as \autoref{it:wf} implying $\subt{\sr X}{\sw X}$; \\

\item for each $i\in 1..n$, $\subt{\Phi(\Y_i)}{\Phi(\X_i)}$ follows from 
conditions $\subt{\sr\Y_i}{\sr\X_i}$ and $\subt{\sw\Y_i}{\sw\Y_i}$ and \autoref{lem:subt}; \\

\item for each $i\in 1..n$, $\sw\X\equiv\rowtype{a_1:\X_1,\ldots,a_n:\X_n}$ is a stronger 
condition thatn $\subt{\sw\X}{\rowtype{a_i:\X_i}}$, therefore using an identical reasoning 
as for case \myrule{C-AttrUpd}, we conclude that
$$\typesau{\Phi(\Xr),\Phi(\Gamma)}{\Phi(X).a_i:=e_i}{\Phi(\Y_i)}$$
\end{itemize}
We can now apply rule \myrule{C-ObjLit} since all its premises are verified, and conclude.

\end{proof}

\subsection{Type ascription and type inference soundness for recursive types}
\label{app:ascription-recursive}
For clarity reasons, we did not consider recursive types in the main body of the 
paper. In this section we show how to extend the type ascription and soundness
proof for recursive types.

\begin{algorithm}
 \begin{algorithmic}[1]
 \Procedure{AscribeType}{$\X$}
\State Replace all instances of $X$ in all bounds by fresh $\alpha$ \label{asrec:repl}
\If{any variable $Y$ appears in a $\ub{\svar X}$ or $\lb{\svar X}$} 
\State {ascribe $Y$ first} \label{asrec:other-vars}
\EndIf
 \If{$\strip \X\in\mathcal C'$}{ strip receivers in $\ub{\svar X}$, $\lb{\svar X}$}\EndIf\label{asrec:strip}
 \ForEach {$\svar{X}$}
   \If{$\ub{\svar{X}} = \emptyset$} $\Phi(\svar{X}) \gets \defaultt$ \label{asrec:default}
   \Else
   \State $\Phi(\svar{X}) \gets {\sf glb}(\ub{\svar{X}})$ \Comment{Fails if no glb} \label{asrec:glb}
   \ForEach {$L \in \lb{\svar{X}}$}
     \If{$\nsubt{L}{\Phi(\svar{X})}$} fail \label{asrec:lb}
     \EndIf
   \EndFor
   \EndIf
 \EndFor
 \If{$\Phi(\sr{X}) = \intt \vee \Phi(\sr{X}) = \defaultt$}
   \State $\Phi(X) \gets \Phi(\sr{X})$ \label{asrec:base}
 \ElsIf{$\Phi(\sr{X})$ is method type} \label{asrec:method}
   \If{$\notmethod{\X} \in \mathcal{C'}$} fail \label{asrec:nsubt}
   \EndIf
   \State $\Phi(X) \gets \Phi(\sr{X})$ \label{asrec:unattached}
 \Else \Comment{$\Phi(\sr{X})$ must be a row}
 \State $\rho \gets \objt{\Phi(\sr{X})}{\Phi(\sw{X})}$ \label{asrec:object}
 \If{$\protocon{X} \in \mathcal{C'}$} \label{asrec:proto}
   \If{$\notprotocon{X} \in \mathcal{C'}$} fail \label{asrec:notproto}
   \EndIf
   \State $\Phi(X) \gets \rho^{\prototypal{\Phi(\smr{X})}{\Phi(\smw{X})}}$ \label{asrec:ascribe-proto}
 \ElsIf{$\concrete{X} \in \mathcal{C'}$}
   $\Phi(X) \gets \rho^\notprotoconc$ \label{asrec:nc}
 \Else\ $\Phi(X) \gets \rho^\notprotoabs$  \label{asrec:na}
 \EndIf
   \If{$\Phi(X)$ contains $\alpha$} 
   {$\Phi(X) \gets \mu\alpha.\Phi(X)$} \label{asrec:rec}
   \EndIf
 \EndIf
 \EndProcedure
 \end{algorithmic}
 \caption{Type ascription including recursive types.}
 \label{alg:type-ascription-rec}
\end{algorithm}

\autoref{alg:type-ascription-rec} shows how to extend \autoref{alg:type-ascription} to handle
recursive types. We added line \ref{asrec:repl} to introduce recursive type variables, and
line \ref{asrec:rec} to construct a recursive type if needed.

\paragraph{Soundness of type inference.}
Extending the proofs of \autoref{lem:ascription} and \autoref{lem:wf}
to recursive types is immediate.
For \autoref{lem:subt}, 
if $\Phi(\X)=\kw{default}$ or $\Phi(\Y)=\kw{default}$, the result is immediate.
Otherwise, since by equirecursivity when $\tau$ does not contain $\alpha$,
$\mu\alpha.\tau \equiv \tau$, we can suppose without loss of generality
that both $X$ and $Y$ are both assigned recursive types,
e.g., $\Phi(X)$ is of the form $\mu\alpha.\sigma$ and $\Phi(\Y)$ is of the form 
$\mu\beta.\tau$. Using the proof of  \autoref{lem:subt} for non-recursive types
and the definition of type ascription, we
can conclude that for all $\alpha,\beta$, $\subt\sigma\tau$. In particular with 
$\alpha\equiv\mu\alpha.\sigma$ and $\beta\equiv\mu\beta.\tau$, we get
$\subt{\sigma[\mu\alpha.\sigma/\alpha]}{\tau[\mu\beta.\tau/\beta]}$, then using 
equirecursivity, we conclude $\subt{\mu\alpha.\sigma}{\mu\beta.\tau}$.

Once those lemmas are proved, extending the proof of \autoref{thm:soundness} is immediate.

\subsection{Soundness of the extended type system}

In this section, we extend the language syntax and typing judgement to account
for run-time expressions, present a small-step operational semantics, and
introduce stores and store typings.  As all variables in JavaScript are
essentially mutable, the syntax and typing judgement in
Figures~\ref{fig:term_syntax} and~\ref{fig:typing} do not include reference
syntax; for clarity, we add it here.
Much of this approach was inspired by the type soundness proof in Appendix B
of~\cite{SJSTechReport}, although we chose a small-step semantics to better
account for both ``stuck'' behavior as well as non-terminating computation.

\begin{figure}
\[ \begin{array}{lrcl}

Expressions & e & ::= & \ldots \mid v \mid \deref{e} \\
Val & v & ::= & n \mid l \mid \mynull \mid \rErr \\
Loc & l & \in & SLoc \cup HLoc \\
\\
Store & \sigma & \in & Loc \rightarrow StoreVal \\
StoreVal & sv & \in & Val \cup Obj \cup Fun \\
\\
Obj & o & ::= & \rObj{am, r} \\
Fun & f & ::= & \rFun{x : \tau, e} \\
\\
Proto & r & ::= & l \mid \mynull \\
AttrMap & am & \in & Attr \rightarrow Val \\

\end{array} \]
\caption{Runtime components of~\cite{SJSTechReport}.}
\label{fig:runtime-components}
\end{figure}

\begin{figure*}
\begin{mathpar}
\fbox{$\steps{\sigma, e}{\sigma', e'}$}
\and
\myinferLeft{SS-VarUpd}{
} {
    \steps{\sigma, \ass{l}{v}}{\override{\sigma}{l}{v}, v}
}
\and
\myinferLeft{SS-LetVar}{
    l \in SLoc \setminus dom(\sigma)
} {
    \steps{\sigma, \letin{x : \tau}{v}{e}}
          {\override{\sigma}{l}{v}, \override{e}{x}{l}}
}
\and
\myinferLeft{SS-Fun}{
    l \in HLoc \setminus dom(\sigma)
} {
    \steps{\sigma, \fun{x : \tau}{e}}{\override{\sigma}{l}{\rFun{x : \tau, e}}, l}
}
\and
\myinferLeft{SS-Obj}{
    l \in HLoc \setminus dom(\sigma) \quad
    o = \rObj{\emptyset, \mynull}
} {
    \steps{\sigma, \obj{\cdot}}{\override{\sigma}{l}{o}, l}
}
\and
\myinferLeft{SS-Attr}{
    v = lookup(\sigma, \sigma(l), a)
} {
    \steps{\sigma, l.a}{\sigma, v}
}
\and
\myinferLeft{SS-AttrNull}{ }{
    \steps{\sigma, \mynull.a}{\sigma, \rErr}
}
\and
\myinferLeft{SS-AttrUpd}{
    \sigma(l) = \rObj{am, p} \quad
    a \in dom(am)
} {
    \steps{\sigma, \ass{l.a}{v}}{\override{\sigma}{l}{\rObj{\override{am}{a}{v}, p}}, v}
}
\and
\myinfer{SS-AttrUpdNull}{ }{
    \steps{\sigma, \ass{\mynull.a}{v}}{\sigma, \rErr}
}
\and
\myinferLeft{SS-MCall}{
    l_1 = lookup(\sigma, \sigma(l), a) \\
    \sigma(l_1) = \rFun{x : \tau, e} \\
    l_2 \in SLoc \setminus dom(\sigma)
} {
    \steps{\sigma, l.a(v)}{\override{\sigma}{l_2}{v}, \override{\override{e}{x}{l_2}}{\this}{l}}
}
\and
\myinferLeft{SS-MNull}{ }{
    \steps{\sigma, \mynull.a(v)}{\sigma, \rErr}
}
\and
\myinferLeft{SS-Proto}{
    l \in HLoc \setminus dom(\sigma) \quad
    o = \rObj{[a_1 \mapsto v_1, \ldots, a_n \mapsto v_n], l_p}
} {
    \steps{\sigma, \proto{\obj{a_1 : v_1, \ldots, a_n : v_n}}{l_p}}
          {\override{\sigma}{l}{o}, l}
}
\and
\myinferLeft{SS-ProtoNull}{
    l \in HLoc \setminus dom(\sigma) \\
    o = \rObj{[a_1 \mapsto v_1, \ldots, a_n \mapsto v_n], \mynull}
} {
    \steps{\sigma, \proto{\obj{a_1 : v_1, \ldots, a_n : v_n}}{\mynull}}
          {\override{\sigma}{l}{o}, l}
}
\and
\myinferLeft{SS-Context}{
    \steps{\sigma, e}{\sigma', e'}
} {
    \steps{\sigma, \econ{e}}{\sigma', \econ{e'}}
}
\\\\
\fbox{$v = lookup(\sigma, \rObj{am, r}, a)$}
\and
\myinferLeft{OL-Local}{
    a \in dom(am)
} {
    am(a) = lookup(\sigma, \rObj{am, r}, a)
}
\and
\myinferLeft{OL-Proto}{
    a \notin dom(am) \\
    v = lookup(\sigma, \sigma(l), a)
} {
    v = lookup(\sigma, \obj{am, l}, a)
}
\end{mathpar}
\caption{Substitution-based small-step operational semantics of~\cite{SJSTechReport}.}
\label{fig:small-step}
\end{figure*}

\begin{figure}
\[ \begin{array}{rcl}

\multicolumn{3}{l}{EvalCtx} \\

E & ::= & [\cdot] \mid x=E \mid            \letin{x : \tau}{E}{e} \\ 
& \mid & \proto{\obj{a_1 : v_1, \ldots, a_i : E, \ldots, a_n : e_n}}{e} \\

& \mid & \proto{\obj{a_1 : v_1, \ldots, a_n : v_n}}{E} \\ 
& \mid & E.a \mid E.a=e \mid v.a=E \\

& \mid & \rFun{x : \tau, E} 
\end{array} \]
\caption{Evaluation context for small-step semantics.}
\label{fig:eval_context}
\end{figure}

In Figure~\ref{fig:runtime-typing}, we extend the typing judgement with a store
typing $\Sigma$ in order to type the run-time components introduced in
Figure~\ref{fig:runtime-components}.

\begin{figure*}
\[ 
\begin{array}{lrcl}
\textit{types} & \tau & ::= & \ldots \mid \reft{\tau} \\
\textit{store typing} & \Sigma & \in & Loc \rightarrow \tau
\end{array}\]
\begin{mathpar}
\fbox{$\types{R, \Gamma, \Sigma}{e}{\tau}$} \and
\myinferLeft{T-Loc}{ }{ \types{R, \Gamma, \Sigma}{l}{\reft{\Sigma(l)}} }
\and
\myinferLeft{T-Deref}{
    \types{R, \Gamma, \Sigma}{e}{\reft{\tau}}
} {
    \types{R, \Gamma, \Sigma}{\deref{e}}{\tau}
}
\and
\myinferLeft{T-Int}{ }{ \types{R, \Gamma, \Sigma}{n}{\kw{int}} }
\and
\myinferLeft{T-Var}{ }{ \types{R, \Gamma, \Sigma}{x}{\reft{\Gamma(l)}} }
\and
\myinferLeft{T-VarUpd} {
    \types{R, \Gamma, \Sigma}{e_1}{\reft{\tau}} \\
    \types{R, \Gamma, \Sigma}{e_2}{\tau}
} {
    \types{R, \Gamma, \Sigma}{\ass{e_1}{e_2}}{\tau}
}
\and
\myinferLeft{T-VarDecl}{
  \types{R, \override{\Gamma}{x}{\tau_1}}{e_1}{\sigma_1} \\
  \wf{\tau_1} \\
  \subt{\sigma_1}{\tau_1} \\
  \types{R, \override{\Gamma}{x}{\tau_1}}{e_2}{\tau}
} {
  \types{R, \Gamma, \Sigma}{\letin{x}{e_1}{e_2}}{\tau}
}
\and
\myinferLeft{T-This}{
} {
  \types{\nu, \Gamma, \Sigma}{\this}{\nu}
}
\and
\myinferLeft{T-Meth}{
  \types{\nu, \override{\Gamma}{x}{\tau_1}, \Sigma}{e}{\tau_2} \\
  \wf{\tau_1}
} {
  \types{R, \Gamma, \Sigma}{\fun{x:\tau_1}{e}}{\reft{\metht{\nu}{\tau_1}{\tau_2}}}
}
\and
\myinferLeft{T-Null}{ \subt{\tau}{\objt{\emptyrow}{\emptyrow}^\notprotoabs}}{ \types{R, \Gamma, \Sigma}{\mynull}{\reft{\tau}} }
\and
\myinferLeft{T-Attr}{
  \subt{\tau_1}{\objt{\rowtype{\f:\tau}}{\emptyrow}^\notprotoabs} \\
  \types{R, \Gamma, \Sigma}{e}{\reft{\tau_1}} \\
  \tau \neq \reft{\metht{\cdot}{\tau_2}{\tau_3}}
} {
  \types{R, \Gamma, \Sigma}{e.\f}{\tau}
}
\and
\myinferLeft{T-AttrUpd}{
  \typesau{R, \Gamma, \Sigma}{\tau_b.\f:=e}{\tau} \\
  \types{R, \Gamma, \Sigma}{e_1}{\reft{\tau_b}}
} {
  \types{R, \Gamma, \Sigma}{\ass{e_1.\f}{e}}{\tau}
}
\and
\myinferLeft{T-MCall}{
  \types{R, \Gamma, \Sigma}{e_1}{\reft{\tau_1}} \\
  \types{R, \Gamma, \Sigma}{e_2}{\tau_2} \\
  \subt{\tau_1}{\objt{\rowtype{\f:\reft{\metht{\cdot}{\tau_3}{\tau}}}}{\emptyrow}^\notprotoconc} \\
  \subt{\tau_2}{\tau_3}
} {
  \types{R, \Gamma, \Sigma}{\app{e_1.\f}{e_2}}{\tau}
}
\and
\myinferLeft{T-Emp}{
} {
  \types{\R, \Gamma, \Sigma}{\obj{\cdot}}{\reft{\objt{\emptyrow}{\emptyrow}^\prototypal{\emptyrow}{\emptyrow}}}
}
\and
\myinferLeft{T-RTObj} {
    \types{\R, \Gamma, \Sigma}{\proto{\obj{a_1 : v_1, \ldots, a_n : v_n}_{\tau}}{v}}{\reft{\tau}}
} {
    \types{\R, \Gamma, \Sigma}{\rObj{[a_1 \mapsto v_1, \ldots, a_n \mapsto v_n ], v}}{\tau}
}
\and
\myinferLeft{T-RTMeth} {
    \types{\R, \Gamma, \Sigma}{\fun{x : \tau_1}{e}}{\reft{\tau}}
} {
    \types{\R, \Gamma, \Sigma}{\rFun{x : \tau_1, e}}{\tau}
}
\and
\myinferLeft{T-ObjLit}{
  \wf{\rho^q} \\
  \rho = \objt{\meta{r}}{\meta{w}} \\
  \qual = \prototypal{\meta{mr}}{\meta{mw}} \\
  \types{\R, \Gamma, \Sigma}{e_p}{{\reft{\objt{\meta{r_p}}{\meta{w_p}}}^{\prototypal{\meta{mr_p}}{\meta{mw_p}}}}}
    \\
  \meta{w} = \rowtype{\f_1:\tau_1,\ldots,\f_n:\tau_n} \\
  \forall i \in 1..n.~\typesau{\R, \Gamma, \Sigma}{\ass{\rho^\qual.\f_i}{e_i}}{\sigma_i} \wedge
  \subt{\sigma_i}{\tau_i} \\
  {\meta{r_p}\cup\meta{w}}={\meta{r}} \\
  \subt{\meta{mr}}{\meta{mr_p}} \\
  \subt{\meta{mw}}{\meta{mw_p}}
} {
  \types{\R, \Gamma, \Sigma}{\proto{\obj{\f_1 : e_1, \ldots, \f_n : e_n}_{\rho^\qual}}{e_p}}{\reft{\rho^\qual}} \\
}
\and
\myinferLeft{T-AttrUpdV}{
  \subt{\tau_b}{\objt{\rowtype{\f: \tau_f}}{\rowtype{\f}}^\notprotoabs} \\
  \types{\R, \Gamma, \Sigma}{e}{\tau} \\
  \tau \neq \reft{\metht{\tau_x}{\tau_y}{\tau_z}} \\
  \subt{\tau}{\tau_f}
} {
  \typesau{\R, \Gamma, \Sigma}{\tau_b.\f := e}{\tau}
}
\and
\myinferLeft{T-AttrUpdM}{
\subt{\tau_b}{\objt{\rowtype{\f :
\reft{\metht{\cdot}{\tau_1}{\tau_2}}}}{\rowtype{\f}}^\notprotoabs} \\
  \types{\R, \Gamma,
  \Sigma}{e}{\reft{\metht{\objt{\meta{r_r}}{\meta{w_r}}^{\notprotoconc}}{\tau_1}{\tau_2}}} \\
  \tau_b = \rho^\qual \quad \qual = \prototypal{\meta{mr}}{\meta{mw}} \\
  \subt{\meta{mr}}{\meta{r_r}} \quad \subt{\meta{mw}}{\meta{w_r}}
} {
  \typesau{\R, \Gamma, \Sigma}{\tau_b.\f := e}{\reft{\metht{\objt{\meta{r_r}}{\meta{w_r}}^{\notprotoconc}}{\tau_1}{\tau_2}}}
}
\\\\
\fbox{$\types{~}{\sigma}{\Sigma}$}
\and
\myinferLeft{S-Types} {
    \forall l \in dom(\sigma).~\types{\emptyset, \emptyset, \Sigma}{\sigma(l)}{\Sigma(l)}
} {
    \types{~}{\sigma}{\Sigma}
}
\end{mathpar}
\caption{The typing judgement of Figure~\ref{fig:typing} extended with run-time expressions and explicit references.}
\label{fig:runtime-typing}
\end{figure*}

\begin{lemma}[Canonical forms]
\label{lem:canonical_forms}
The following hold:

    \begin{enumerate}

    \item If $v$ is a value of type $\intt$, then $v = n$.
    \item If $v$ is a value of type $\reft{\tau}$, then $v = l$ or $v = \mynull$.

    \end{enumerate}

\begin{proof}

    For the first part, according to the grammar in
    Figure~\ref{fig:runtime-components}, values may either be $n, l, \mynull$,
    or $\rErr$.  The desired result follows immediately from the first case.
    The second and third case cannot occur, as shown by inversion of the typing
    relation.  The final case also cannot occur, as $\rErr$ is not well typed.
    The second part is similar.

\end{proof}
\end{lemma}

\begin{lemma}[Substitution]
\label{lem:substitution}
The following hold:

\begin{enumerate}

\item If $\sigma$ is well typed at $\Sigma$, $l$ at $\tau$, and $v$ at $\tau$,
then $\override{\sigma}{l}{\tau}$ is well typed at $\Sigma$.

\item If $\types{\R, (\Gamma, x : \tau_x), \Sigma}{e}{\tau}$ and $\types{\R,
\Gamma, \Sigma}{v}{\tau_v}$, and $\subt{\tau_v}{\tau_x}$, then $\types{\R, \Gamma,
\Sigma}{\override{e}{x}{v}}{\tau}$.

\item If $\types{\nu, \Gamma, \Sigma}{e}{\tau}$ and $\types{\cdot, \Gamma,
\Sigma}{e_\nu}{\tau_\nu}$, then $\types{\cdot, \Gamma,
\Sigma}{\override{e}{\this}{e_\nu}}{\tau}$.

\end{enumerate}

\begin{proof}

The first result follows from the definition of $\types{~}{\sigma}{\Sigma}$ and
the hypotheses.  The second and third proceed by straightforward induction on
the typing relation.

\end{proof}
\end{lemma}

\begin{lemma}[Field lookup]
\label{lem:field_lookup}
If $\rObj{am, v}$ is well typed at $\tau = \objt{r}{w}^k$, $\tau$ is well
formed, and $\types{}{\sigma}{\Sigma}$, then

\begin{enumerate}

\item If $\subt{\tau}{\objt{a : \tau_a}{\cdot}^k}$, then $lookup(\sigma,
\rObj{am, v}, a)$ exists and is well typed as a subtype of $\tau_a$.

\item If $\subt{\tau}{\objt{\cdot}{a : \tau_a}^k}$, then $lookup(\sigma,
\rObj{am, v}, a) = am(a)$ and is well typed as a subtype of $\tau_a$.

\end{enumerate}

\begin{proof}

The proof goes by induction on the structure of the typing relation.  If $a \in
\dom{w}$, then inverting the typing relation shows that $am(a)$ exists and is
well typed as a subtype of $w[a]$.  Otherwise, $a \in \dom{r}$, and, again
inverting the typing judgement, we have that $v = l$ and $\sigma(l) =
\rObj{am_p, v_p}$, which is well typed at $\objt{r_p}{w_p}^{k_p}$, and the
result follows from the IH.

\end{proof}
\end{lemma}

\begin{lemma}[Method call]
\label{lem:method_call}

If
\begin{itemize}

\item{H1:} $\types{~}{\sigma}{\Sigma}$, and
\item{H2:} $l$ is well typed at $\reft{\tau_1}$, and
\item{H3:} $\subt{\tau_1}{\objt{a : \reft{\metht{\cdot}{\tau_2}{\tau_3}}}{a}^\notprotoconc}$,

\end{itemize}
then
\begin{itemize}

\item{G1:} $lookup(\sigma, \sigma(l), a) = l'$, and

\item{G2:} $\sigma(l') = \rFun{x : \tau_2, e}$, and

\item{G3:} $\types{\R, \Gamma, \Sigma}{\rFun{x : \tau_2,
e}}{\metht{\nu}{\tau_2}{\tau_3}}$, and

\item{G4:} $\subt{\tau_1}{\nu}$.

\end{itemize}

\begin{proof}

\begin{enumerate}

\item From inverting (H3), we have that $\tau_1$ is an object type.  

\item With that, inverting (H2) and (H1) shows that $\Sigma(l) = \tau_1 =
\objt{r}{w}^\prototypal{mr}{mw}$.  

\item \label{itm:rw_sub_mrmw} With this new information, inverting (H3) via
\myrule{S-ProtoConc} shows that $\subt{r}{mr}$ and $\subt{w}{mw}$

\item Applying Lemma~\ref{lem:field_lookup} yields (G1), and $l'$ is well typed
at a subtype of $\reft{\metht{\cdot}{\tau_2}{\tau_3}}$.

\item This implies (G2) via inverting the subtyping relation and the typing
relation.

\item (H1) and (G2) together imply (G3).

\item Inverting the typing relation in (2) leads back to
\myrule{T-ObjLit}, which shows that $\typesau{\R, \Gamma,
\Sigma}{\ass{\tau_1.a}{l'}}{\tau_a}$ and $\subt{\tau_a}{\tau_1}$.

\item \label{itm:mrmw_sub_rw} Working backwards to \myrule{T-AttrUpdM} shows that $\nu =
\objt{r_r}{w_r}^\notprotoconc$ and $\subt{mr}{r_r}$ and $\subt{mw}{w_r}$.

\item (\ref{itm:rw_sub_mrmw}) and (\ref{itm:mrmw_sub_rw}) show that
$\subt{r}{r_r}$ and $\subt{w}{w_r}$.

\item Hence, by \myrule{S-ProtoConc}, $\subt{\tau_1}{\nu}$, satisfying (G4).

\end{enumerate}

\end{proof}
\end{lemma}

\begin{theorem}[Progress]

    For all receiver typing contexts $R$, typing contexts $\Gamma$, store types
    $\Sigma$, types $\tau$, 
        stores $\sigma_1$, and closed expressions $e_1$, if
        \begin{enumerate}
    
    \item $\types{~}{\sigma_1}{\Sigma}$
    \item $\types{R, \Gamma, \Sigma}{e_1}{\tau}$
    
    \end{enumerate}
        then either $e_1$ is a value or
    $\exists \sigma_2, e_2.~\steps{\sigma_1, e_1}{\sigma_2, e_2}$.

\begin{proof}

The proof proceeds by induction on the structure of the typing relation.  We
show case for method application; other cases are either immediate or similar.

\mypara{Method call (\myrule{T-Meth}).}
From the second hypothesis, we have that $\types{\R, \Gamma,
\Sigma}{e_1.a(e_2)}{\tau}$.  By inversion of the typing relation, it follows
that:
\begin{itemize}

\item{H1:} $\types{\R, \Gamma, \Sigma}{e_1}{\reft{\tau_1}}$
\item{H2:} $\types{\R, \Gamma, \Sigma}{e_2}{\tau_2}$
\item{H3:} $\subt{\tau_1}{\objt{\rowtype{\f:\reft{\metht{\cdot}{\tau_3}{\tau}}}}{\emptyrow}^\notprotoconc}$
\item{H4:} $\subt{\tau_2}{\tau_3}$

\end{itemize}

The remainder of this case is guided by the structure of $e_1$ and $e_2$.  If
$e_1$ or $e_2$ are expressions, then the goal follows from the IH and
\myrule{SS-Context}.  If $e_1$ is a value, then it follows from the canonical
forms lemma (Lemma~\ref{lem:canonical_forms}) and (H4) that $e_1 = l$ or $e_1 =
\mynull$.  For the latter, \myrule{SS-MNull} applies.  For the former, our goal
follows from \myrule{SS-MCall} if we can show the following:
\begin{itemize}

\item{G1:} $l_1 = lookup(\sigma, \sigma(l), a)$
\item{G2:} $\sigma(l_1) = \rFun{x : \tau, e}$

\end{itemize}
From the hypotheses, we have that $\types{\R, \Gamma,
\Sigma}{\sigma(l)}{\tau_1}$.  (H3) and inversion of the subtyping relation
show that $\tau_1 = \objt{r_1}{w_1}^k$ and $r_1[a] \equiv
\reft{\metht{\cdot}{\tau_3}{\tau}}$.  Hence, $\sigma(l) = \rObj{am, r}$ and
$am(a) = l_1$ (by inversion of the typing relation on $\sigma(l)$), and
$lookup(\sigma, \sigma(l), a) = l_1$ (G1) by \myrule{OL-Local}, and $\types{\R,
\Gamma, \Sigma}{\sigma(l_1)}{\tau_a}$ and
$\subt{\tau_a}{\reft{\metht{\cdot}{\tau_3}{\tau}}}$.  Hence, (G2) follows from
inversion of the typing relation (on $\sigma(l_1)$) and the subtyping relation,
which establishes that dereferencing a location typed at a subtype of an erased
method type yields a function closure.

\end{proof}
\end{theorem}

\begin{theorem}[Preservation]

    For all receiver types $R$, typing contexts $\Gamma$, store types $\Sigma$,
    types $\tau$,
        expressions $e_1$ and $e_2$, and stores $\sigma_1$ and $\sigma_2$,
        if
        \begin{itemize}
    
    \item{H1:} $e_1$ is closed,
    \item{H2:} $\types{R, \Gamma, \Sigma}{e_1}{\tau}$,
    \item{H3:} $\types{}{\sigma_1}{\Sigma}$, and
    \item{H4:} $\steps{\sigma_1, e_1}{\sigma_2, e_2}$,

    \end{itemize}
        then there either exist $R_2, \Gamma_2, \Sigma_2$ such that
        \begin{itemize}

    \item{G1:} $\types{}{\sigma_2}{\Sigma_2}$
    \item{G2:} $\types{R_2, \Gamma_2, \Sigma_2}{e_2}{\tau}$
    \item{G3:} $e_2$ is closed

    \end{itemize}
        or $e_2 = \rErr$.

\begin{proof}

The proof proceeds by induction on the structure of the small-step relation.

\mypara{\myrule{SS-VarUpd}}
We have that $e_1 = \ass{l}{v}$.  Inverting (H2) shows that $l$ is well typed
at $\reft{\tau}$ and $v$ at $\tau$.  (G1) follows from the substitution lemma
(Lemma~\ref{lem:substitution}).  (G2) and (G3) are immediate.

\mypara{\myrule{SS-LetVar}}
We have that $e_1 = \letin{x : \tau}{v}{e}$.  For (G1), we take $\Sigma_2 =
\override{\Sigma_1}{l}{\tau}$, and the result follows from inverting (H2) and
Lemma~\ref{lem:substitution} (substitution).  For (G2), we take $e_2 =
\override{e}{l}{v}$, and the result follows from (H2) and
Lemma~\ref{lem:substitution} (substitution).  (G3) is immediate.

\mypara{\myrule{SS-Fun}}
We have that $e_1 = \fun{x : \tau_x}{e}$.  Inverting (G2) and applying
\myrule{T-RTMeth} shows that $\rFun{x : \tau_x, e}$ is well typed at $\tau =
\reft{\tau'}$.  It then follows from \myrule{T-Deref} that $l$ is well typed at
$\tau$.  For (G1), we take $\Sigma_2 = \override{\Sigma_1}{l}{\rFun{x : \tau,
e}}$ and the result follows from Lemma~\ref{lem:substitution} (substitution).
For (G2) and (G3), the result is immediate.

\mypara{\myrule{SS-Obj}}
We have that $e_1 = \obj{\cdot}$.  The result is immediate.

\mypara{\myrule{SS-Attr}}
We have that $e_1 = l.a$.  The result follows from inverting (H2) and
Lemma~\ref{lem:field_lookup}.

\mypara{\myrule{SS-AttrNull}}
We have that $e_1 = \mynull.a$ and the result is immediate.

\mypara{\myrule{SS-AttrUpd}}
We have that $e_1 = \ass{l.a}{v}$ and $\sigma(l) = \rObj{am, p}$.  Inverting
(H2) yields two cases: method and non-method update.
\begin{enumerate}

\item In a non-method update, (H2) inverts via \myrule{T-AttrUpd} with
\myrule{T-AttrUpdV}: $l$ is well typed at $\subt{\tau_b}{\objt{a :
\tau_f}{a}^\notprotoabs}$ and $\subt{\tau}{\tau_f}$.  (G1) follows from
Lemma~\ref{lem:substitution} (substitution), with \myrule{T-RTObj} used to type
the run-time object.  (G2) and (G3) are immediate.

\item In a method update, (H2) inverts via \myrule{T-AttrUpd} and
\myrule{T-AttrUpdM}.  It follows from (H1) that $\rObj{am, p}$ is well typed by
\myrule{T-RTObj} (and, transitively, by \myrule{T-ObjLit}).  Combined with
\myrule{T-AttrUpdM} allows us to also type $\rObj{\override{am}{a}{v}, p}$,
which satisfies (G1).  (G2) and (G3) are immediate.

\end{enumerate}

\mypara{\myrule{SS-AttrUpdNull}}
We have that $e_1 = \ass{\mynull.a}{v}$, and the result is immediate.

\mypara{\myrule{SS-MCall}}
We have that $e_1 = l.a(v)$, $l_1 = lookup(\sigma, \sigma(l), a)$, and
$\sigma(l_1) = \rFun{x : \tau_x, e}$.
\begin{enumerate}

\item Inverting (H2) allows us to apply Lemma~\ref{lem:field_lookup}, which
establishes that $\rFun{x : \tau_x, e}$ is well typed as a subtype of
$\metht{\cdot}{\tau_3}{\tau}$, and inverting the subtyping relation shows that
$\tau_x = \tau_3$.

\item (G1) follows from Lemma~\ref{lem:substitution} (substitution).

\item Typing $\override{e}{x}{l_2}$ follows from (H2) and (2) and
Lemma~\ref{lem:substitution} (substitution).

\item Typing $\override{\override{e}{x}{l_2}}{\this}{l}$, and hence (G2),
follows from Lemma~\ref{lem:method_call} and Lemma~\ref{lem:substitution}
(substitution).

\item (G3) is immediate.

\end{enumerate}

\mypara{\myrule{SS-MNull}}
We have $e_1 = \mynull.a(v)$ and the result is immediate.

\mypara{\myrule{SS-Proto}}
We have $e_1 = \proto{\obj{a_1 : v_1, \ldots, a_n : v_n}}{l_p}$, and the result
follows from Lemma~\ref{lem:substitution} (substitution).

\mypara{\myrule{SS-ProtoNull}}
Similar to \myrule{SS-Proto}.

\mypara{\myrule{SS-Context}}
We have $e_1 = E[e]$ and the result follows from application of the induction
hypothesis.

\end{proof}
\end{theorem}

\makeatletter{}

\section{Performance data on Octane}\label{sec:performance}

\autoref{fig:crossovers} gives complete
performance data---space consumption and running time, respectively---for the six Octane programs we studied.

\begin{figure}
\includegraphics[trim=0 0 0 0,clip,width=\textwidth]{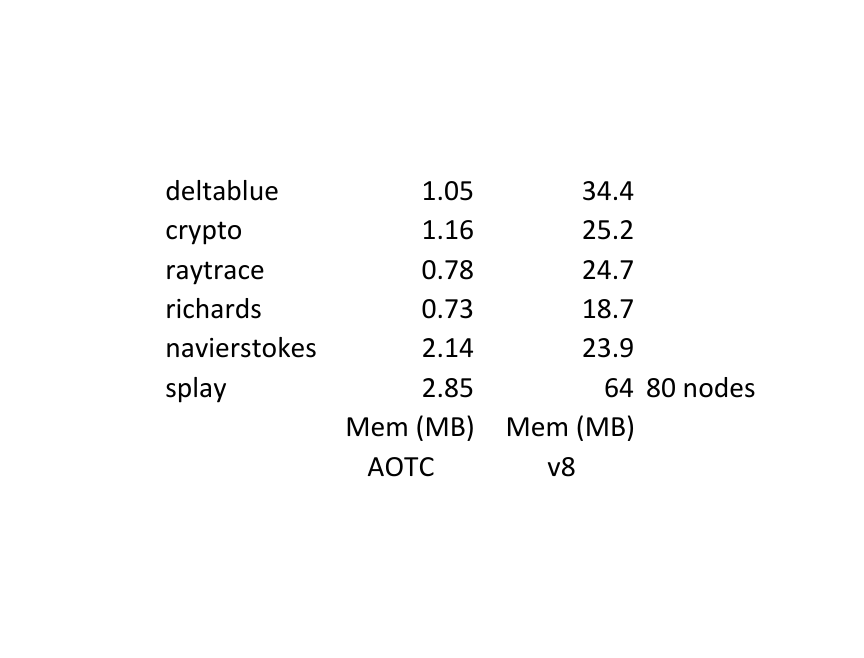}
\caption{Space usage of our AOTC system vs. the v8 runtime.} \label{fig:space}
\end{figure}

\begin{figure}
 \includegraphics[trim=0 0 0 0,clip,width=\textwidth]{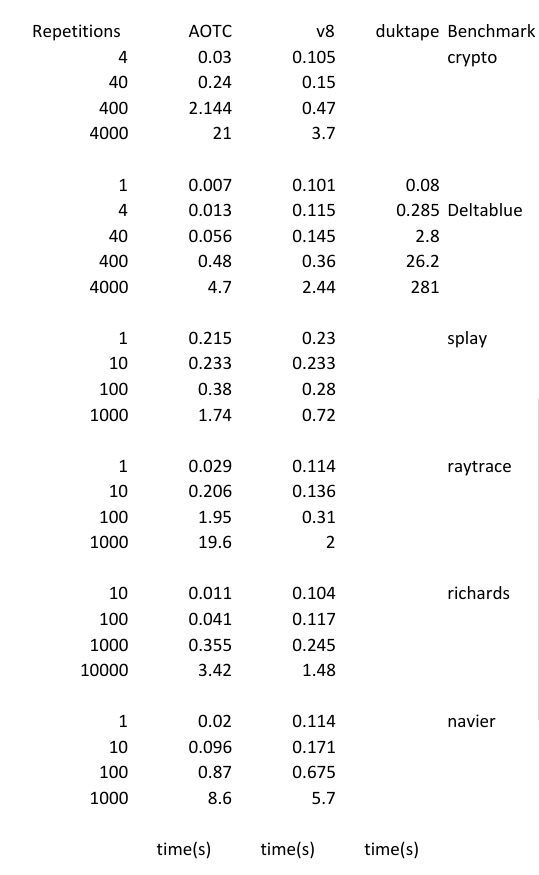}
 \caption{Time and crossover behavior of our AOTC system vs. the v8 runtime. We ran duktape only on Deltablue benchmark.}
\label{fig:crossovers}
\end{figure}

\fi
\end{document}